\documentclass[]{elsarticle}

\usepackage{amsmath,amssymb,graphicx,color,array,amsfonts}

\usepackage{epsfig}
\usepackage{url}

\begin{document}

\begin{frontmatter}

\title{Tunable transport with broken space-time symmetries}
\author{Sergey Denisov$^{a,b*}$}
\author{Sergej Flach$^c$}
\author{Peter H\"{a}nggi$^{b,d,e}$}
\address{$^a$ Sumy State University, Rimsky-Korsakov Street 2, 40007 Sumy, Ukraine}
\address{$^b$ Institut f\"ur Physik, Universit\"at Augsburg,
Univerit\"atsstr.1, 86135 Augsburg, Germany}
\address{$^c$ New Zealand Institute for Advanced Study, Centre for Theoretical Chemistry and Physics,
Massey University, Private Bag 102 904 NSMCS, 0746 Auckland, New Zealand}
\address{$^d$Center for Phononics and Thermal Energy Science, School of Physics Science and Engineering,
Tongji University, 200092 Shanghai, China}
\address{$e$ Nanosystems Initiative Munich, Schellingstr. 4, D-80799 M\"{u}nchen, Germany}
\cortext[cor1]{Corresponding author. Tel.: +49-821-598-3228 \\
E-mail address: sergey.denisov@physik.uni-augsburg.de}

\begin{abstract}
Transport properties of particles and waves in spatially periodic structures that are driven by external
time-dependent forces
manifestly depend on the space-time symmetries of the corresponding equations of motion.
A systematic  analysis of these symmetries
uncovers the conditions necessary for obtaining directed transport. In this work we give a unified introduction into
the symmetry analysis and demonstrate its action on the motion in one-dimensional periodic, both in time and space, potentials.
We further generalize the analysis to quasi-periodic drives, higher space dimensions,
and quantum dynamics.
Recent experimental results on the transport of cold and ultracold atomic ensembles in ac-driven
optical potentials are reviewed as illustrations of theoretical considerations.
\end{abstract}

\vskip 1cm
\end{frontmatter}

\tableofcontents
\newpage

\section{Introduction}
\label{sec1}

The last decade has witnessed a large number of studies on nonequilibrium,
fluctuation-induced transport in  potentials and confinements of different shapes
and geometries, where ubiquitous fluctuations  included colored or
idealized white noise or/and regular time-periodic components
\cite{HB_LNP1996,astumian1997,astumian2002,r02pr}. The main subject of these studies, the so-called {\it ratchet} effect,
occurs due to the violation of the thermal equilibrium  so that
the Second Law of Thermodynamics no longer applies. With no restriction imposed by the Second Law
there are no constraints on the appearance
of a steady transport, even in the absence of constant forces or gradients.
The issue was recognized  already back in 1912,
when Marian von Smoluchowski posed the question concerning the (im)possibility to extract  work
from a ratchet-toothed wheel placed into a heat bath \cite{s12pz}. Fifty ears latter,
it had been revived and popularized by Richard Feynman \cite{feynman1970}, and since the late 1990s  the studies of the ratchet phenomenon proceeded
along many research tracks in  statistical mechanics, condensed matter, chemistry and  biophysics \cite{astumian2002,r02pr,nori2005,hm09rmd}.

The majority of  ratchet studies typically focuses on  systems acting in
a noisy and viscous environment, so that the resulting dynamics emerges  to be essentially stochastic and
dissipative. The corresponding evolution, being hampered by the thermal noise and/or other sorts of random nonequilibrium fluctuations,
is characterized by  space-time correlations that are restricted to short ranges. This extends also to quantum ratchets, see
Refs.~\cite{ggh98,gh98,gh2001,linke,kharpai}.
A large variety of strongly dissipative  stochastic models was put forward in order to describe intracellular transport
and to serve as  blue-prints for synthetic molecular motors,  microfluidic pumps, and colloidal separators.
Most of them can be unified under the  name `\textit{Brownian motors}` \cite{astumian2002,r02pr,nori2005}.
A recent comprehensive review  by one of the authors \cite{hm09rmd} provides an up-do-date  information on
such Brownian machinery.

Fast progress in experimental manipulations with cold- and ultra-cold atoms \cite{Grynberg2001,Bloch2008}
has created a new testing ground for the ratchet concept.
The corresponding physical systems, ensembles of atoms kept in optical or
magnetic confinements \cite{Morsch2006,Windpassinger2013},  evolve without being subjected to strong external noise.
The  resulting dynamics  appears to be coherent on time  scales much larger than the typical time scale set by the characteristic frequency
of the potential.
As it is known from the theory \cite{gutz}, the evolution of a fully coherent system may be governed  by several
co-existing invariant manifolds, e.g. multiple attractors
in the limit of weak dissipation \cite{Ott1992, suz92}, regular and chaotic regions in the
classical Hamiltonian limit \cite{suz92} and  eigenstates in the case of
quantum  evolution \cite{Stockmann1999}.  Asymptotic  regimes
appear as interference patterns  to
which different invariant manifolds contribute simultaneously. Even the presence of weak  fluctuations
cannot erase the interference effects completely \cite{risken1989}.
The functioning of Hamiltonian and weakly dissipative ratchets, therefore, is essentially different from that of
Brownian motors, in which the presence
of strong ambient noise induces an effective averaging over the system phase space and erases the memory on initial states, or,
more generally, on initial preparations \cite{HB_LNP1996,r02pr}.

Various interpretations of the ratchet effect have been introduced, including
Maxwell daemons, harmonic mixing,  nonlinear response, to name but a few among others  \cite{HB_LNP1996,astumian1997,r02pr,astumian2002,hm09rmd}.
However, often evaluation of a particular ratchet system does not allow for an in-depth analytical treatment
but calls for direct numerical simulations \cite{yfzo01epl,dhm2009}.
The focus of this review,  the \textit{symmetry analysis} \cite{fyz00prl,dfoyz02pre},
allows to predict and control the rectification outcome avoiding  computational studies of the system dynamics.
The latter can become rather cumbersome when chaotic motion, and thus a  sensitivity to initial conditions, are at work.
The analysis can
be performed systematically on various levels
of description, ranging from  microscopic equations of motion to kinetic equations governing
the evolution of  probability distributions.
To be more specific, the symmetry analysis does produce a list of symmetries which
\textit{prevent} rectification.
The corresponding symmetries include operations on the system variables and time.
A proper choice of control parameters,
especially of the  driving field, leads to the destruction  of all relevant `no-go' symmetries which forbid rectification.
The review focuses on that relationship between the appearance of  \textit{directed transport}
and the \textit{symmetries of the equations} describing the evolution of the system.

The structure of the review is as follows: Section \ref{sec2} introduces the main ideas which underpin  the symmetry
analysis. In Section \ref{sec3} we apply the symmetry analysis in the context of  classical one-dimensional transport,
while in Section \ref{sec4} we step into higher dimensions and discuss multi-directional currents
and the generation of vortices.
We next discuss the extension of the ratchet concept to the case of coherent quantum dynamics in Section \ref{sec5}.
In all these sections, the theoretical analysis  is
supplemented with the outcomes of numerical studies and recent results from the field of cold- and ultracold-atom experimental physics.
Finally, in Section \ref{sec6} we briefly touch upon applications of the analysis to other systems,
such as Josephson junctions and spins, and conclude the review with a discussion  on
issues that constitute promising  directions for further applications of the symmetry analysis.

\section{Symmetries in a nutshell}
\label{sec2}

Consider a set of differential equations, which governs the evolution of a  system,
\begin{eqnarray}
\dot{\Psi}= \mathcal{F}_{\boldsymbol \theta}(\Psi,t) = \mathcal{F}_{\boldsymbol \theta}(\Psi,t+T), \label{system}
\end{eqnarray}
where $\Psi$ is a vector describing the state of the system. Examples are a finite set of coordinates and momenta,
$\Psi = \{x,y,...,v_x,v_y,...\}$, or a vector in a Hilbert space describing a state of a quantum system,
or a probability density function whose evolution is governed by the Fokker-Planck equation.
Here $\dot{\Psi}$ denotes the derivative of $\Psi$ with respect to time.
The vector function $\mathcal{F}_{\boldsymbol \theta}(\Psi,t)$ is explicitly time-periodic with period $T$
and may depend on a set of control parameters ${\boldsymbol \theta} = \{\theta_1, \theta_2,...,\theta_D\}$.
Consider an observable $A(t) = \mathcal{A}[\Psi(t)]$, where $\mathcal{A}$ is an operator or functional over
the phase space of the system. Is it possible to tune the control parameter(s)
${\boldsymbol \theta}$ in such a way
that the evolving system yields a nonzero value of the time-averaged observable,
$\overline{A}= \lim_{t \rightarrow \infty} (1/t) \int_{0}^{t} A(s) d s \neq 0$?
To answer the question, we follow the {\it symmetry analysis protocol} \cite{dfoyz02pre}:

\begin{enumerate}
  \item Identify all possible transformations, $\widehat{S}: \{\Psi,t\} \rightarrow \{\Psi', t'\}$, which
            change the sign of the observable, $\tilde{A}(t') = \mathcal{A}[\Psi'(t')] =$ $-A(t)=-
            \mathcal{A}[\Psi(t)] $, and at the
            same time leave  equation (\ref{system}) invariant. Then
            the  time evolution of the observable, $\tilde{A}(t') = \mathcal{A}[\Psi'(t')]$, is supported by a solution
of (\ref{system}), i.e. a certain trajectory in the system phase space.
              Transformations may include, for example,
            time reversal, $t \rightarrow t' = -t $,
            a time shift, $t \rightarrow t' = t + \tau$, similar operations on the space variables,
            or/and  other operations on system variables, including, for example, permutations \cite{den2007}.

\item   Consider  contributions from the original
            trajectory yielding an observable  $ \, A(t) \, $ and the symmetry-related trajectory,
            which yields $\tilde{A}(t') = -A(t) \, $. If both are parts of  the same long trajectory,
            or if they belong to different trajectories, which have identical statistical weights, the average expectation value
            of the observable vanishes exactly, $\overline{A}=0$ .

\item   Choose control parameters in such a way as to destroy \textit{all} the symmetries. Then, in general, the rectification effect would appear,
with $\overline{A}\neq 0$, in agreement with Curie's principle \cite{c1894}.
\end{enumerate}


\subsection{Symmetries of periodic functions}
\label{Symmetries of periodic functions}

Consider two periodic functions,
\begin{equation}
\label{PerF}
E(t+T) = E(t), ~~ f(x+L) = f(x),
\end{equation}
with the temporal period  $T=2\pi/\omega$ and spatial period $L=2\pi/k$, and both with zero mean
\begin{equation}
\label{ZerMean}
\int_0^{T} E(t) dt = \int_0^{L} f(x) dx = 0 \, .
\end{equation}

The functions can be  {\it symmetric} around certain values
of their arguments,
\begin{equation}\label{SymmFunc}
    E_s(t - t') = E_s(-t-t'), ~~ f_s(x-x') = f_s(-x-x').
\end{equation}
After proper shifts the functions, being expanded in Fourier series,
reduce to a cosine series.

The functions can be {\it antisymmetric} around certain values
of their arguments,
\begin{equation}\label{A-SymmFunc}
  E_a(t - t') = -E_a(-t-t'), ~~ f_a(x-x') = -f_a(-x-x').
\end{equation}
After proper shifts, the Fourier expansions of the functions reduce to a sine series.

The functions can be  {\it shift-symmetric}, in that case they change their signs after the shift of the argument
by half of the period\footnote{In the context of nonlinear oscillations the shift-symmetric property
is also referred to as `anti-periodicity', see Ref.~\cite{Freire2013} and references therein.},
\begin{equation}\label{Sh-SymmFunc}
E_{sh}(t+T/2) = -E_{sh}(t), ~~ f_{sh}(x+L/2) = -f_{sh}(x).
\end{equation}
The Fourier expansion of shift-symmetric functions contains
odd harmonics only.

Henceforth, we mark these three symmetries  by labels ``$s$'', ``$a$'' and
``$sh$'' respectively.  A periodic function of zero-mean may have neither  of the symmetries,
exactly one of them, or all three. For example,
$ E(t) = \cos(t)$ possesses all three symmetries, for it is symmetric around zero,
antisymmetric around $t=\pi/2$, and  evidently shift-symmetric.
Function $ \, f(x) = \cos(x) +
\cos(3x + \Delta)$  is  shift-symmetric, and in addition becomes symmetric and antisymmetric
for $\Delta = 0,\pi$. Finally, function $ E(t) =
\cos(t) + \cos(2t + \theta) \, $ does not possess any of the  symmetries except for
$ \theta = 0, \pi \, $ (symmetric) and $\theta =\pm \pi/2 $ (antisymmetric).
Differentiation preserves shift symmetry and exchanges symmetry with antisymmetry (and vice versa).

\subsection{Symmetries of quasiperiodic functions}
\label{Symmetries of quasiperiodic functions}

A quasiperiodic function $E(t)$ can be viewed as a function that depends on $N$ different variables $t_i$. The latter are
linear functions of the variable $t$ but with different coefficients, i.e. frequencies:
\begin{equation}
E(t) \equiv \tilde{E}(t_1,t_2,...,t_N),\;\frac{d t_i}{d t}= \omega_i\;.
\label{quasi1}
\end{equation}
The frequencies are mutually incommensurate,  with all the ratios $\omega_i/\omega_j$ being irrational numbers when $i \neq j$.
The function $\tilde{E}$ is periodic in every of these variables,   $\tilde{E}(t_1,t_2,...,t_i+2\pi,...,t_N) =
\tilde{E}(t_1,t_2,...,t_i,...,t_N)$,  for all $i$.

A quasiperiodic function $E(t)$ is {\it symmetric} if
\begin{equation}
\tilde{E}_s(-t_1,-t_2,...,-t_N) = \tilde{E}_s(t_1,t_2,...,t_N),
\label{quasi2}
\end{equation}
and {\it antisymmetric} if
\begin{equation}
\tilde{E}_a(-t_1,-t_2,...,-t_N) = -\tilde{E}_a(t_1,t_2,...,t_N).
\label{quasi3}
\end{equation}
Note that these two symmetry operations necessarily involve simultaneous actions on all $N$ variables $t_i$.

Finally, the quasiperiodic function $E(t)$ can be {\it shift-symmetric} on a  subset of indices $\{i,...,j\}$, if
\begin{equation}
\tilde{E}_{sh}(t_1,...,t_i+\pi,...,t_j+\pi,...,t_N) = -\tilde{E}_{sh}(t_1,...,t_i,...,t_j,...,t_N).
\label{quasi3a}
\end{equation}
Note that shift-symmetry may involve actions on  one, two,  and so on up to all $N$ variables.

Topologically, the function  $\tilde{E}(t)$ evolves on a $N$-dimensional torus, $\{ t_1,...,t_N \}$.
The incommensurability of the frequencies guarantees that,
in the course of time, the torus surface  is
scanned in an ergodic manner \cite{Ott1992, suz92}, and will be covered with uniform density in the asymptotic limit
$t \rightarrow \infty$.

\subsection{An example: Hamiltonian ratchets}
\label{Hamiltonian ratchets}

Let us consider a dissipationless, point-like particle of mass $m$, which moves in a
spatially periodic potential. In addition, the potential is periodically modulated in time, e.~g. because
the particle is electrical charged and  exposed to
a time-periodic spatially-homogenous electric field.
The position of the particle
is governed by the following equation of motion \cite{Ott1992}:
\begin{equation}
\label{HamRatchet}
   m {\ddot x}  = \cos(x) + E(t).
\end{equation}
The driving function $E(t)$ is  periodic, $E(t+T)=E(t)$, and has zero mean,
$\langle E(t) \rangle = (1/T)\int_0^T E(\tau)d\tau = 0$.
Let us find the conditions when the driving field sets the particle into motion with a nonzero
average velocity.

The relevant
observable is
\begin{equation}
\label{velocity}
\bar{v} = \langle \dot{x}\rangle = \lim_{t \rightarrow \infty} x(t)/t.
\end{equation}

We  search for
transformations of the system variables, $\hat{S}:\{x,t\} \rightarrow \{\tilde{x},\tilde{t}\}$,
that change the sign of the observable, $\bar{v}(\tilde{x},\tilde{t}) = -\bar{v}(x,t)$, while leaving the equation of
motion (\ref{HamRatchet}) invariant.

There are two possibilities to change the sign of $\dot{x}$.  Namely, either to change the sign of $x$ and, in case of a need,
perform additional shifts,
$ \tilde{x} = -x+\chi$, $\tilde{t} = t+\tau$, or change the sign of time, $\tilde{t} = -t+\tau$, with an additional spatial shift
$\tilde{x} = x+\chi$, if needed.
Thus we have two possible transformations:
\begin{eqnarray}
\label{SpaceInvClass1d}
&&{\hat S}_x[\chi,\tau]:~ x \rightarrow \tilde{x} = -x + \chi \, , \quad t \rightarrow \tilde{t} = t+\tau\, , \\
\label{TimeRevClass1d}
&&{\hat S}_t[\chi,\tau]:~ x \rightarrow \tilde{x} = x + \chi  \, , \quad t \rightarrow \tilde{t} = -t+\tau\, ,
\end{eqnarray}

Transformation ${\hat S}_x$ changes the sign of $\ddot{x}$ on the lhs of Eq.~(\ref{HamRatchet}).
Additional shifts in time and space have to be performed, such that the rhs terms change their signs as well.
The force term
will change sign by setting $\chi= \pi$.
The drive must be \textit{shift-symmetric}, with  $\tau=T/2$, $E(t+T/2) = -E(t)$, and
only in this case symmetry ${\hat S}_x$ will hold.  In order to break this symmetry, we can simply
choose a function
$E(t)$ which is not shift-symmetric (note, however, that we could also choose a more complicated ratchet potential which alone could
violate this symmetry).
The simplest choice of  a driving function that violates shift-symmetry is a bi-harmonic combination,
\begin{equation}
\label{biharm}
E(t) = E_1 \cos(\omega t) + E_2 \cos(2\omega t + \theta).
\end{equation}

\begin{figure}
\begin{center}
\includegraphics[angle=0,width=0.85\textwidth]{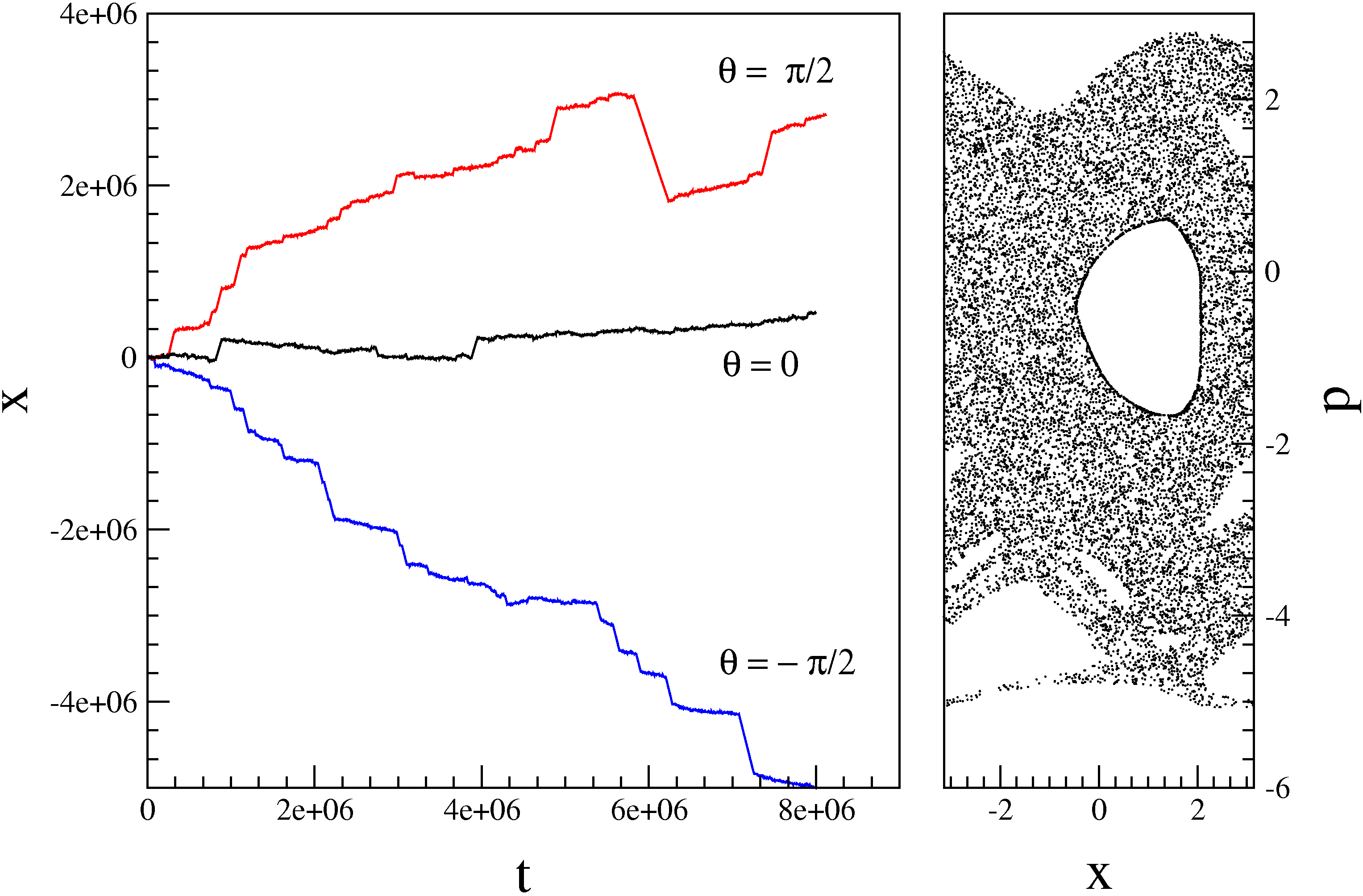}
\caption{(color online) Left panel: Dependence $x(t)$ vs $t$ for Eq.~(\ref{HamRatchet}), with $m=1$  and initial conditions
$\{x = \pi,\dot{x} = 0, t =0\}$. Right panel:  Poincar\'{e} section for $\theta = -\pi/2$ plotted at  time instances $t = 0, T,...,10^4T$.
The parameters of the driving function, Eq.~(\ref{biharm}), are $E_1=E_2=2$ and $\omega = 2$.} \label{figHam1}
\end{center}
\end{figure}

Transformation ${\hat S}_t$ does not change the sign of  $\ddot{x}$.
Reversal of time does not affect the first term  on the rhs of Eq.~(\ref{HamRatchet}), $\cos(x)$, either.
However, the reversal  affects  the driving term, $E(t)$.
Then, for the symmetry ${\hat S}_t$
to hold, the driving function should be symmetric,
$E(-t-\tau) = E(t)$.
To break this symmetry we have to choose a function $E(t)$ which is not symmetric. For the bi-harmonic function
(\ref{biharm}) this holds with $\theta \neq 0,\pm\pi$. Therefore, a bi-harmonic driving function
alone can destroy both symmetries.

The dynamics of the system (\ref{HamRatchet}) allows both for regular (periodic and
quasiperiodic) solutions and chaotic trajectories embedded
into the chaotic layer around the line $\dot{x} = 0$ \cite{suz92}. Note that ergodicity
holds in the layer.  Therefore the average
velocity will be the same for all trajectories from the chaotic layer, \textit{at the asymptotic limit} $t\rightarrow \infty$.
If any of the symmetries, ${\hat S}_x$ or ${\hat S}_t$, holds, then, when it applied to a chaotic trajectory, it will
generate  a new trajectory, also embedded inside the layer and therefore  chaotic.
Consequently the average velocity vanishes inside the  layer.

In the absence of both symmetries,  with a properly set parameter $\theta$,
a directed transport within the stochastic layer will emerge. This was observed in numerical
simulations, see, for example, Ref.~\cite{fyz00prl}.  Fig.~\ref{figHam1} depicts several trajectories launched from the same initial point,
$\{x(0),\dot{x}(0)\}$, chosen in the layer area and then propagated in time\footnote{Attention should be paid to the choice of the
long-run propagation algorithm, which needs to be symplectic in the case of Hamiltonian systems,
see Refs. \cite{gh2001,fyz00prl,Goychuk2000}.}.
The numerical results are in a full  agreement with the theory. Asymptotic
transport is negligible when $\theta = 0$ and  symmetry ${\hat S}_t$ is present. There is a
steady transport in positive (negative) direction when $\theta = \pi/2~(-\pi/2)$ and both key symmetries are broken.
This is a typical performance  of \textit{Hamiltonian ratchets}\footnote{A Hamiltonian ratchet is
an ac-driven Hamiltonian system which is able to exhibit regimes of directed transport in absence of a bias
\cite{r02pr,gh2001,dhm2009,Goychuk2000,dkos00ap,sokd01prl}.} \cite{gh2001,fyz00prl,Goychuk2000,sokd01prl,df01pre}, see for a further discussion in  Section \ref{Hamiltonian ratchets revisited}.

\section{Directed transport in one spatial dimension}
\label{sec3}
In this section we  consider the general case of  a point-like particle moving in a time-modulated spatially periodic potential $V$,
\begin{equation}
\label{2.1}
   m {\ddot x}+\gamma {\dot x} =  g(x,t)~,~~~~~ g(x,t)=-\partial_x V(x,t) ~.
\end{equation}
Zero-mean force function $ \, g(x,t) \, $
is periodic in both $x$ and $t$:
\begin{equation}
   g(x,t)  =  g(x,t+T) = g (x+L,t)~,~~
   \int_{0}^T\int_{0}^L g(x,t) \, dt \, dx =  0~.
\label{2.1a}
\end{equation}

Equation  (\ref{2.1}) can be rewritten as a set of three
autonomous differential equations of first order,
\begin{eqnarray}
\dot{x}=p/m,~\label{eq21a}
\\
\dot{p} = -\gamma p/m + g(x,\Omega),~\label{eq21b}
\\
\dot{\Omega} = \omega, \label{eq21c}
\end{eqnarray}
where $\omega = 2\pi/T$.
Therefore, the phase space of the  system is three-dimensional.

The function  $g(x,t)$ can be continuous or discontinuous, smooth or non-smooth, etc.
Examples of discontinuous functions are  trains of delta-like peaks
\cite{sokd01prl,mdhi02prl,cbcs05prl}, while  piecewise-linear
sawtooth potentials \cite{r02pr,smhn04epjb,sav2} constitute a good example of non-smooth functions.
The symmetry analysis applies
equally well to all situations though  below we will focus mainly on the
smooth functions consisting of a few  harmonics only. This is a reasonable choice
in the context of manipulations with cold atoms in  optical potentials because it allows to build an experimental
set-up with a minimal number of lasers and acousto-optical modulators \cite{Morsch2006,jgr05prl,gbr05prl}.

\subsection{Symmetry analysis }

Transport in the system (\ref{2.1}) is characterized by the
velocity of the particle, $\dot x$. The total current is
defined by means of averaging over the phase space:
\begin{equation}
   J(t) =  \langle \dot x \rangle_\Gamma ~.
\label{2.4}
\end{equation}
Here $\langle \dots \rangle_\Gamma$ stands for the phase space
averaging with some  distribution function, which needs to be specified in order to define the problem entirely
(we will discuss this issue in  Sec. \ref{Issue of initial conditions}).
The asymptotic current is given by:
\begin{equation}
\label{2.5}
  J = \langle J(t) \rangle_t~,~~\left\langle \ldots \right\rangle_t
  \equiv \lim_{t - t_0 \to \infty }
 \left[ \frac{1}{t-t_0} \int_{t_0}^{t} {\rm d}t' ( \ldots ) \right]~.
\end{equation}

Our aim now is to figure out the necessary conditions for
$J$ to acquire non-zero values. To this end, we
first will find the sufficient conditions which  prevent from this and then will
discuss how to violate them.

Let assume that there is a trajectory which contributes  to $J$, i. e. it has been taken into account when
performing the averaging over the phase space, Eq.~(\ref{2.4}).
Suppose now that there is a transformation of the
equation of motion (\ref{eq21a}-\ref{eq21c}), which maps parts of
the phase space onto each other. The symmetry
operation transforms the trajectory into another
one -- but with opposite velocity. If at
least one  symmetry exists, then the contributions of the
trajectory and its symmetry-related twin  will cancel
each other after the averaging
provided that both trajectories contribute to the quantity given by Eq.~(\ref{2.4}) \textit{with equal
statistical weights}. Finally, if the symmetry-based cancelation mechanism
works for all  trajectories,
one may conclude that the asymptotic current $J$ is strictly zero.
Evidently, a nonzero current can only appear when all the
symmetries are broken.  There might be other specific mechanisms, aside of  the
trajectory-by-trajectory cancelation,  which prohibit the appearance of a directed
current. However, we again remind that the absence of the symmetries is
a \textit{necessary} but not \textit{sufficient}  condition for the rectification
effect to appear.

  As before, there are only two possibilities to change  the sign of the velocity $\dot{x}$.
The corresponding symmetry transformations were obtained in the previous section:
\begin{eqnarray}
\label{SpaceInvClass1da}
  &&{\hat S}_x[\chi,\tau]:~ x \rightarrow -x + \chi\, , \quad t \rightarrow t+\tau
\, ,\label{SpaceRevClass1d} \\
  &&{\hat S}_t[\chi,\tau]:~ x \rightarrow x + \chi  \, , \quad t \rightarrow -
t+\tau \, , \label{TimeRevClass1da}
\end{eqnarray}
where $\chi$ and $\tau$ are some appropriate shifts,
which depend on the shape of the function $g(x,t)$ and the friction strength parameter $\gamma$.

Now we briefly address different  dynamical regimes of the system  (\ref{2.1}),
which differ with respect to the friction strength $\gamma$ and mass $m$.
\\[1\baselineskip]

\begin{itemize}
\item
{\it Hamiltonian case, $m > 0, \gamma = 0$}.
The corresponding equation of motion is
\begin{equation}
\label{hami_gen}
   m {\ddot x} =  g(x,t).
\end{equation}

In this case both symmetries ${\hat
S}_x$ and ${\hat S}_t$ can be present, as it has been shown in Section \ref{Hamiltonian ratchets}. Namely,
\begin{eqnarray}
\label{2.9a} \widehat S_x ~~~\mbox{holds if} ~~~~&& g(-x-\chi,t+\tau )=-g( x,t) \, ,\\
\label{2.9b} \widehat S_t ~~~\mbox{holds if} ~~~~&& g(x+\chi,-t-\tau)=g(x, t) \, .
\end{eqnarray}
Note that $\chi$ and $\tau $ are used to shift the function to the relevant inversion points in time or space.
These parameters can take, in principle,  any values.
However, the shifts which are not accompanied by a sign change, e.g. $\tau$ for
(\ref{2.9a}) and $\chi$ for (\ref{2.9b})
 are restricted to $\tau =\pm T/2$ or $0$, and $\chi=L/2$ or $0$, correspondingly.
\\[1\baselineskip]

\item
{\it Dissipative case, $m, \gamma > 0$}.
The corresponding equation of motion is
\begin{equation}
\label{diss_gen}
  m {\ddot x}+\gamma {\dot x} =  g(x,t).
\end{equation}

In this case  the time-reversal symmetry
${\hat S}_t$ is broken
by the simultaneous presence of the dissipative and inertia terms in the equation of motion.
Symmetry ${\hat S}_x$, however,  can hold:
\begin{eqnarray}
\widehat S_x ~~~\mbox{holds if} ~~~~&& g(-x-\chi,t+\tau )=-g( x,t) \, ,
\end{eqnarray}
\\[1\baselineskip]

\item
{\it Overdamped limit, $m = 0, \gamma > 0$}.
The corresponding equation of motion is
\begin{equation}
\label{over_gen}
  \gamma{\dot x} =  g(x,t).
\end{equation}

Here both symmetries, ${\widehat
S}_x$ and ${\widehat S}_t$,  can be present. The conditions
for the symmetry $\widehat S_x$ to hold  remain the same as for the Hamiltonian case, Eq.~(\ref{2.9a}),
while the condition for  ${\widehat S}_t$
is modified:
\begin{eqnarray}\label{over_x}
\widehat S_x ~~~\mbox{holds if} ~~~~&& g(-x-\chi,t+\tau )=-g( x,t) \, ,\\
\label{over_t}
\widehat S_t ~~~\mbox{holds if} ~~~~&& g(x+\chi, -t-\tau)= -g(x, t) \, .
\end{eqnarray}

The  presence of the time-reversal antisymmetry  in the overdamped limit  was first
noticed in Ref. \cite{fyz00prl}  and  was explained then in Ref. \cite{r01prl}.
\end{itemize}


The presence of either of the symmetries
(\ref{SpaceRevClass1d})-(\ref{TimeRevClass1da}) in the Hamiltonian or the dissipative cases guarantees that for any
trajectory with a nonzero velocity the  phase space contains
its image, i.e., another trajectory with the velocity opposite to the velocity of the original
trajectory. Assuming that both trajectories contribute equally to the system evolution,  we arrive at the conclusion that the overall
current produced by the system equals zero\footnote{In the overdamped case the situation is more tricky, see Refs.~\cite{fyz00prl,dfoyz02pre,r01prl}.
We will address this issue
in Sec. \ref{From Langevin dynamics to the Fokker-Planck equation}}. Thus, in order to obtain  a nonzero directed current, we should
choose a function $g(x,t)$  which does not satisfy either of the symmetries.

Finally, it is noteworthy that all relevant symmetries can be re-formulated in terms of
the original potential, $V(x,t)$, by noting that $g(x,t)= -\partial V(x,t)/\partial x$ and taking into account the modifications
of the symmetries of a function by the differentiation operation, see Sec. \ref{Symmetries of periodic functions}.

\subsection{The role of initial conditions}
\label{Issue of initial conditions}
As we discussed in the introduction, the phase space of a coherent, noise-free system may consist  of several invariant manifolds.
The co-existence of many invariant manifolds is the general case for ac-driven Hamiltonian systems, so the corresponding phase space
is conventionally called  'mixed phase space' \cite{suz92}. The co-existence of attractors is a typical situation in
weakly-damped systems \cite{Ott1992}, and, sometimes, it is the case for  overdamped systems \cite{bhk94epl}.
Because of this  co-existence, the asymptotic characteristics of a system  depend on the form of the initial distribution
function, ${\mathcal F}(x_{0}, p_{0}, \Omega_{0} = t_0)$, and
the overall asymptotic current should be calculated as
\begin{equation}
\label{average_J}
\tilde{J} = \int \!\!\ {\rm d} x_0 \,  {\rm d}p_0 \, {\rm
d}\Omega_0 \ {\mathcal F} ( x_0, \, p_0 , \, \Omega_0) \ J ( x_0, \,
p_0,\, \Omega_0 ) \, ,
\end{equation}
where $J ( x_0, \, p_0,\, \Omega_0 )$ is the asymptotic current, Eq.~(\ref{2.5}), produced by the trajectory launched from point
$\{ x_0, \, p_0,\, \Omega_0 \} $.

Every invariant manifold can be characterized by its asymptotic average velocity, which might be different
from zero even when  one of the
basic symmetries, Eqs.~(\ref{SpaceInvClass1d}-\ref{TimeRevClass1d}), is present. A
symmetry transformation can map  two manifolds with nonzero opposite velocities onto each other.
By initiating more trajectories on the manifold with the average velocity $\upsilon \neq 0$
than on its symmetry-related counterpart, with the velocity $-\upsilon$, one will detect a non-zero current
in the situation when the corresponding equations of motions are perfectly symmetric.
The presence of a symmetry guarantees
zero-current output only when  symmetry-related trajectories contribute equally to the overall current (\ref{average_J}).
That happens when
the initial distribution function, ${\mathcal F}(x_{0}, p_{0}, \Omega_{0})$, also satisfies
the relevant symmetry.  Since the phase
space of the system given by Eqs.~(\ref{eq21a} - \ref{eq21c}), $(x, p, \Omega)$, is compact along the
variables $x ~\mathsf{mod}~ L$ and $\Omega~ \mathsf{mod}~ T$, a reasonable  choice of the initial
distribution function is  the product of the uniform distributions over the variables $x_0 ~\mathsf{mod}~ L$
and $\Omega_0~ \mathsf{mod}~ T$, and a
symmetric distribution with respect to the momentum variable, ${\mathcal F}(x_0, -p_0,
t_0$) = ${\mathcal F}(x_0, p_0, t_0$). The latter can be, for example, the Maxwell distribution  \cite{fyz00prl,sokd01prl}.
Such initial distributions possess both fundamental symmetries,
Eqs.~(\ref{SpaceRevClass1d}, \ref{TimeRevClass1d}), for any values of  $\chi$ and $\tau$.

\subsection{Special cases}

\subsubsection{Additive driving}
\label{Additive driving}

First we consider force functions of the type $g(x,t)=f(x)+E(t)$. It is
a sum of two independent, space- and time-dependent functions\footnote{In the ratchet literature this
is often coined {\it tilting} or {\it rocking} ratchet \cite{r02pr,hm09rmd}.}:
\begin{equation}
m {\ddot x}+\gamma {\dot x} = f(x)+E(t)~,~~E(t)=E(t+T),~~f(x)=f(x+L). \label{2.21}
\end{equation}
Functions
$E(t)$ and $f(x)$ both have zero means.  Therefore one can classify
them with respect to their symmetry properties, as described in Section \ref{Symmetries of periodic functions}.
We list in Table
\ref{tabl1} the requirements for the
functions $f(x)$ and $E(t)$  to satisfy either of the two basic symmetries.
\begin{table}[t]
\begin{center}
\includegraphics[angle=0,width=1.0\textwidth]{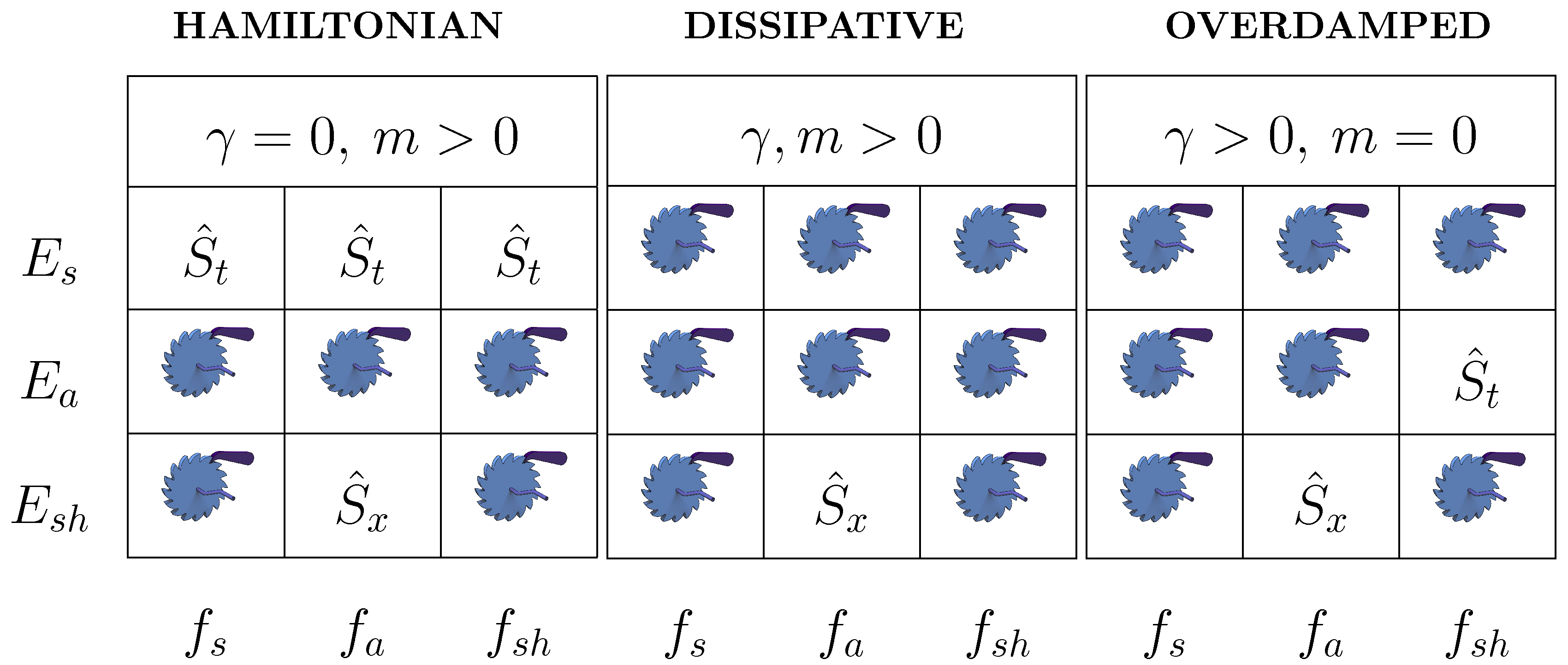}
\caption{(color online) Conditions for  the fundamental
symmetries, Eqs.~(\ref{2.9a}-\ref{2.9b}), to hold in the case of  additive driving, Eq.~(\ref{2.21}).
Ratchet symbol indicates the cases when both  symmetries are broken.
Note that any other combination, not included in the tables,
also corresponds to the case of broken symmetries.}
\label{tabl1}
\end{center}
\end{table}
Any choice of the functions $f(x)$ and $E(t)$, which does not fall into one of  the listed cases,  leads to the symmetry breaking.
The appearance of the directed current in systems of the type  (\ref{2.21}) has been verified
both analytically, numerically and experimentally
\cite{r02pr, hm09rmd}. In  the dissipative regime, the following choice of potential force
\cite{bhk94epl,lbh95epl,jkh96prl,m00prl,fyz00prl,bs00pre}:
\begin{equation}
f(x)=f_1 \sin{x} + f_2 \sin{(2x+\Delta)},
\label{u2}
\end{equation}
guarantees the violation of   symmetry $\hat S_x$ if $f_2\neq
0$, $\Delta \neq 0, \pm \pi$. Therefore, the presence of a bi-harmonic ratchet potential is a sufficient condition
to obtain directed current when $\gamma > 0$ and $m \neq 0$, see the only entry in the corresponding table.
Note, however,  that this is not yet enough to guarantee the violation of the time-reversal symmetry $S_t$
in the Hamiltonian and overdamped limits.

The bi-harmonic driving function,
\begin{equation}
E(t)=E_1 \cos{\omega t} +E_2  \cos{(2\omega t+\theta)}. \label{e2}
\end{equation}
allows one to violate the fundamental  symmetries in the Hamiltonian, dissipative, and overdamped limits,
by tuning  parameter $\theta$.
Moreover,  even in the case of a single-harmonic potential,
$f_2 = 0$ in Eq.~(\ref{u2}), the bi-harmonic driving alone can break
both symmetries  \cite{jkh96prl,m00prl,fyz00prl,dfoyz02pre,bs00pre,dkos00ap,df01pre,dkuf01pd}.



\begin{table}[t]
\begin{center}
\includegraphics[angle=0,width=1.\textwidth]{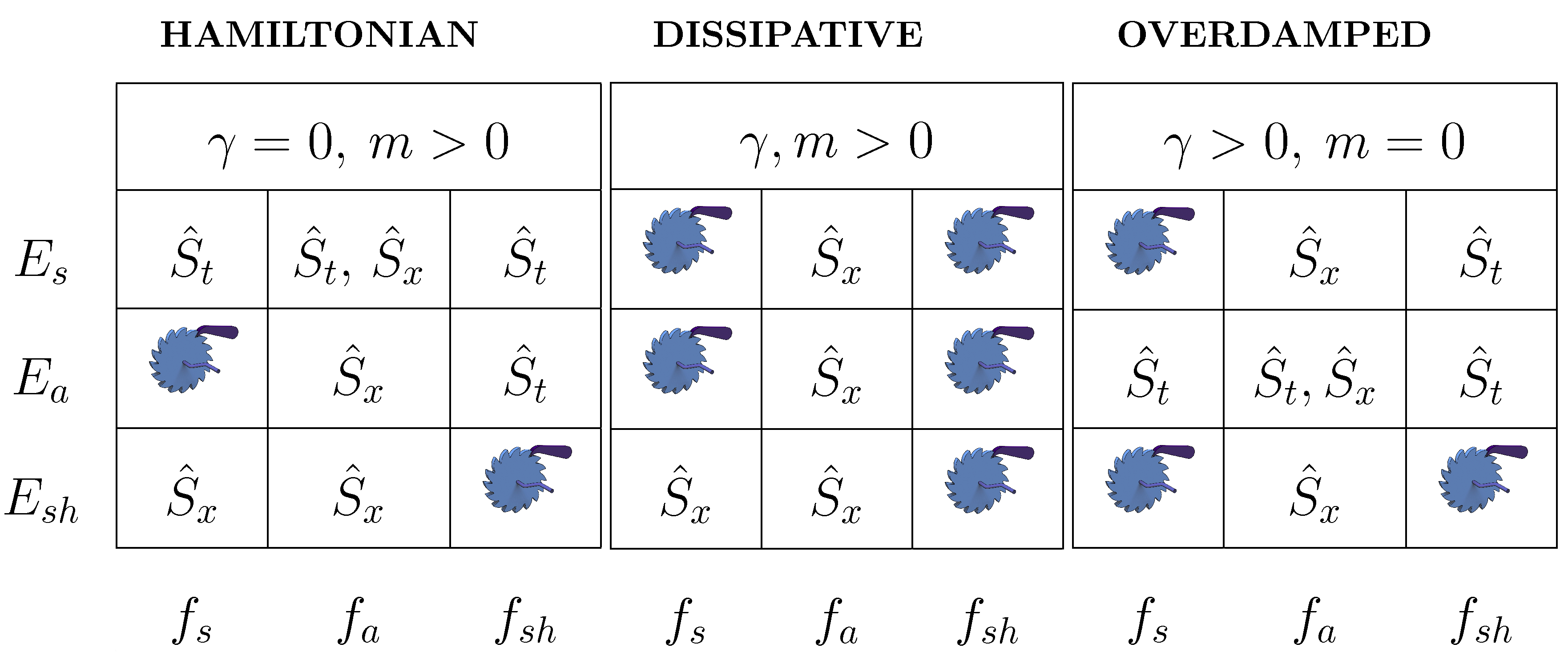}
\caption{(color online) Conditions for  the fundamental
symmetries, Eqs.~(\ref{2.9a}-\ref{2.9b}), to hold in the case of   multiplicative driving, Eq.~(\ref{2.22}).
Ratchet symbol indicates the cases when both  symmetries are broken.
Note that any other combination, not included in the tables,
also corresponds to the case of broken symmetries.}
\label{tabl2}
\end{center}
\end{table}

\subsubsection{Multiplicative driving}
\label{Multiplicative driving}

Another frequently used  setup corresponds to a spatial potential with its amplitude
periodically modulated in time\footnote{
In the literature often referred to as  {\it pulsating} or {\it flashing} ratchet \cite{r02pr,hm09rmd}.},
\begin{equation}
m {\ddot x}+\gamma {\dot x} = f(x)E(t),~~f(x)=f(x+L),~~E(t)=E(t+T). \label{2.22}
\end{equation}
Because of the condition (\ref{2.1a}), at least one the
functions, $f(x)$ and $E(t)$, has to be of zero mean.
For example, while the function $E(t)$ may have a nonzero constant component,
the average potential,  $g(x,t)=\sin(x)[E_0 + \sin (\omega t)]$, remains unbiased.

The corresponding symmetry conditions are given in Table \ref{tabl2}. Note that when, for example, function $f(x)$ is antisymmetric,
equation  (\ref{2.22}) remains invariant  with respect to the symmetry $\hat S_x$  for any choice of the driving function $E(t)$. This, and a
larger overall number of possible symmetries as compared to the previous case of additive driving, are the consequences of multiplicative driving.

A minimal setup, which allows to control all
relevant symmetries, consists of bi-harmonic functions, with $f(x)$
and $E(t)$ given by Eqs.~(\ref{u2}, \ref{e2}). This combination has been used in the experimental realization of ac-driven quantum ratchets
with a Bose-Einstein condensate \cite{weitzScience}.
We will review these experiments in more details in Section \ref{Experiments with Bose - Einstein condensate}.

The overdamped limit of multiplicative ratchets has been extensively studied, see Refs.
\cite{ap92crasp, r02pr}. One of the popular choices was
the driving field in the form of dichotomous function, with $E(t) = 1~
\mathrm{or} ~0$, so the potential is in either "on" or "off" position.
\cite{r02pr}.
The Hamiltonian limit of the multiplicative set-up  has been studied  by using
different modifications of the kicked-rotor model \cite{Ott1992, suz92}.
In these studies the driving field $E(t)$ was represented by a train of delta kicks so that the propagation in time
was reduced to iterations of a two-dimensional map \cite{dkos00ap,sokd01prl,mdhi02prl,cbcs05prl}.

\subsubsection{Traveling potentials}
\label{Traveling potentials}


A particular setup, named {\it traveling potential ratchet}
\cite{r02pr}, has the force function of the form
\begin{equation}
g(x,t)=\sum_{k}f_{k}[x+a_{k}(t)],
\end{equation}
where $f_{k}(x)$ and $a_k(t)$ are periodic functions of zero mean.
If $f_{k \neq 1} = 0$ and the function $a_1(t)\equiv a(t)$ is periodic in time, then, by using
the transformation $x \rightarrow \tilde x = x + a(t)$, the equation of
motion (\ref{2.1}) can be cast in the following form:
\begin{equation}
\label{modulation}
m {\ddot {\tilde x}}= -\gamma {\dot {\tilde x}} +
 f(\tilde x)- \gamma \dot{a}(t)-m \ddot{a}(t).
\end{equation}
where both the driving functions, $\dot{a}(t)$ and $\ddot{a}(t)$, have zero means.
The asymptotic current is the same in both frames,  $J=
\langle \dot x \rangle_{\Gamma} = \langle \dot {\tilde x}
\rangle_{\Gamma}$. Therefore,  it suffices to  perform the analysis in the
new frame, $\{\tilde x, t\}$. By taking into account that the
differentiation operation transforms  symmetric functions into
antisymmetric ones  while leaving  the shift-symmetric
property invariant, we immediately reduce the symmetry analysis of Eq.~(\ref{modulation})
to the already considered case of additive driving, Eq.~(\ref{2.21}). The
conditions for the symmetry $\widehat{S}_x$ to hold remain the same: $f(x)$ should be antisymmetric,
and $a(t)$ shift-symmetric. In the Hamiltonian limit,  the same is true for the  time-reversal symmetry,
$\widehat{S}_t$,  which now requires $a(t)$ to be symmetric.
The overdamped limit is instructive - for the time-reversal (anti)symmetry to hold,
$a(t)$ has to be now symmetric instead of being antisymmetric as before.
The  force
function $g(x,t)=\cos [x-E(t)]$, with $E(t)=E_{1}\cos(\omega
t)+E_{2}\cos(2\omega t+\theta)$, has been used for the realization
of rocking ratchets in optical potential with cold atoms  \cite{ss-prg03prl,jgr05prl,gbr05prl,r05contp} and glass microspheres \cite{avm11}.
We will review these experiments  in Sections~\ref{Experiments with cold atoms} and \ref{OverDampedSect},
respectively.

The choice $a_{k}(t)=\upsilon_{k}t$ corresponds to a so-called
{\it genuine traveling potential ratchet} \cite{r02pr} and
mimics the dynamics of a particle interacting with several propagating waves
\cite{suz92}. Each wave drags the particle in the direction of the wave propagation and
the asymptotic current appears as a result of competition
between the waves.
This setup was frequently used in  quantum pumps \cite{Niu1990, Altshuler1999}, and in its
simplest form, with $k=1$, was proposed by Thouless to pump adiabatically electrons \cite{Thouless1983}.
For more efficient non-adiabatic pumping  we refer to recent research activities  \cite{PRKohler05,Strass05}.
Indeed, consider the traveling-wave potential,
$g(x, t)= \cos(x - \varOmega t)$. It can be rewritten as
$g(x, t) = \sin(x)\cos(\varOmega t) + \cos(x)\sin(\varOmega t)$ \cite{Altshuler1999}.
Applying the symmetry analysis, it is easy to see that all relevant
symmetries are broken when $\varOmega \neq 0$.

\subsubsection{Gating ratchets}
\label{Gating ratchets}

Finally, we consider  a \textit{gating ratchet} \cite{bm05c},
\begin{equation}
g(x,t)=f(x)a(t)+ E(t),
\label{2.23}
\end{equation}
Without loss of generality,
we assume that $f(x)$ is a function of zero mean while $a(t)$ may have a constant component
(the inverse situation can be analyzed in the same manner),
\begin{equation}
f(x)=f \sin (x),~~a(t)=1+\varepsilon \sin( {\omega t}),
~~E(t)=E \sin {(\omega t+\theta)}. \label{gait}
\end{equation}
By setting  $\theta \neq 0, \pm \pi$, one can break all  the symmetries in all the
limits with respect to the friction strength.  Note that, in this case, there is no need for a
second harmonic neither for the force term nor for the drive function.
For more analytical and numerical results on the overdamped
limit of gating ratchets we refer the reader to Ref.~\cite{bm05c}.
The gating ratchet setup has also been implemented in experiments with cold atoms \cite{glbr08}.

\subsection{Extension to quasiperiodic driving}
\label{Extension to quasiperiodic driving}
As an example we consider
the rocking ratchet  (\ref{2.21}) with the driving function
$E(t) = \widetilde{E}(\omega_1t, \omega_2t,...,\omega_Nt)$ \cite{np02epjb,fd04appb}.

First we write the corresponding equations of motion in the following form:
\begin{eqnarray}\label{3-2}
m\ddot{x} + \gamma \dot{x} - f(x) - E(\phi_1,\phi_2,...,\phi_N) =0
\;,\\
\dot{\phi}_i = \omega_i, ~ i = 1,2,...,N \;,\nonumber
\end{eqnarray}

It is straightforward to identify the symmetries of the system. There are two types of them,
\begin{eqnarray}
\nonumber
{\hat S}^q_x[\chi,\{i,j,...,m\}]:  x \rightarrow -x + \chi,~~
\phi_{i,...,j} \rightarrow  \phi_{i,...,j}+\pi, \\
~~~~~~~~~~\;\; {\rm if} \; \{f_{a}, E_{sh,\{i,j,...,m\}} \},
\label{symqp1}
\end{eqnarray}
\begin{eqnarray}
{\hat S}^q_t[\chi]:  x \rightarrow x + \chi, \;\; t\rightarrow -t, ~~\phi_i\rightarrow -\phi_i,
\;\; {\rm \qquad \ if} \; \{E_s,\;\gamma = 0\}.
\label{symqp2}
\end{eqnarray}
Symmetry ${\hat S}^q_t[\chi]$ is a particular case of the symmetry  (\ref{TimeRevClass1d}) with $\tau = 0$.
Symmetry ${\hat S}^q_x[\chi,\{i,j,...,m\}]$ branches into  a whole set of symmetry operations,
defined by a subset of indices $\{i,...,j\}$.

A force function $f(x)$ can also be quasiperiodic,
with $M$ spatial harmonics. Generalization of the symmetry analysis
to this case is straightforward \cite{fd04appb}.


\subsection{Dynamics}
The symmetry analysis tells when a current is absent or \textit{may be} present. However it neither
specifies current's value nor its sign. These characteristics can be obtained from reasonable perturbation approaches
-- at the vicinities of symmetry points in the control parameter space, or by simply performing numerics.
In the following subsections we discuss the microscopic dynamical mechanisms underlying the
rectification processes in Hamiltonian and dissipative systems.

\subsubsection{Hamiltonian ratchets revisited}
\label{Hamiltonian ratchets revisited}

Let us return to the model introduced in Sec. \ref{Hamiltonian ratchets}.
Both key symmetries, $\widehat{S}_{x}$ and $\widehat{S}_{t}$,
can be violated by using a proper driving  function, $E(t)$ \cite{fyz00prl,dfoyz02pre,df01pre,dkuf01pd}.
A setup with a single-harmonic potential force and a bi-harmonic drive,
\begin{equation}
m {\ddot x}  = \cos(x) + E(t),
\label{pot_ham}
\end{equation}
\begin{equation}
E(t)=E_{1}\cos(\omega t)+E_{2}\cos(2\omega t+\theta),
\label{driv_ham}
\end{equation}
constitutes the simplest choice.
For $E_{2}\neq 0$ and $\theta
\neq 0, \pi$, both key symmetries are broken so that the appearance of a non-zero current can be expected.
Before proceeding further with the analysis of the rectification process, we
will briefly discuss some general properties of  Hamiltonian chaos \cite{Ott1992, suz92}.

\begin{figure}
\begin{center}
\includegraphics[angle=0,width=0.7\textwidth]{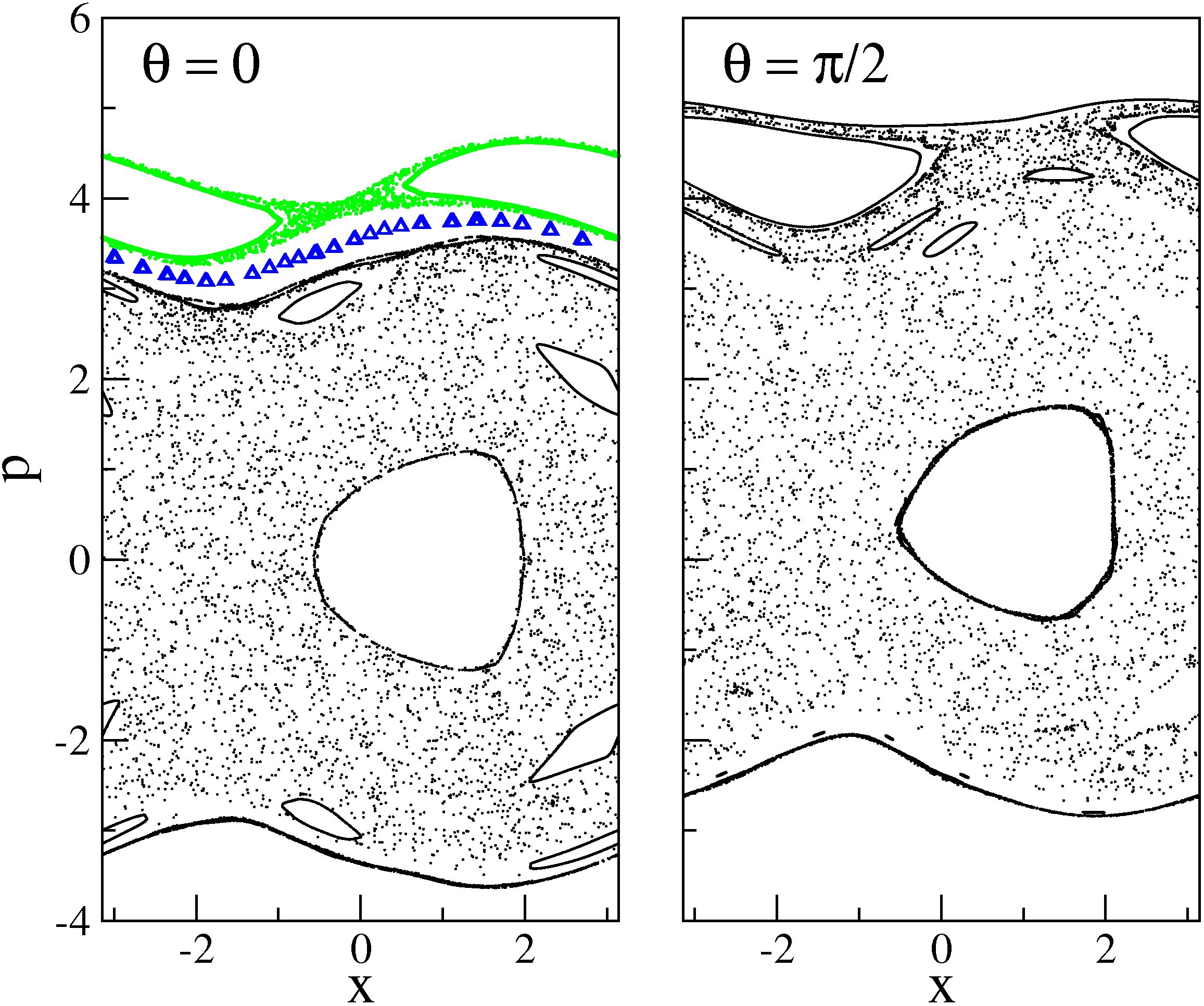}
\caption{(color online) Poincar\'{e} maps of the Hamiltonian system (\ref{pot_ham}, \ref{driv_ham}) for  different values of $\theta$.
Due to the time-reversal symmetry $\hat{S}_t[0,0]$, Eq.~(\ref{over_t}),  the Poincar\'{e} map for $\theta = 0$ (left panel) is
symmetric at the line $p=0$.
Triangles mark a ballistic invariant manifold, a regular torus with the average velocity $\upsilon \approx 1.64$.
Bright (green) dots mark another ballistic invariant manifold, a thin chaotic layer enclosing a ballistic regular island.
The velocity of the orbits inside the island are equal to the velocity of the periodic orbit at the center of the island,
$\upsilon = 4$. This is also the average velocity of all orbits inside the ballistic chaotic layer.
 Symmetry violation results in the overlap between the ballistic resonance and the main stochastic
layer as they merge into a  single chaotic  manifold (right panel).
This effect leads to the appearance of a strong positive current within the newly created stochastic layer, see Fig.~\ref{figHam3}.
The  parameters are  $E_1 = E_2 = 2$ and $\omega=2$.} \label{figHam2}
\end{center}
\end{figure}

A standard way to visualize dynamics of the Hamiltonian system (\ref{pot_ham}, \ref{driv_ham}) is to use the stroboscopic
Poincar\'{e} map \cite{Ott1992}. This can be done by propagating the system in time and plotting the
values of coordinate and momentum on the stripe $\{x ~\mathsf{mod} ~L, p\}$
at equidistant instants of time $t_n = nT$, $n = 1,2,...$.
By choosing different initial points in the $\{x, p\}$ plane, we will collect a large set of points
$\{x(t_n) ~\mathsf{mod} ~L$, $p(t_n)\}$, thus resolving the structure of the phase space.
Note that the result will  depend on the parameter $\theta$.
Typical Poincar\'{e} maps  are depicted on Fig.~\ref{figHam2}. They reveal that the phase space of the Hamiltonian
system is indeed 'mixed', i. e. it consists of
different  invariant manifolds, that are chaotic layers, regular
islands, tori, etc \cite{suz92}. Each manifold  is characterized by its average
velocity $\upsilon_{i} = \langle \upsilon_{i}(t) \rangle_t$. A manifold velocity might be nonzero even  when all
relevant symmetries hold. For example, if we launch the particle  from one of the points marked  by triangles on the left panel
of Fig.~\ref{figHam2}, the particle will move ballistically with near constant positive velocity. Therefore
the quest for a nonzero current
by a Hamiltonian ratchet is intimately related to the choice of initial
conditions.
If the set of initial conditions overlaps with different manifolds
then the corresponding asymptotic current  should be calculated as a weighted sum over
the velocities of the contributing manifolds, $\tilde{J} = \sum_i p_i \upsilon_i$, where
$p_i$ is the fraction of the ensemble that overlaps with the $i$-th manifold, $\sum_i p_i = 1$.

Among many different invariant manifolds, the  chaotic layer
around the line $p=0$, see Fig.~\ref{figHam2}, is of
particular importance. This manifold is a result of the destruction of the
separatrix of the non-driven system (\ref{pot_ham}) when the latter is exposed to the drive.
The chaotic layer typically overlaps substantially with an initial ensemble of particles
of low-kinetic energies, for example, an ensemble  with a low-temperature Maxwell velocity distribution  \cite{fyz00prl,sokd01prl}.
Therefore, the main chaotic layer is  the most relevant region of the phase space in the context of
cold- and ultra-cold atom experiments \cite{jgr05prl, gbr05prl}.
The chaotic layer is an ergodic manifold \cite{Ott1992, suz92} and its average
velocity, $\upsilon_{ch}$, is the same for all trajectories initiated within the layer. Thus, the asymptotic chaotic current,
Eq.~(\ref{2.5}),  is $\tilde{J} = \upsilon_{ch}$.
When one of the two fundamental symmetries is present, i. e. when  Eq.~(\ref{2.1}) is invariant under either  $ {\widehat
S}_{x}$ or ${\widehat S}_{t}$,  the corresponding transformation
maps every trajectory from the layer  onto another one, also belonging to
the layer but having opposite velocity.
Due to mixing \cite{suz92}, both trajectories can be considered as
parts of a single infinitely long trajectory. The presence of the symmetry implies that the
asymptotic velocity of any trajectory, initiated within the chaotic layer, is strictly
zero and thus $\upsilon_{ch} = 0$. When both
symmetries are broken,  like in the case of the set-up (\ref{pot_ham}, \ref{driv_ham}) with  $\theta \neq 0, \pm \pi$,
we expect the appearance of a directed  current within the layer, $\upsilon_{ch} \neq 0$.

There are two possibilities to estimate $\upsilon_{ch}$.
The first  is straightforward: Calculate $\upsilon_{ch}$ by propagating  numerically  a very long trajectory.
We have already resorted to this idea in Section  \ref{Hamiltonian ratchets}.
However, this brute-force  approach has several drawbacks.  First, it is a time-consuming task.
The trajectory has to be long enough in order to
sample the chaotic layer, and it is hard to predict when the
average velocity, $\bar{\upsilon} =  x(t)/t$, will saturate to the asymptotic value $\upsilon_{ch}$ within a given accuracy.
Often, in order to get a
clue about even the direction of the chaotic transport, one has to run  trajectory for  times which are
six to seven  orders of magnitude larger then the period of the driving $T$. Second, a numerical scheme for the propagation of
Hamiltonian systems has to account for the symplectic nature of the dynamics. Otherwise accumulations of numerical round-offs
will lead to wrong results. Standard schemes, like the Runge-Kutta forth-order algorithm, are not suitable for this task  and
one should use symplectic integrators instead\footnote{An easy-to-use symplectic scheme which performs extremely well when used
to propagate Hamiltonian ratchets of the type (\ref{pot_ham}, \ref{driv_ham}) \cite{fyz00prl,dfoyz02pre, df01pre}, is the Verlet
(also coined `leap-frog') algorithm \cite{Swope1982}. One can use more sophisticated  symplectic integrators \cite{Leimkuhler2005}.}.

\begin{figure}
\begin{center}
\includegraphics[angle=0,width=0.95\textwidth]{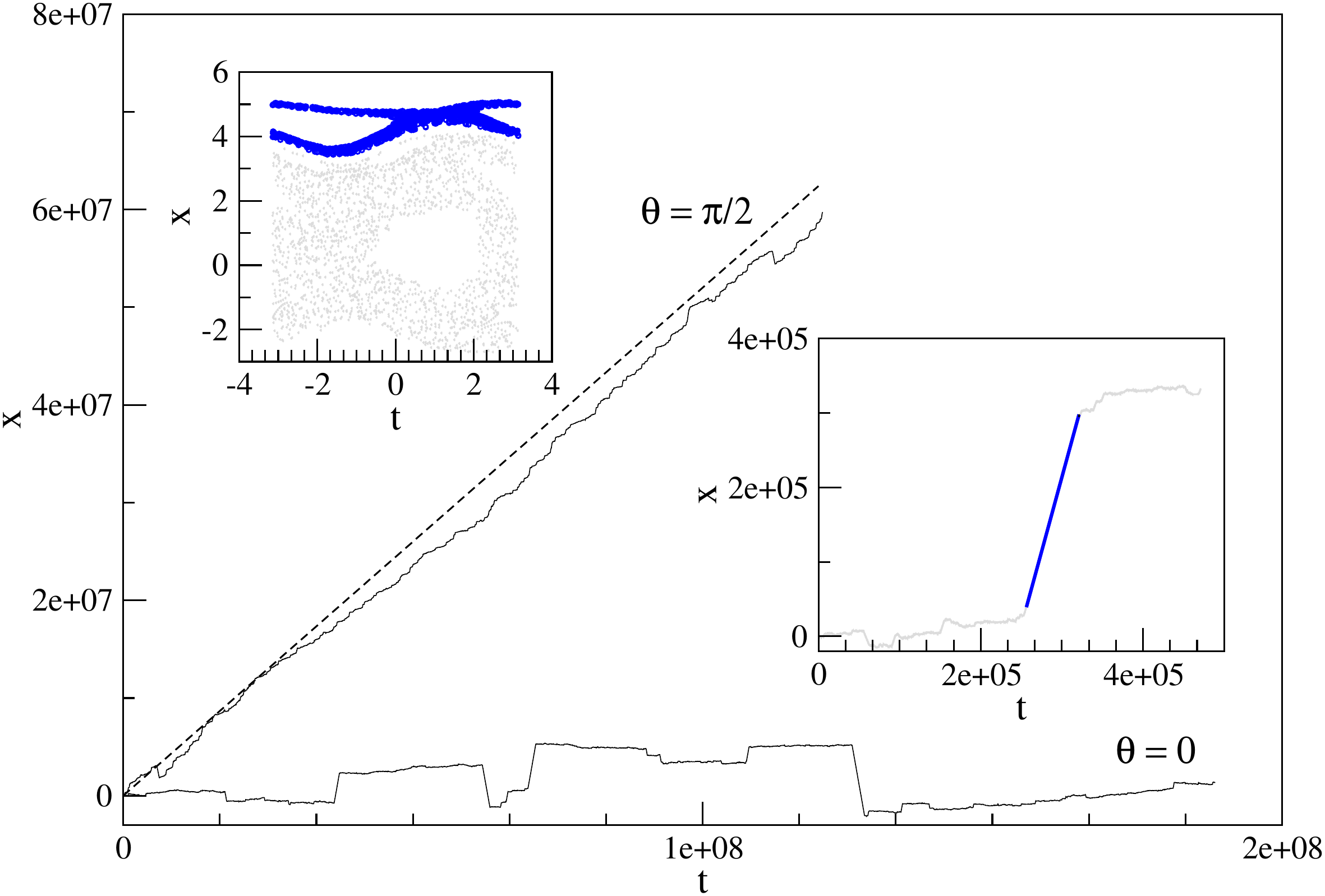}
\caption{(color online) Left panel: $x(t)$ versus $t$ for the Hamiltonian system  (\ref{2.1}, \ref{driv_ham}).
The dashed line corresponds to $x = \upsilon_{ch}t$, with $\upsilon_{ch}$ obtained by using the sum-rule, Eq.~(\ref{sum_rule})
Right lower inset: A zoomed part of the trajectory with the thick blue line marking a single ballistic flight
with velocity $\upsilon = 4$. The gray part corresponds to a diffusive-like dynamics within  the chaotic sea. Left upper inset:
The Poincar\'{e} map corresponding to the trajectory part shown on the right inset.
Blue dots  correspond to the ballistic flight and gray dots to the diffusion-like part of the trajectory.
The parameters are the same as in Fig.~\ref{figHam2}.} \label{figHam3}
\end{center}
\end{figure}

An alternative approach is based on the so-called `sum rule'  \cite{sokd01prl}. This method is capable of estimating the asymptotic  current
with a high accuracy while avoiding long-time numerical propagation. The sum rule expresses the averaged velocity in the \textit{chaotic} layer
in terms  of kinetic characteristics of its adjoining \textit{regular} components. Below we  give a brief recipe  while referring
interested readers to Refs. \cite{sokd01prl, Schanz2005} for more detailed explanations.

\begin{itemize}

\item Launch a trajectory from a point which belongs to a chaotic layer. A good choice is to use initial momentum value close to zero, and the coordinate value close to a maximum of the potential.
It may happen that we will have to perform several short trial runs in order to be sure that we really got into the chaotic layer.
Next propagate the trajectory  and  plot the corresponding Poincar\'{e} map.
Continue until  the chaotic layer structure becomes visible, i.e. its boundaries  and inner regular islands are resolved
(typically $10^{3...4}$
periods $T$ are sufficient).

\item The obtained chaotic layer is  confined  between two  tori, upper $\Gamma_{u}$ and lower
$\Gamma_{l}$ ones, and occupies a finite area $A_{ch}$ on the Poincar\'{e} map. The layer may also enclose
different regular islands of  areas $A_{i}$, see `holes' on Fig.~\ref{figHam2}. Each island is characterized by an averaged velocity  $\upsilon_{i}$.
These velocities can be estimated by launching short trajectories from the interior of the corresponding island.
It is not a time-consuming task since the dynamics within islands is  regular (quasiperiodic);

\item Calculate  the average kinetic energies of the border tori,
$\bar{K}_{u,l} = \langle p_{u,l}^{2}(t)/2 \rangle_{t}$. This can be done by launching trajectories
from the initial points that are outside the layer but close to the  layer boundaries;

\item Finally, calculate the asymptotic current by using the following expression \cite{sokd01prl}\footnote{
Note the absence of the factor $L$ in the numerator of the original expression in Refs. \cite{sokd01prl, Schanz2005}.
When $L=1$ (which was the case considered in the papers), this absence causes dimensionality problems only.
However, in  the general case, when $L\neq1$, it leads to incorrect results.},
\begin{equation}\label{sum_rule}
J=\frac{L \cdot [\bar{K}_{u} - \bar{K}_{l}]-
\sum_{i}A_{i}\upsilon_{i}}{A_{ch}}.
\end{equation}

\end{itemize}

As an illustration, we apply the sum rule to the system shown on Fig.~\ref{figHam2}b. The area of the chaotic
layer is  $A_{ch} \approx 41.1 \pm 0.03$\footnote{We have measured  areas of the chaotic region and regular islands by filling them with
rectangles, circles and ellipses. We used  a free-ware Linux 2d plotting program, \textit{Grace}, \url{http://plasma-gate.weizmann.ac.il/Grace/}}.
The area of the central island, with $\upsilon_1 = 0$, and the ballistic island at the top of the layer (of the
velocity $\upsilon_2 = 4$) are $A_1= 4.4 \pm 0.01$ and $A_1= 2.8 \pm 0.01$ correspondingly. Kinetic energies of the boundary
tori are $T_u = 10.20 \pm 0.01$ and $T_l = 5.05 \pm 0.01$. Finally, the sum rule (\ref{sum_rule})
yields $\upsilon_{ch} = 0.521 \pm 0.006$ \footnote{We neglected the contributions from smaller islands and therefore
the actual relative error is larger.}. The obtained value is in a good agreement with the result of the numerical
propagation, see the left panel of Fig.~\ref{figHam3}.

Although being characterized by a uniform invariant density in the asymptotic limit  \cite{sokd01prl, Schanz2005}, the chaotic layer is not
uniform on finite time scales. The dynamics of a Hamiltonian system is qualitatively different in different regions of the chaotic sea
though all these regions  belong to the same, overall ergodic  manifold.
In particular, the dynamics is nearly regular at the vicinity of embedded regular
islands. These regions of the layer are structured by  \textit{cantori} \cite{suz92}, which form partial barriers
for chaotic trajectories. A trajectory, which entered the region enclosed by a cantorus,
can be trapped in the vicinity of the corresponding island for a very long time  \cite{suz92, szk93n}.
During this \textit{sticking} event \cite{Meiss1986},
the trajectory reproduces the dynamics of the orbits located inside the island \cite{szk93n, df01pre}.
If the corresponding island is transporting, i. e. $\upsilon_i \neq 0$,  the sticking event produces a ballistic flight,
see lower right panel  of Fig.~\ref{figHam3}. A strong  current appears when the set of regular islands, submerged into the chaotic layer,
is asymmetric, i. e. when there are islands with nonzero velocities which do not have symmetry-related twins.
This leads to the violation of the  balance between ballistic flights
in opposite  directions and results in the appearance of a strong current  \cite{df01pre}. Functioning of Hamiltonian ratchets
relies therefore  on the harvesting of long temporal correlations that extend over the time scales
much larger than the period of the driving $T$. Ballistic flights are also  responsible  for
anomalous diffusion \cite{szk93n}, when the mean square
displacement, $\sigma(t)= \langle x^{2}(t)\rangle - \langle
x(t)\rangle^{2}$, grows  algebraically, $\sigma(t) \propto
t^{\mu}$,  with the scaling exponent $\mu
> 1$ \cite{zk95pre}. Naturally, directed current and superdiffusion represent two facets of the peculiar Hamiltonian
kinetics \cite{dkuf01pd, Denisov2004}.
In some limits, however, it may become difficult to separate ballistic flights from non-ballistic chaotic
diffusion.
This is a typical situation  in the regimes of strong driving, when $E_1, E_2 \gg 1$.
With the increase of the driving amplitude(s) the chaotic layer starts to inflate,
absorbing more and more  ballistic islands.
The chaotic sea is becoming structured by a network of cantori \cite{suz92},
so that the chaotic dynamics within the layer is far outside of
the dichotomous description ``unbiased chaotic diffusion vs  ballistic flights''.

While the symmetry analysis does not predict sign and value of a ratchet current, it
allows to predict  when the sign of the current will invert.
Namely,
either of the following transformations, \begin{eqnarray}
\theta \rightarrow \theta + \pi, ~t \rightarrow t+T/2, ~x
\rightarrow -x + \pi,  \label{Ham_tr1} \\
\theta \rightarrow -\theta, ~t \rightarrow -t, ~x
\rightarrow x,  \label{Ham_tr2}
\end{eqnarray}
reverses the current, $J \rightarrow -J$. Therefore, in order to change  the direction of induced transport, it is enough  to
change the sign of $\theta$ or to  shift it by $\pi$.
Evidently, the current is a periodic function of the phase shift $\theta$,  $J(\theta +2\pi) = J(\theta)$. Thus it
can be expanded into a Fourier series, $J(\theta)=\sum J_k \exp(ik \theta)$.
It follows from symmetry (\ref{Ham_tr2}) that the expansion
consists of sine terms only. In addition, symmetry (\ref{Ham_tr2}) predicts that
the Fourier series consists of odd harmonics only, so that $J(\theta)=J_{1}\sin(\theta) + J_{3}\sin(3 \theta)+ ...$.
Assuming that the  first  term of the expansion  dominates,
we arrive at a simple expression,
\begin{eqnarray}
J(\theta) \propto J_{0}\sin(\theta). \label{current_expans}
\end{eqnarray}
This dependence was detected
in the experiments with ac-driven cold \cite{ss-prg03prl}  and ultracold \cite{weitzScience}
atom ratchets, see Sects.~~\ref{Experiments with cold atoms} and ~\ref{Experiments with Bose - Einstein condensate}, respectively.

\subsubsection{Transport with dissipation}
\label{Transport with dissipation}

In the case of finite dissipation, $\gamma/m < \infty$,  the only  symmetry transformation to take care about is $\hat S_x$,
see Table \ref{tabl1}. One way to violate this symmetry  is to use, for example,  the rocking set-up with a potential force $f(x)$
that is not antisymmetric. A bi-harmonic combination, Eq.~(\ref{u2}), will do in this case \cite{fyz00prl,jkh96prl,m00prl,bs00pre}.

In the dissipative regime, similar to the Hamiltonian limit, the asymptotic current is determined by the transport properties of
invariant manifolds. There can be several manifolds \cite{fghy96, Rodrigues2009} but  typically
their number is much smaller than that in the Hamiltonian case.
The are two kinds of invariant manifolds existing in the  phase space of a dissipative system, that are
\textit{attractors} and \textit{repellers} \cite{Ott1992}. In the asymptotic limit $t\rightarrow \infty$, the dynamics of a
noise-free dissipative system  is determined by the attractors only. Typically, there is only a few attractors coexisting
in the system phase space when the dissipation strength is not too small, $\gamma/m \gtrsim 0.01$. In the simplest case, when
only one attractor exists in phase space, all  trajectories end up on this manifold independent of their starting points.
Therefore, similar to the  case of Hamiltonian dynamics,
in order to obtain information on the asymptotic transport, one should study  the transport
properties of the system attractor(s).

There are two types of attractors in periodically driven dissipative system, namely periodic limit cycles
and chaotic attractors\footnote{There is also a possibility
to produce a quasiperiodic attractor,
but for that one  needs either  a driving force with  non-zero constant component  \cite{Ott1992} or
 quasiperiodic  driving  \cite{np02epjb},  see Section \ref{Extension to quasiperiodic driving}}.
An attractor is characterized by its average velocity, $\upsilon_A = \langle v_A(t) \rangle_t$.
A limit-cycle attractor is locked by the driving field $E(t)$, and its transport properties are fully determined by a pair of
co-prime integers, $m$ and $n$, $x(t+mT)=x(t) + Ln$, $p(t+mT) = p(t)$, so that the average velocity  is $\upsilon_A = nL/mT$.
If this is the only attractor of the system, its velocity defines the asymptotic current, $\tilde{J} = \upsilon_A$.
In the multi-attractor case, the phase space  contains several coexisting attractors, all of different types and different velocities.
Each attractor has its own \textit{basin of attraction} \cite{Ott1992}. That is a part of the phase space,
$\varSigma_A$, such that a trajectory
launched from a point
$(x_0, \, p_0 , \, \Omega_0) \in \varSigma_A$  ends up, after some transient, on
the corresponding attractor. Basins of different attractors are often entangled and form complex fractal-like
structures \cite{Ott1992, Aguirre2009}.

\begin{figure}
\begin{center}
\includegraphics[angle=0,width=0.85\textwidth]{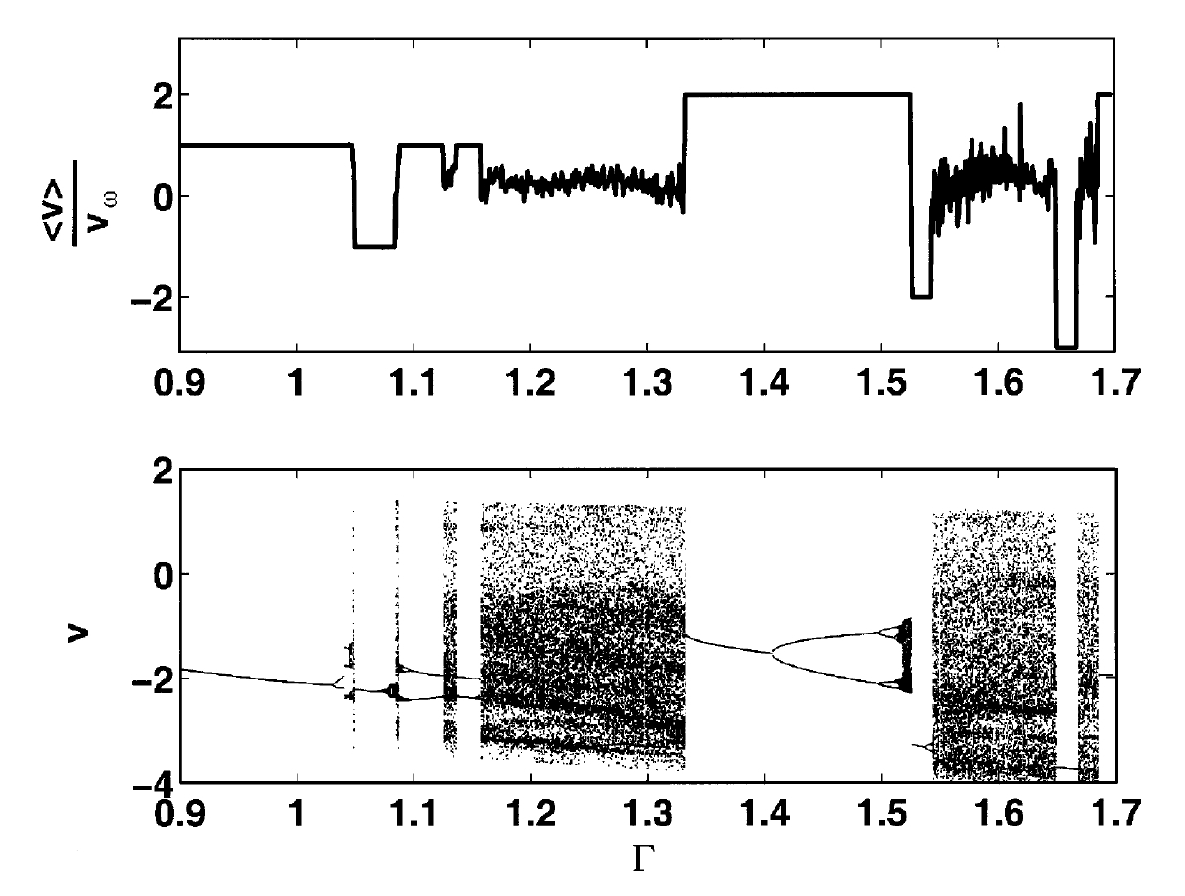}
\caption{(top) Asymptotic velocity of a dissipative ratchet vs driving
amplitude, $E_1 \equiv \Gamma$.
The equations of motion are given by (\ref{2.21},\ref{u2},\ref{e2}).
The velocity is measured in units of $\upsilon_\omega = L/T = \omega$;
(Bottom) Bifurcation diagram, i. e. stroboscopic values of the particle velocity, $v(t_i) = p(t_i)/m$, plotted at
the instants of time $t_i = (i + 1/2)T$, $i=1,2,...$.
Note the bifurcation from the chaotic attractor to the limit cycle at the point $\Gamma \approx 1.65$, which
leads to a current reversal. The parameters are $m=1.1009$, $f_1=1$, $f_2=0.5$, $\Delta = \pi/2$, $\gamma = 0.1109$, $\omega = 0.67$,
and $E_2 = 0$. The initial conditions are  $\{x(0)=0, p(0)=1.1009\}$. Adapted from Ref. \cite{Larrondo2003}.} \label{larrondo}
\end{center}
\end{figure}

Let us consider first a situation when   symmetry
$\widehat{S}_x$ holds. In the single-attractor case, the
corresponding transformation maps the attractor
onto itself, and, therefore, $\upsilon_A = 0$.
There are two alternatives in the multi-attractor case. Namely, the transformation  (i) either maps an attractor $A$ onto itself,
and therefore this attractor again is  non-transporting or (ii) it maps an attractor onto its symmetry-related twin
$A'= \widehat{S}_x:A$, such that $\upsilon_A = - \upsilon_{A'}$.
The symmetry also enforces the basins of attraction of two symmetry-related attractors
to be mapped onto each another by the  same transformation. Therefore,  a distribution of initial conditions
${\mathcal F}(x_{0}, p_{0}, t_{0})$ which occupies the same volume in both basins (note that it must not even
be strictly symmetric point by point),
will yield zero  current \cite{fyz00prl}.

By violating  symmetry  $\widehat{S}_x$ we remove the above constraints and expect a nonzero current \cite{r02pr, hm09rmd}.
In the case of  a single limit-cycle attractor, the transition to  the ratchet regime  must necessarily
involve a bifurcation \cite{Ott1992}, i. e. a sudden change of the attractor structure, since it is impossible to tune continuously  between zero
and any nonzero rational numbers $n/m$.
Therefore, this bifurcation will happen only after some finite parameter tuning.
So it may happen that for these restrictive topological reasons  the overall current will stay zero
even though all symmetries have already been  broken.

In the case when two symmetry-related limit cycles coexist,
the transition to the ratchet regime can be smooth.
A violation of the symmetry will cause the
desymmetrization of the  basins of attraction so that the averaging with a symmetry-respecting  distribution function,
${\mathcal F}(x_{0}, p_{0}, \Omega_{0})$, Eq.~(\ref{average_J}), will result in a finite asymptotic current.
Further increase of an asymmetry parameter will lead to a bifurcation after which  one of the attractors disappears \cite{fyz00prl}.
In the case of a single chaotic attractor, the transition to the ratchet regime  can be continuous as well.
Variations of the asymmetry parameter(s) may cause different  bifurcations, with a
birth of a new attractor and death of an old one, via, for example, a period-doubling bifurcation or
an inverse tangent bifurcation \cite{Ott1992}.
There is a multitude of possible  bifurcation scenarios \cite{J.M.T.Thompson1994},  and each of them can show up
upon parameter variations. A bifurcation changes the attractor structure, thus leading to  sudden changes
of the current, up to current reversals \cite{m00prl, bs00pre, m03pa, Larrondo2003, Celestino2011}, see Fig.~\ref{larrondo}.

\subsubsection{Transport in the overdamped limit}
\label{OverDampedSect}

The symmetries in the overdamped limit, $m = 0$, were elaborated in great detail in Refs. \cite{fyz00prl, r01prl}, while
a big variety of  overdamped ratchet models is reviewed in Refs. \cite{r02pr, hm09rmd}.
Below we only briefly outline this limit by using the rocking ratchet set-up introduced in Section \ref{Additive driving},
as a model.

The dynamics of a particle is described by the following equation,
\begin{equation}\label{eq_overdamped}
 \gamma \dot{x}=f(x) + E(t).
\end{equation}
The asymptotic evolution, similar to the above considered general dissipative case, is governed by attractors \cite{Ott1992}, and
symmetry  $\widehat{S}_x$ acts here in the same manner as before.
Surprisingly, time-reversal symmetry $\widehat{S}_t$ Eq.~(\ref{over_t}) is acting again,
which appears to be odd.
The time reversal operation will map an attractor;  i.e., a  manifold, which attracts trajectories from
a part (or whole) of the
phase space, onto a repeller \cite{Ott1992}, an unstable manifold which repels trajectories and seemingly does not
influence the asymptotic dynamics on large time scales. Therefore, the line of reasoning used before for Hamiltonian
systems does not apply here: Contributions from
two symmetry-related  manifolds do not cancel each other because their statistical weights are different.
The resolution of this paradox is that the time-reversal symmetry, also termed  'supersymmetry' in Refs. \cite{r01prl, r02pr},
imposes  certain restrictions not directly on the equations of motion themselves but on
their asymptotic solutions.
We will address this issue  in  subsection \ref{From Langevin dynamics to the Fokker-Planck equation},
when discussing the approach based on the Fokker-Planck equation.

\begin{figure}
\begin{center}
\includegraphics[angle=0,width=0.85\textwidth]{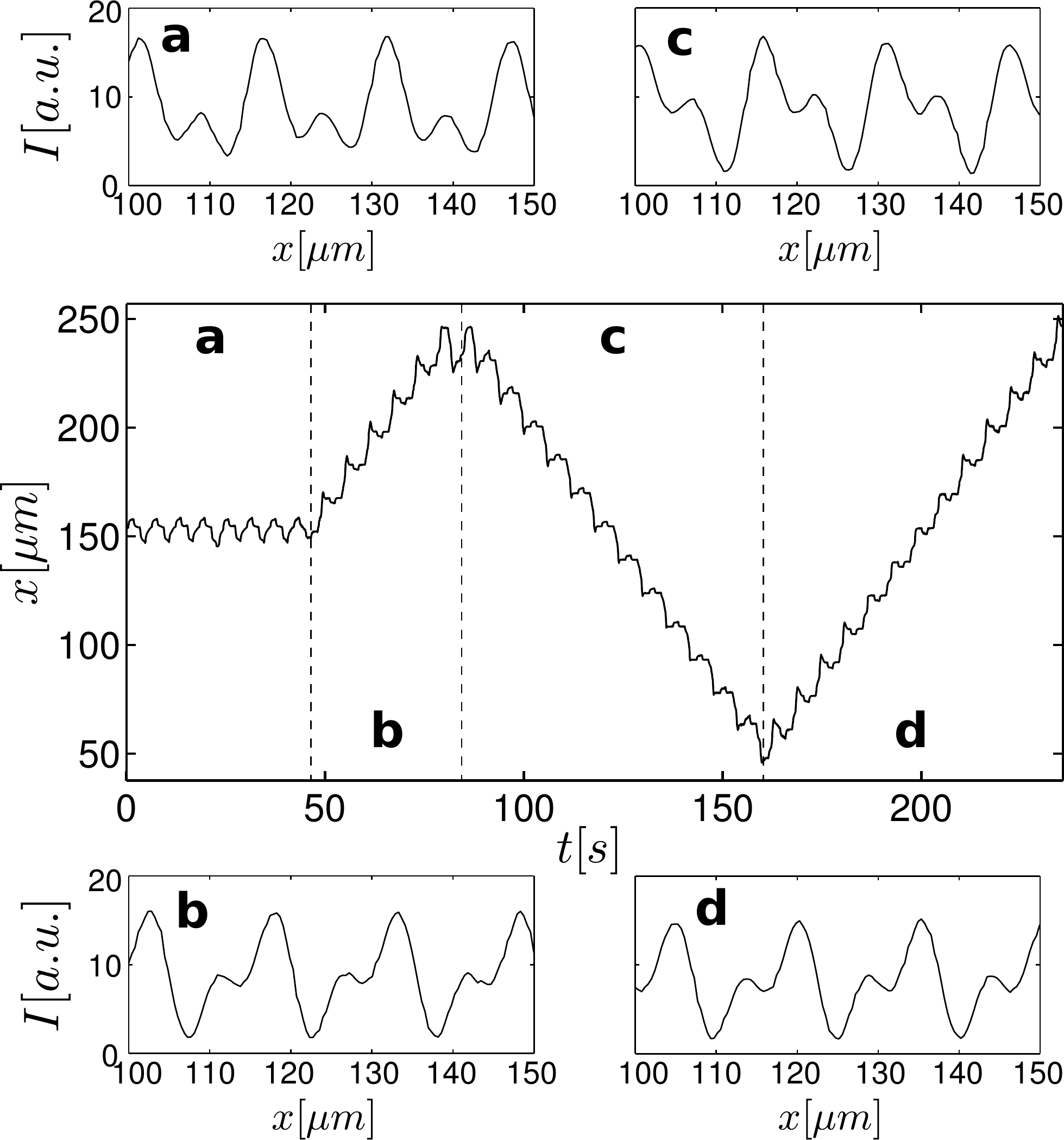}
\caption{Central part: Monitored position of a glass microsphere vs time obtained with optical realization of the overdamped rocking ratchet,
Eqs.~(\ref{eq_overdamped},\ref{mateos_force},\ref{mateos_drive}) \cite{avm11}.
The dashed lines indicate instants of time
when  the parameter $\Delta$ of the optical force, Eq.~(\ref{mateos_force}), was changed to a new value.
The four time regions correspond to (a) $\Delta = \pi/2$, (b) $\Delta = 0$, (c) $\Delta = \pi$,
and (d) $\Delta = 0$.  Top and bottom plots show  the profiles of optical forces corresponding to different different values of $\Delta$.
Adapted from Ref.~\cite{avm11}.} \label{mateos}
\end{center}
\end{figure}

Recently, time-reversal symmetry $\widehat{S}_t$ was used   for the real-time control of  microsphere transport
in a fluid  \cite{avm11}. The bi-harmonic potential force,
\begin{equation}\label{mateos_force}
f(x)= f_1\cos(x) + f_2\cos(2x + \Delta),
\end{equation}
has been  realized with a periodic pattern of fringes produced by interference of several
laser beams. A rocking ratchet was introduced by means of phase modulations of the beams, Eq.~(\ref{modulation}), which procedure
resulted in the appearance of a three-state tilting force,
\begin{equation}\label{mateos_drive}
E(t)=
\begin{cases}
E_0 & \text{if}\ \ 0\leq t< \tau_1,\\
0 & \text{if}\ \ \tau_1\leq t< \tau_1+\tau_0,\\
-E_0 & \text{if} \ \ \tau_1+\tau_0\leq t< T-\tau_0,\\
0 & \text{if} \ \ T-\tau_0\leq t< T,
\end{cases}
\end{equation}
with the period $T = 2(\tau_0+\tau_1)$. The function $E(t)$ has zero mean and possesses all three possible symmetries,
Eqs.~(\ref{SymmFunc}, \ref{A-SymmFunc}, \ref{Sh-SymmFunc}). It is symmetric around the point
$t = \tau_1/2$, antisymmetric around the point $t = \tau_1 + \tau_0/2$, and, consequently, is shift-symmetric under the shift
$\tau = \tau_1 + \tau_0 = T/2$.
The bi-harmonic form of the potential force (\ref{mateos_force}) violates symmetry $\widehat{S}_x$ by default.
Symmetry $\widehat{S}_t$, Eq.~(\ref{over_t}),  holds when $\Delta = \pm \pi/2$. Any other choice of $\Delta$ violates the time-reversal symmetry
and may lead to the appearance of net transport of the microsphere. In addition, the shift of $\Delta$ by $\pm\pi$ should
reverse the direction of the motion.
The obtained experimental results perfectly validated all these predictions,
see Fig.~\ref{mateos} (and Ref.~\cite{avm11} for more details).

\subsection{Transport with noise and fluctuations}

Many applications require
statistical descriptions of transport properties, especially when the system of interest is coupled
to a heat bath so that the system evolution is essentially stochastic. Thermal noise will change the dynamics of a deterministic dissipative system
by allowing the latter to explore the phase space outside the attractor(s).
This results in a self-averaging of the dynamics over the phase space so that the system asymptotic
state is no longer localized on the attractor(s) but should be described  by a certain distribution function, $P(x,p,t)$.
This function  depends on the temperature of the heat bath $\mathcal{T}$,
$P(x,p,t;\mathcal{T})$. The symmetries of the equations of motion of the noise free case will be recovered in corresponding
evolution equations for the distribution function $P$.

We recapitulate that in the Hamiltonian noise-free limit the phase space is mixed,
and regular islands of a Hamiltonian system enclose their 'cores', marginally stable
elliptic periodic orbits. These orbits are characterized by  velocities $\upsilon = nL/mT$, given by  pairs of co-prime
integers, $n$ and $m$. Weak dissipation (with still no noise) transforms the chaotic layer into a set of limit-cycle attractors,
which are located inside the former regular islands \cite{fghy96, dfoyz02pre}.
The velocities of these limit cycles are equal to the velocities of their predecessors, the elliptic orbits,
while their basins of attraction form complex fractal-like structures in the region of the phase space
which was  occupied by the Hamiltonian chaotic layer.
Additional thermal noise induces jumps  between the
attractors \cite{Feudel2002}, and in between two consecutive jumps the trajectory
may stick to one of attractor for a relatively long time.
If  the  coupling to the bath is weak, the trajectory  mainly explores the phase-space region corresponding to the former chaotic layer.
The joint effect of weak damping, weak noise and complex geometric structure  of the attractor basins results in
velocity probability distribution which is far from the conventional  Maxwell's distribution \cite{graham,dfoyz02pre}.
While it is tempting to perform direct numerical integrations of stochastic differential equations like the
Langevin equation, such approaches appear to be notoriously hard, since all kinds of transient effects,
convergence rates and relaxation times, and other technical issues, will typically make this approach
not very efficient.

Statistical approaches, based on the direct evaluation of the distribution function, $P(x,p,t)$,
can lead to the needed asymptotic average characteristics of the system.  Below we explain how the symmetry analysis can
be generalized within two complementary statistical formalisms. We will consider first  the  method based on the kinetic Boltzmann equation,
an approach frequently used  in  condensed matter physics \cite{LifPit1981}.
Next we will review an  approach based on
the Fokker-Planck equation,  a popular tool  of computational statistical physics \cite{risken1989}. As an illustration we use the
tilting-ratchet setup  with  bi-harmonic potential and driving,
\begin{eqnarray}
\label{kinetic1} V(x) & = & V_1\cos (x) + V_2 \cos (2x + \Delta) , \\
\label{kinetic2} E(t) & = & E_1 \cos ( \omega t ) + E_2  \cos ( 2\omega t + \theta) \, ,
\end{eqnarray}

\subsubsection{Molecular chaos assumption and the Boltzmann equation }

The  kinetic Boltzmann equation for the case of additive drive
reads \cite{yfzo01epl}
\begin{eqnarray}
\hat{{\mathcal L}} P \equiv \partial_t P
+ \dot{x} \partial_x P +
\dot{p}  P = {\mathcal J}(P,F)~,
\label{KE-1}
\\
 \dot {x}=p,~~\dot{p}=g(x,t)=-V'(x)+E(t)~,
\label{KE-1a}
\end{eqnarray}
where $P = P(x,p,t) $ is the nonequlibrium distribution function, $F=F(x,p) $
is some \textit{equilibrium} distribution function,
\begin{equation}
F(x,p)\equiv F_x(x)F_p(p)=\frac{e^{-p^2/2}}{\sqrt{2\pi}}\cdot \frac{e^{-V(x)}}{\Xi},~
\Xi=\int_0^{2\pi}e^ {-V(x)} dx,
\label{EQD}
\end{equation}
and ${\mathcal J}(P,F)$ is a collision integral. We use a collision integral of the type \cite{LifPit1981}
\begin{equation}
  {\mathcal J}(P,F) = - \gamma (P-F) \, .\label{KE_coll}
\end{equation}
The dissipation constant $\gamma$ determines the characteristic relaxation time, $\tau = \gamma^{-1}$.
The dissipative linear equation (\ref{KE-1}) has a unique limit-cycle solution,
$P_A(x,p,t+T) = P_A(x,p,t)$. The asymptotic current is given by the attractor velocity,
\begin{eqnarray}
\label{jdc}
J  =  \langle (p/m) \cdot P_A(x,p,t) \rangle_{x,p,t} = \\
\frac{1}{T \sqrt{2 \pi} \Xi}
\int_0^{T}  \int_{-\infty}^{\infty} \int_0^{2\pi} (p/m) \cdot P_A(x,p,t) dt~dp~dx. \nonumber
\end{eqnarray}

In the overdamped limit,  where $\gamma^{-1}$  sets the shortest timescale of the system dynamics,
the limit-cycle solution  can be obtained by expanding the  function $P_A(x,p,t)$
into a power series over $\gamma^{-1} $. In the case $V_2 = 0$ and $E_2 \neq 0$, the
first non-zero term of the expansion appears in seventh order of $\gamma^{-1}$ \cite{yfzo01epl},
\begin{equation}
J \simeq -\frac{45 }{4\gamma^7 } \frac{I_1(V_0)}{I_0(V_0)} E_2E_1^2
        \cos( \theta) \,  .\label{BE_pertur}
\end{equation}
where $I_n(z)$ is the modified Bessel function of $n$-th order \cite{Abramowitz1965}. The average current is proportional to the term
$\cos (\theta )$, reflecting the fact that the current disappears when  $\theta = \pm \pi/2$, i. e. when
symmetry $\widehat{S}_t$ holds, see Table 1.

The dissipative case cannot be handled analytically in the general case.
However,  equation (\ref{KE-1}) can be solved numerically by using  the expansion of the distribution
function $P_A(x,p,\Omega)$ into a series over Hermitian
polynomials in $p$-space and into Fourier series in the $x$-space.
Then, after a proper basis truncation, one ends up
with a system of ordinary linear differential  equations for the expansion coefficients $A_{n,m}(t)$, where
$n$ and $m$ denotes the order of the Hermitian polynomial and Fourier component, correspondingly.
The set of equations can be  propagated numerically, and, after some transient evolution, it will converge
to a limit-cycle  attractor, $A_{n,m}(t+T) = A_{n,m}(t)$.
This corresponds to the attractor solution,  $P_A(x,p,t)$,
from which one could calculate the asymptotic current by using Eq.~(\ref{BE_pertur})\footnote{There is an alternative approach
based based on the so-called method of characteristics. The expansion method is optimal for the case of moderate dissipation,
$ \, \gamma/\omega \ge 0.01 $,  but both methods demonstrate a  good agreement in the range  $ \, \gamma/\omega
\gtrsim 10^{-2} \div 10^{-1}$ \cite{yfzo01epl}.}.

\begin{figure}[t]
\begin{center}
\includegraphics[angle=0,width=0.65\textwidth]{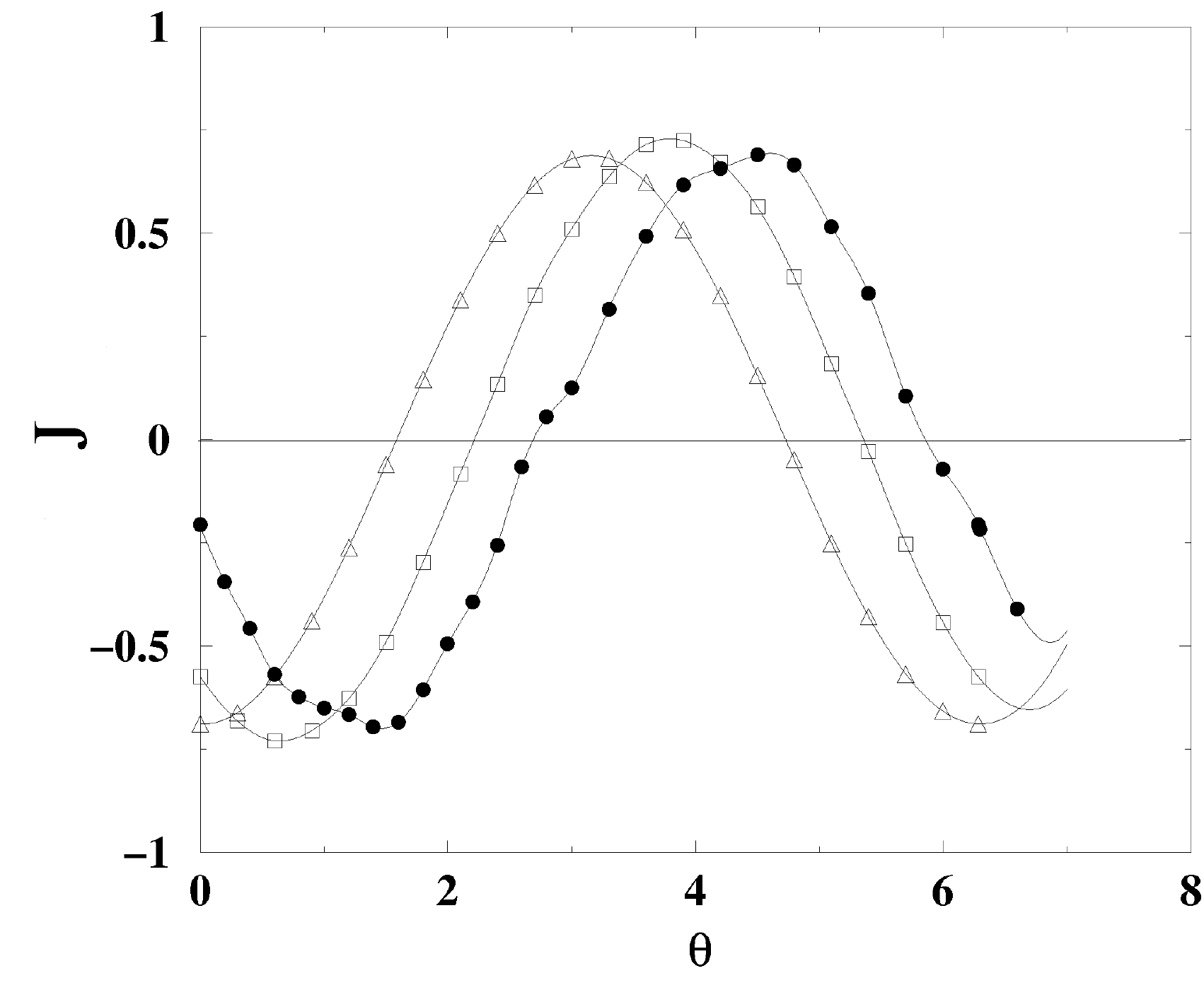}
%
\caption{Dependence of the asymptotic current $J$, Eq.~(\ref{jdc}), on the phase shift
$\theta$ for different values of dissipation parameter: $\gamma=0.3$ (filled circles),
$\gamma=1$ (squares), and $\gamma=4$ (triangles). Note that the
current values are scaled by a factor of 4.86 ($\gamma=1$) and
181.6 ($\gamma=4$). The parameters of equations (\ref{kinetic1} - \ref{KE-1a})
 are $V_1 = 6$, $V_2 = 0$, $E_1 = -2.6$, $E_2 = -2.04$, $\omega = 0.85$. Adapted from Ref.~\cite{yfzo01epl}.
}
\label{fig3-KU}
\end{center}
\end{figure}

In Fig.~\ref{fig3-KU} we show the dependence of the asymptotic current $J$ on
$\theta$ for  different values of the dissipation strength.
The results demonstrate the crossover between
the overdamped and dissipative cases. In the case of strong dissipation, $\gamma = 4$,  the asymptotic current
vanishes when $ \theta\ $ is close to  $ \pm \pi/2$, i.e. when the symmetry
${\widehat S}_t$ is restored at the overdamped limit.
When the dissipation is weak, $\gamma = 0.3$,
ratchet current drops to zero at the vicinities of points $\theta =  0, \pm \pi$,
when the symmetry ${\widehat S}_t$ is restored at the Hamiltonian limit.
It is noteworthy that the transition from the overdamped limit
to the limit of weak dissipation leads to an overall current
enhancement  of more than two orders of magnitude.
Another interesting result is that the dependence
$J(\theta)$ preserves its smooth, sine-like form so that
the effect of dissipation can be quantified with a phase-lag $\theta_0$ and some overall  prefactor,
\begin{equation}
\label{lag}
J \sim J^0_{\gamma} sin[\theta - \theta_0(\gamma)],
\end{equation}
where $\theta_0 = 0$ in the  Hamiltonian limit and $\theta_0 = \pm \pi/2$ in the overdamped limit.


\subsubsection{From Langevin dynamics to the Fokker-Planck equation}
\label{From Langevin dynamics to the Fokker-Planck equation}

The coupling to a heat bath can be accounted for by introducing  noise terms in the rhs
of Eq.~(\ref{2.1}),
\begin{equation}
\label{langevin}
   m {\ddot x}+\gamma {\dot x} =  g(x,t) +\xi(t),
\end{equation}
where $\langle \xi(t) \xi(t')\rangle = 2\gamma k_B \mathcal{T} \delta(t-t')$.
Thermal noise $\xi(t)$ has all possible symmetries in the statistical sense. From any particular realization of the noise, $\xi(t)$, we can obtain
another one, $\xi'(t)$, by applying to the former either of the transformations, Eqs.~(\ref{SpaceRevClass1d}, \ref{TimeRevClass1d}).
The realization $\xi'(t)$ can be taken as another noise realization produced by the same heat bath.
In this sense the  symmetry properties of Eq.~(\ref{langevin}) are completely the same
as of its deterministic predecessor, Eq.~(\ref{2.1})

The transport properties of the system (\ref{langevin}) can be evaluated in terms of the probability distribution function, $P(x,p,t)$, by
using the Fokker-Planck equation (FPE). Below we briefly review main results. The interested reader can find a detailed analysis
in  Ref.~\cite{dhm2009}.

For the dissipative cases the FPE reads\cite{risken1989}
\begin{equation}
\Big\{\frac{\partial}{\partial t}+ \frac{\partial}{\partial x} \frac{p}{m} -
\frac{\partial}{\partial p} [\frac{\gamma}{m} p - g(x,t)]-
\gamma m k_B \mathcal{T} \frac{\partial^{2}}{\partial
p^{2}}\Big\}P(x,p,t)=0, \label{fpe1_under}
\end{equation}
where $k_B$ is the Boltzmann constant.
The respective FPE for the overdamped limit, when
inertia is negligible, $m=0$, reads \cite{risken1989}
\begin{equation}
\{\gamma \frac{\partial}{\partial t}+\frac{\partial}{\partial x} g(x,t)- k_B \mathcal{T}
\frac{\partial^{2}}{\partial x^{2}}\}P(x,t) = 0. \label{fpe1_over}
\end{equation}

Equations (\ref{fpe1_under}) and ({\ref{fpe1_over}) are
linear, dissipative, and preserve the norm, $\int P dx dp$ for Eq.~(\ref{fpe1_under}) and $\int P dx$ for Eq.~(\ref{fpe1_over})
\cite{Jung}. In addition, the equations possess discrete time- and
space-translation symmetries so that  operations $x \rightarrow
x + L$ and $t \rightarrow t + T$ leave the equations invariant. For
given boundary conditions and a fixed norm, any initial
distribution, $P(...,0)$, will converge to a single time-periodic
attractor solution, $P_A(...,t)=P_A(...,t+T)$.
The current can be calculated from a spatially periodic solution of
the type $P(x,...)=P(x+L,...)$ \cite{r02pr}.

The averaged asymptotic current on the attractor $P_A(x, p, t)$ is given by Eq.~(\ref{jdc}) for the dissipative case,
and by \cite{r02pr}:
\begin{eqnarray}
&&J = {\gamma}^{-1}\langle g(x,t) \cdot P_A(x,t)\rangle_{T,
L}, \label{cur1_over}
\end{eqnarray}
in the overdamped limit.

It is easy to see that the FPE for the general dissipative case, Eq.~(\ref{fpe1_under}),
inherits all the symmetries of the corresponding equations of motions, including those for the Hamiltonian limit,
$\gamma = 0$.
However, the time-reversal symmetry for the overdamped limit, Eq.~(\ref{over_t}),  cannot be detected with Eq.~(\ref{fpe1_over}).
This symmetry  imposes certain restrictions not directly on symmetries the equation itself but on its solutions.
Consider the case of rocking ratchet,
Eq.~(\ref{kinetic1}). The corresponding attractor solution,
$P_A(x,t)$,  is periodic both in time and space. We define an operator
\begin{equation}
\widehat{W} = \frac{-\frac{\partial}{ \partial x}}{\gamma \frac{\partial}{\partial t}
-\frac{\partial ^2}{\beta \partial x^2} + E(t) \frac{\partial}{\partial x}}
\;.
\end{equation}
Then the attractor solution is the solution of a Lippmann-Schwinger-type integral
equation \cite{Weinberg1995},
\begin{equation}
P_A(x, t) = 1 + \widehat{W}f(x) P_A(x,t) \;.
\label{ls}
\end{equation}
Both operators, $\partial/\partial
t$ and $\partial / \partial x$, are anti-Hermitian in the space of $x,t$-periodic
functions.
Provided the conditions for $\hat{S}_t$ hold, i.e.
$E(t)$ is antisymmetric, see Tab.~1, the operator
$\widehat{W}$ has the following properties:
\begin{equation}
\widehat{W}^{\dagger} = - \widehat{W}(-t)\;,\;\widehat{W}(x+x_0) = \widehat{W}(x)\;.
\label{propertyT}
\end{equation}
Expanding Eq.~(\ref{ls}) in a formal series in $f(x)$ we obtain
\begin{equation}
P_s = 1 + \sum_{n=1}^{\infty} (\widehat{W} f(x))^n\;.
\end{equation}
With Eq.~(\ref{cur1_over}) we finally  obtain for the average current
\begin{equation}
J  = \sum_{n=1}^{\infty} \int f(x) (\widehat{W} f(x))^n dx\;dt\;.
\label{series}
\end{equation}
Since all terms in (\ref{series}) are real-valued, we conclude
that all integrals with even $n$ vanish when the symmetry $\widehat{S}_t$ holds, i. e.
the function  $f(x)$ is shift-symmetric,
and these terms contain odd powers of $f$ only.
All integrals with odd $n$ vanish because of the conditions
(\ref{propertyT}). Thus we conclude that indeed the average current exactly vanishes when $\hat{S}_t$ holds.

Equations (\ref{fpe1_under}, \ref{fpe1_over}) can be solved numerically,
by expanding $P(x,p,t)$ into the Fourier series over $x$ and $t$,
and into the Hermite polynomial series over $p$. This will result in a system of linear
algebraic equations for the expansion coefficients, which can be solved then by using
standard numerical diagonalization  routines \cite{dhm2009}.

\subsection{Experiments with cold atoms}
\label{Experiments with cold atoms}
Fast progress in the field of experimental manipulations with
cold and ultracold atoms  served  a  new class of  physical systems which  fulfill the conditions
required to observe symmetry-controlled directed transport.
The dynamics of the corresponding systems  is weakly dissipative, or even, on certain time scales, near perfectly
Hamiltonian, thus allowing one to get in touch with classical Hamiltonian and quantum dynamics \textit{in vivo}.
Laser created periodic optical potentials provide a
possibility to explore nonequilibrium  transport induced by
violation of time-space symmetries. In this section we review a series of experiments with rocking cold atom ratchets which have perfectly
validated the  results of the symmetry analysis for classical ratchets in the regime of weak dissipation.
Again, before starting the discussion, we will flash through some basic information on the cold atom optics (for more
information see Ref. \cite{Grynberg2001}).

An atom with nonzero spin has several internal states, determined by the spin projection on a given direction.
The transition between two different states can be induced  by an external electromagnetic field, i.e. by light.
The electromagnetic field also shifts energy levels of the atoms, which leads to an effective interaction between the light and the atom.
A periodic potential for the atom can be created using counter-propagating laser beams which form standing light waves.
The periodic interference pattern introduces a spatial modulation of the energy shift,
thus creating a potential. The parameters of optical potentials can be to tuned by changing the parameters of
the lasers - intensities, relative shifts, etc. \cite{Grynberg2001,Morsch2006}.

In experiments with cold atom ratchets, optical lattices serve two ends:
They (i) create  periodic potentials for atoms and (ii) provide a possibility to reduce  the kinetic energy of the atoms
by  cooling them down to $m$Kelvin temperatures. The latter  was possible due to the \textit{Sisyphus cooling} phenomena \cite{Dalibard1989}.
As an illustration we use  the simplest case when an atom has only two internal states,
a ground state, with the spin projection $J_g = 1/2$, and an
excited state, $J_e = 3/2$.
Imagine now that the atom is placed into an optical potential created by two counter-propagating laser beams,
both with the same wavelength $\lambda$ but of mutually orthogonal polarizations.
The interference between beams results in a $\lambda/2$ - periodic potential with spatially-dependent elliptic polarization.
When in the ground state $J_g = \pm 1/2$, the atom feels a periodic potential
\begin{eqnarray}
V_{\pm}(x) = \frac{V_0}{2}[-2 \pm \cos(k_Lx)],
\label{at_potential}
\end{eqnarray}
where $k_L = \pi/\lambda$, and the potential depth $V_0$ can be tuned by changing the laser
intensity and frequency\footnote{A bi-harmonic potential of the form
$V(x) = \frac{1}{2}[V_1\cos(kx) + V_2\cos(2kx+ \Delta)]$,
with the tunable phase $\Delta$,
can be created by using dispersive properties of multiphoton Raman transitions \cite{rgscw06,weitzScience}}.
If the laser frequency is close to the frequency of the transition $J_g \rightarrow J_e$, the corresponding electromagnetic field
may induce a transition between the two internal states of the atom.
Consider the situation when the atom, initially located at the point $x = 0$, has a positive velocity and
it is in the ground state $J_g = -1/2$. Atom starts to climb the potential slope $V_-(x)$ (dashed line on Fig.~\ref{Fig:Sisyphus}),
by transferring  its kinetic energy into  potential energy. At the potential summit, the light field consists of the $\sigma_+$-component only,
and the probability of the transition $J_g =-1/2 \rightarrow J_g = +1/2$, through two consecutive events of absorption and emission, is maximal.
This transition results in a loss of the  potential energy by the atom.
After the emission  the atom is subjected to the potential $V_+(x)$ (solid line on Fig.~\ref{Fig:Sisyphus}), where it appears at the
bottom of a
potential well. The process repeats again and again, like in the ancient story about king Sisyphus \cite{sis}.
Finally, after several cycles of absorption and emission,
the atom ends up in a state where it does not have  enough energy  to climb the potential slope,
thus remaining localized at the bottom of one of the potential wells.

\begin{figure}[t]
\begin{center}
\includegraphics[width=0.85\linewidth,angle=0]{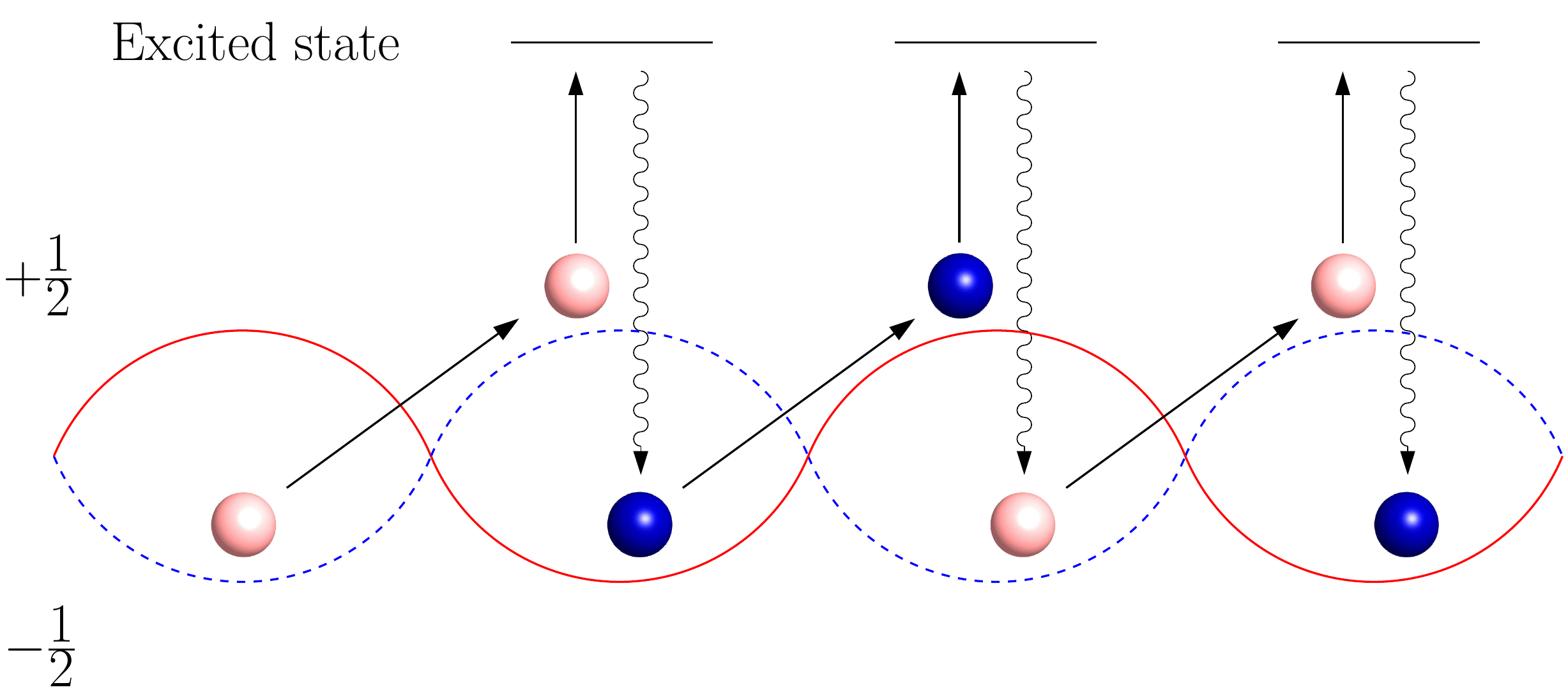}
\label{Figure5} \caption{(color online) Sisyphus cooling of an atom
with two internal states defined by the projections of its spin, $J_g = -1/2$ and $J_e = +1/2$,
in a  lin$\bot$lin optical lattice \cite{Dalibard1989}. The atom (ball) travels
from the left to the right through the lattice and its state is indicated by the
color of the ball, blue for $J_e$ and red for $J_e$. The upwards arrows denote absorptions of
photons, while downwards wave arrows mark the spontaneous emission events.
} \label{Fig:Sisyphus}
\end{center}
\end{figure}

In the Sisyphus cooling machinery, fluctuation and dissipation mechanisms  are closely interwoven as they
are produced by the same chain of stochastic transitions between atom internal states.
The overall effect is quantified by the frequency of the transitions, also called \textit{scattering rate} $\varGamma$.
Its value depends on the intensity of the laser field $I_l$ and the difference between the laser frequency and the
frequency of the transition, $\triangle = \omega_l - \omega_0$,   $\varGamma \propto I_l/\triangle^2$. The
potential height scales as  $U_0 \propto I_l/\triangle$ \cite{Grynberg2001}.
Therefore, it is possible to change the decoherence strength
by changing simultaneously the intensity $I_L$ and the detuning $\triangle$ while keeping their ratio, i. e. the
potential strength, constant.

\begin{figure}[t]
\begin{center}
\includegraphics[width=0.65\linewidth,angle=0]{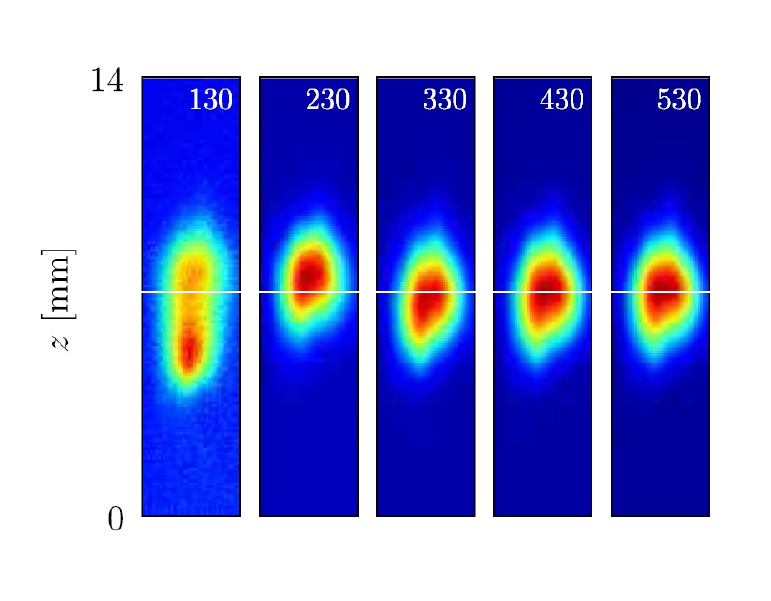}
\label{Figure4e} \caption{(color online) Fluorescence images of the atomic cloud obtained in experiments with a
rocking ratchet setup, Eq.~(\ref{biharm_q}), after  the exposition to the driving during  26 $ms$.
Different panels correspond to different values of the driving frequency $\omega$ (in $k$Hz units).
The white line marks the initial position of the cloud's center of mass.
The phase shift is $\theta = \pi/2$ in all cases. Courtesy of Ferruccio Renzoni.} \label{flour}
\end{center}
\end{figure}

A driving force can be realized  in different ways,
depending on a particular  ratchet setup one wants to implement. The multiplicative driving, $g(x,t)= E(t)f(x)$, can be implemented
by modulating  the intensity of the lasers periodically in time \cite{weitzScience}.
The additive driving, $g(X,t) = f(x) + E(t)$, can be introduced through phase modulations of one of the beams, in a manner similar
the described in Sec. \ref{Traveling potentials}.
The modulations result in a shaking potential $V(x) \propto \cos(kx - \alpha(t))$ so that in the co-moving frame,
$x' = x - \alpha(t)/k$,  an atom of mass $m$ experiences a tilting force $E(t) = -m \ddot{\alpha}(t)/k$.
A bi-harmonic function $\alpha(t)$ will produce a bi-harmonic
driving $E(t)$,  with the same phase shift $\theta$. Finally, due to the periodicity of the driving,
$\alpha(t+T) = \alpha(t)$, the averaged  velocities of an atom  are the same in the lab and co-moving frames,
$\langle \dot{x} \rangle =  \langle \dot{\widetilde{x}} \rangle$. This elegant idea has been implemented in a series of experiments
reported in Refs. \cite{ss-prg03prl,jgr05prl}.
The spatial positions of the atoms were monitored through the direct imaging by using a
charge-couple-device (CCD) camera. From these images, the position of the cloud center of mass was calculated, thus allowing to detected
the transport of the atomic cloud along  the optical lattice \cite{ss-prg03prl}, see Fig.~\ref{flour}.

\subsubsection{Cold atom ratchets with periodic driving}

\begin{center}
\begin{figure}[t]
\begin{center}
\includegraphics[width=0.9\textwidth]{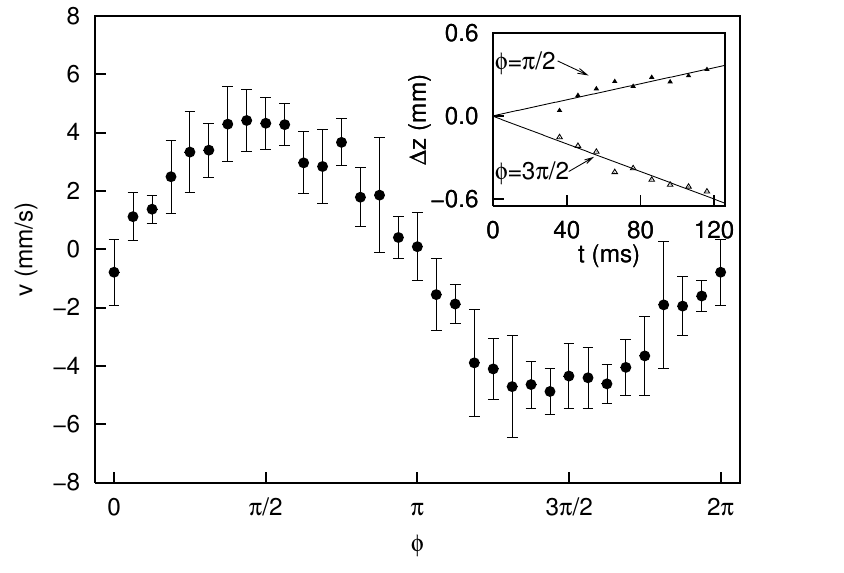}
\end{center}
\caption{Average velocity of the atomic cloud center-of-mass as a
function of the phase $\theta$ (denoted by $\phi$ here). Inset: the displacement of the center-
of-mass as a function of time for two different values of the phase
$\theta$.
The potential is $V(x) = V_0 \cos(kx)$, with the bi-harmonic drive of the form
$E(t) = E_1\cos(\omega t)$ $+E_2\cos(2\omega t + \theta)$. Adapted from Ref. \cite{ss-prg03prl}.} \label{renzoni1}
\end{figure}
\end{center}

The rocking ratchet set-up, used in the experiments with cold atom ratchets \cite{ss-prg03prl,jgr05prl,RenzoniEL2012},
corresponds to a simple-harmonic potential force,
\begin{equation}
\label{biharm_p}
f(x) = f_1 \sin(k_Lx),
\end{equation}
and a bi-harmonic driving function,
\begin{equation}
\label{biharm_q}
E(t) = E_1 \cos(\omega t) + E_2 \cos(2\omega t + \theta).
\end{equation}
The measured dependence of the center-of-mass velocity of the atomic vs $\theta$
has a well-pronounced sine-like shape (\ref{current_expans}), see Fig.~\ref{renzoni1}.
It is important to understand, however,  that the dynamics of atoms in an optical lattice is far from being exactly Hamiltonian.
The decoherence of the atomic dynamics, created by by emission-absorption events,  is always present. Moreover, it plays a key role because it
is responsible for  the cooling of the atoms.  From the classical model it follows that   the overall effect of
dissipation and fluctuations can be absorbed into the phase lag $\theta_0$, Eq.~(\ref{lag}). The effect of decoherence,
induced by the interaction between an atom and the
electromagnetic field of a laser beam,  is then replaced by dissipation and fluctuations, exerted on the classical  particle
by a heat bath, see Eq.~(\ref{langevin}). By looking at the experimental data, Fig.~\ref{renzoni1}, we can conclude that
the decoherence effects were  not able to produce a tangible shift of the sine-like  dependence
(expected at the Hamiltonian limit).

The decoherence quantifier,  the scattering rate $\varGamma$ \cite{Grynberg2001},  is a tunable parameter which can be controlled in experiment.
This naturally leads to  the following question: Would it be possible, by increasing $\varGamma$, to see the  the effect
of dissipation-induced phase lag,
similar to that observed  in computational studies of the stochastic classical model, Eq.~(\ref{lag})?
\begin{figure}[t]
\begin{center}
\begin{tabular}{cc}
\includegraphics[width=0.5\linewidth,angle=0]{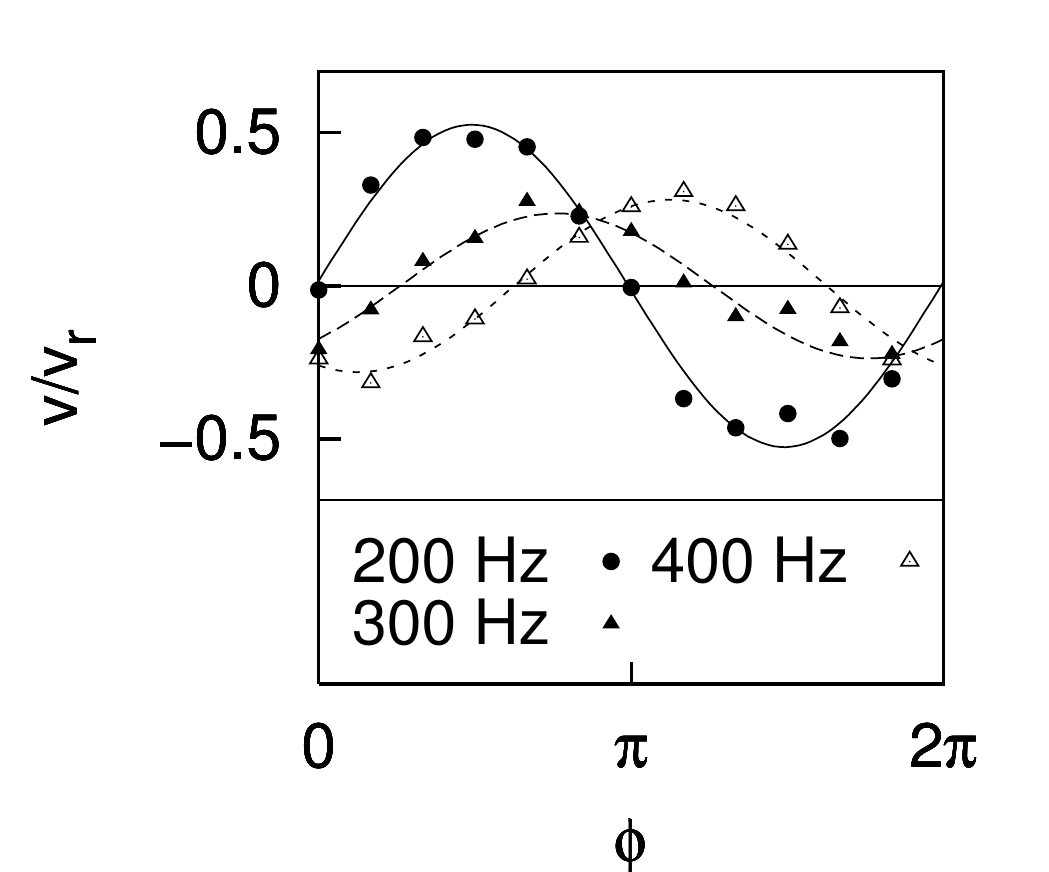}
\includegraphics[width=0.5\linewidth,angle=0]{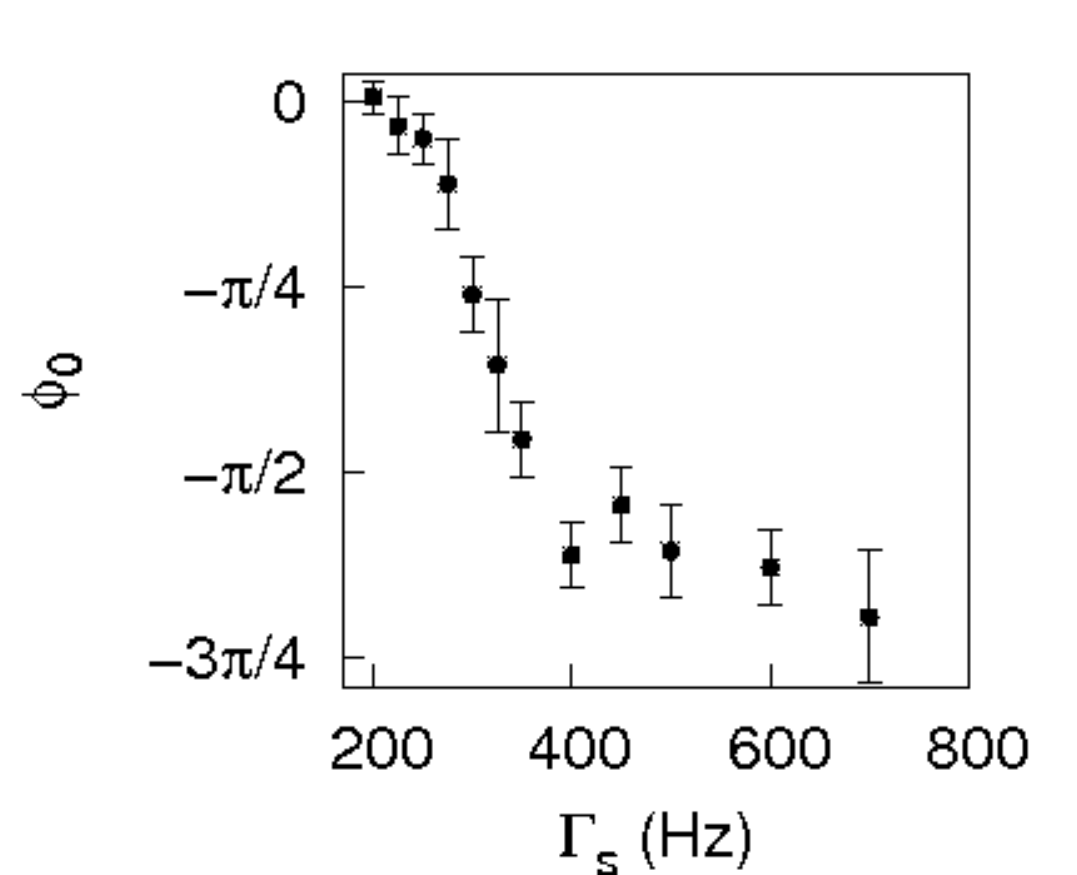}
\end{tabular}
\label{Figure4a} \caption{(left panel) Average velocity of the atomic cloud center of mass as  function of  $\theta$
(denoted by $\phi$ here) for different scattering rates $\varGamma$.
Lines correspond to the fitting obtained with Eq.~(\ref{lag_renz}); (right panel)
Phase lag $\theta_0(\varGamma)$ (denoted by $\phi_0$ here) as a function of the scattering rate.
Adapted from Ref. \cite{gbr05prl}.} \label{renzoni2}
\end{center}
\end{figure}
This question was positively answered with cold atoms \cite{gbr05prl}.
The main result of the  experiments is shown on  Fig.~\ref{renzoni2}.  By fitting the data points obtained for different
values of $\theta$ with the target function
\begin{equation}
\label{lag_renz}
J(\theta;\varGamma)  \sim J^0_{\varGamma}\sin[\theta - \theta_0(\varGamma)],
\end{equation}
the dependence $\theta_0(\varGamma)$ was extracted, see the right panel of Fig.~\ref{renzoni2}.
The phase lag is close to zero for the smallest examined scattering rate,  it lowers down to $\theta_0 = -\pi/2$ with the increase
of the scattering rate, and then even goes below this value, finally approaches  $\theta_0 = -3\pi/4$.
This deviation from the results obtained for the classical ratchet, see  Fig.~\ref{fig3-KU}, which say that
that phase lag cannot go below the threshold $\theta_0 = -\pi/2$ (corresponding to the overdamped limit), is noteworthy.
This discrepancy  demonstrates that the analogy between the classical stochastic model and the quantum system, explored
in the experiments, should not be overstretched. A  proper modeling requires more appropriate numerical approaches, for example,
quantum or quasiclassical Monte-Carlo methods \cite{zoller1, Grynberg2001} (see also Refs.~\cite{kohler1, tanimura} for alternative
approaches to  decoherence-induced phenomena in quantum ratchets).

Directed current of cold atoms depends not  only on $\theta$ but on the other parameters of the driving function as well.
Multiple current reversals have been detected in the experiments upon tuning  amplitudes \cite{jgr05prl} and
frequency of $E(t)$ \cite{Gommers2005},
see  Fig.~\ref{renzoni2b}.   It has  been observed that the  dependence $J(\theta)$, Eq.~(\ref{lag_renz}), preserves
its smooth sine-like upon the variations of the driving frequency \cite{Wickenbrock2011}. Therefore, the dependence of the atomic current
on $\omega$ can again be absorbed into the phase lag, $\theta_0(\varGamma, \omega)$.

\begin{figure}[t]
\begin{center}
\includegraphics[width=0.8\linewidth,angle=0]{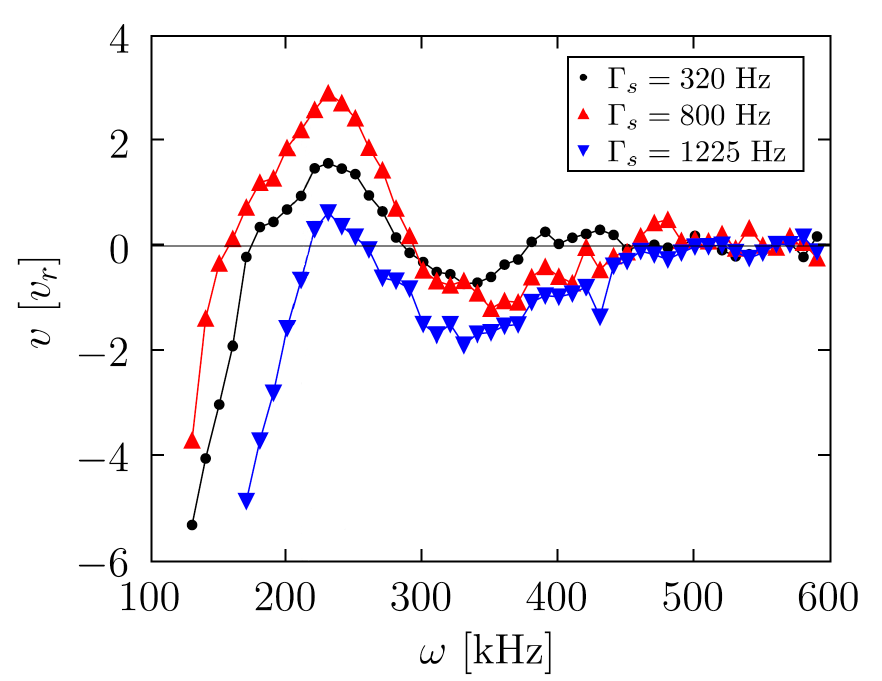}
\label{Figure4b} \caption{(color online) Average velocity of the center of mass of the atomic cloud as  function of the driving
frequency $\omega$ for different scattering rates, $\varGamma_s$. The phase shift is $\theta =  \pi/2$.
The fluorescence images corresponding to $\varGamma_s = 320$ $k$Hz are shown on Fig~\ref{flour}.
Courtesy of Ferruccio Renzoni.} \label{renzoni2b}
\end{center}
\end{figure}

A realization   of the gating ratchet with cold cesium atoms, Eqs.~(\ref{2.23}, \ref{gait}),   was
presented in Ref. \cite{glbr08}. The obtained experimental results again  confirmed the predictions of the symmetry analysis.
Finally, the results obtained with cold atom version of the kicked rotor (a particular example of the multiplicative  setup,
Section  \ref{Multiplicative driving}), were reported in Ref. \cite{Jones2007}.

\subsubsection{Cold atom ratchets with quasiperiodic driving}

Quaisiperiodic driving, Sec. \ref{Extension to quasiperiodic driving}, provides another chance to probe
the symmetry analysis with cold-atom ratchets. In principle, an arbitrary time-dependent driving function $E(t)$ can be generated
by modulating the relative phases of  the laser beams. By using additional  acousto-optic modulators (AOMs),
it is possible to obtain signals with different frequencies, $\omega_1$, $\omega_2$, $\omega_3$, ... \cite{Renzoni2009}.
If the ratio $\omega_2/\omega_1$ is an irrational number then the driving is quasiperiodic.
It is evident,  however, that the ideal quasiperiodicity is a mere mathematical abstraction
which is absent in the real life. Effective cutoffs  appear due to unavoidable parameters fluctuations during
an experiment. Thus, the ratio $\omega_1/\omega_2$ can always be approximated by a rational number
$p/q$, with $p$, $q$ being two co-prime integers. However, as the duration of the experiment is always finite, one could,
by choosing $p$ and $q$ sufficiently large,  obtain a driving which is  quasiperiodic on the time scale of the experiment.
This idea was realized in  another  series of experiments with cold  cesium atoms \cite{gdr06prl}.

\begin{figure}[t]
\begin{center}
\includegraphics[width=0.65\linewidth,angle=0]{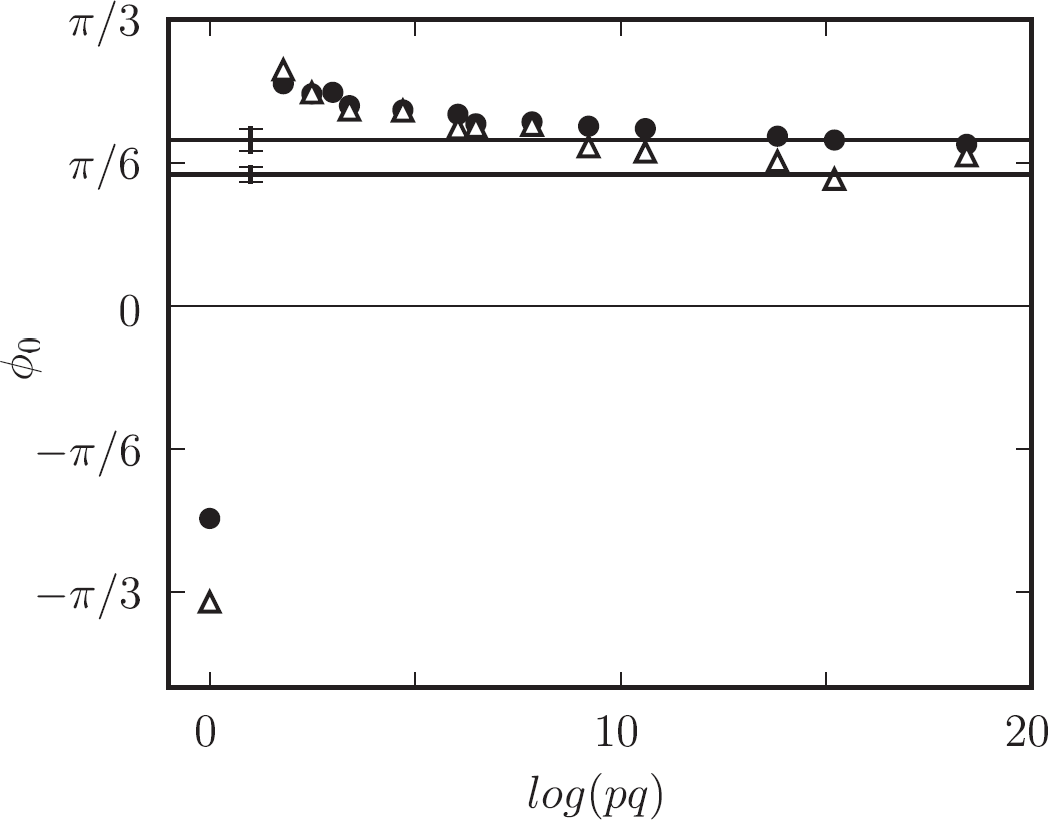}
\label{Figure4c} \caption{Phase lag $\theta_0(\varGamma,\delta)$ (denoted $\phi_0$ here),
Eq.~(\ref{lag_q}), as a function of the product $pq$, where $p$ and $q$ are co-primes determined by the ratio $\omega_1/\omega_2 = p/q$.
The quasiperiodic limit corresponds to $pq \rightarrow \infty$. The two
data sets correspond to different amplitudes of the driving. The phase of the driving at frequency $\omega_2$, Eq.~(\ref{triharm_q}),
is $\delta=\pi/2$.
The phase lag was estimated by fitting the current dependence
on $\theta$ with Eq.~(\ref{lag_q}). The two
horizontal lines indicate the phase shift $\theta_0$ for the bi-harmonic drive,
i.e. in the absence of the driving at frequency $\omega_2$, $E_3 = 0$. Adapted from Ref. \cite{gdr06prl}.} \label{renzoni3}
\end{center}
\end{figure}

The first type of the driving function examined in the experiments  was a sum of three harmonics,
\begin{equation}
\label{triharm_q}
E(t) = E_1 \cos(\omega_1 t) + E_2 \cos(2\omega_1 t + \theta)+ E_3 \cos(\omega_2 t + \delta),
\end{equation}

Let us start the symmetry analysis of this setup from the case when $\omega_1/\omega_2$ is a rational number.
For $E_3=0$ we have a standard bi-harmonic setup, Eq.~(\ref{biharm_q}), with the time-reversal symmetry holding for $\theta = 0,\pm\pi$
in the limit of vanishing dissipation. When $E_3 \neq 0$ and $\delta = 0,\pm\pi$, the time-reversal symmetry
is still present. For $\delta \neq 0,\pm\pi$, the symmetry is broken and the directed current can appear at the former symmetry points
$\theta = 0,\pm\pi$. Thus the third harmonics produces an
additional shift effect of the current dependence  $J(\theta)$. By taking the decoherence effects into account, we arrive at the following
target function \cite{gdr06prl}:
\begin{equation}
\label{lag_q}
J \sim J^0_{\varGamma, \delta} \sin[\theta - \theta_0(\varGamma,\delta)].
\end{equation}
The difference between this expression and that one introduced before, Eq.~(\ref{lag_renz}), is that now
$\theta_0$ depends not only on the strength of the decoherence,  determined by the scattering rate $\varGamma$, but also on the
phase shift of the third harmonics $\delta$.

Let us  turn now to the case of quasiperiodic driving, when $\omega_1/\omega_2$ is an irrational number. As discussed
in Sec. \ref{Symmetries of quasiperiodic functions}, in this case the phases $t_1=\omega_1 t$ and $t_2=\omega_2 t$ can be treated
as independent variables. The tri-harmonic driving, Eq.~(\ref{triharm_q}), is invariant under the transformation
$t_2 \rightarrow -t_2$, for any value  of $\delta$ and for  $\theta = 0,\pm\pi$.   In the Hamiltonian limit,
we should return to the dependence of
the ratchet current given by Eq.~(\ref{lag}). In the quaisperiodic limit,  the third harmonics does not induce any
additional lag, which is determined now by the strength of decoherence solely. This prediction was  confirmed by the results of
the experiments, see Fig.~\ref{renzoni3}.  It was found that upon the approach of the quasiperiodic limit, when the ratio
$\omega_1/\omega_2$ tends to be more and more irrational, the dependence of the
lag on the phase of the third harmonics disappears and the phases $t_1$  and $t_2$ behave as two independent variables.

\begin{figure}[t]
\begin{center}
\includegraphics[width=0.65\linewidth,angle=0]{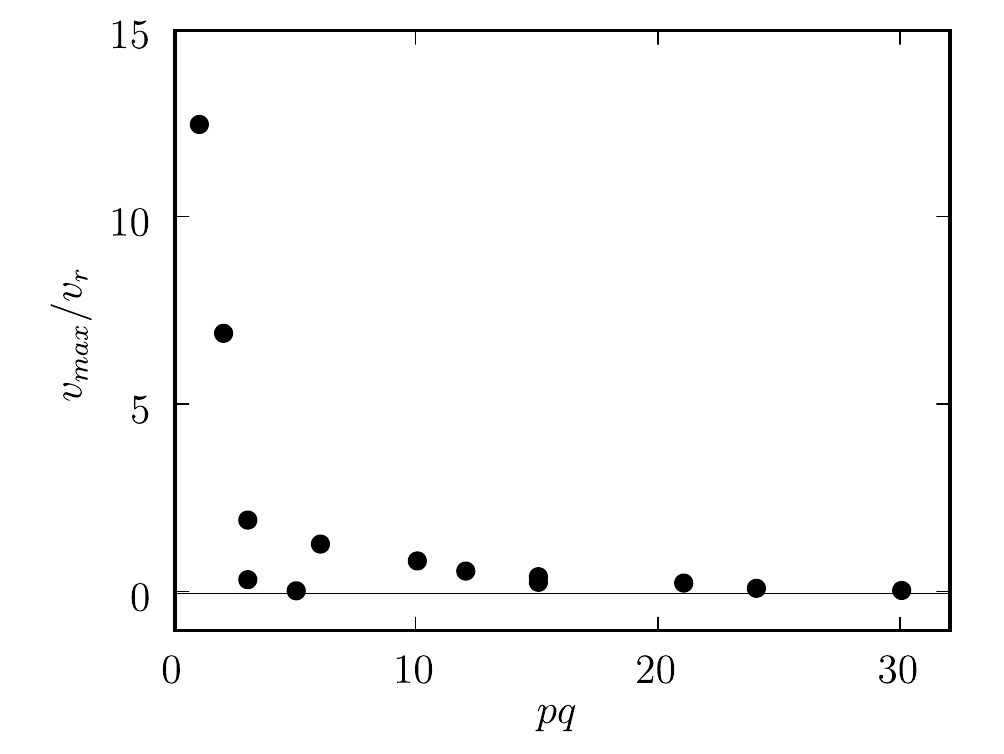}
\label{Figure4d} \caption{Maximum average velocity of the atomic cloud as a function of $pq$, where $p$ and $q$ are
the co-primes defined by the ratio of the driving frequencies,
$p/q=\omega_2/\omega_1$. The maximum velocity  was estimated by fitting the current dependence
on $q$ with the function $\upsilon(q) \sim \sin(q \delta - \pi/2)$. Adapted from Ref. \cite{gdr06prl}.} \label{renzoni4}
\end{center}
\end{figure}

A different form of the driving function,
\begin{eqnarray}
\label{tricky}
E(t)= E_1 \cos(\omega_2 t+\delta)
[ a\sin(\omega_1 t) + b \sin(2\omega_1 t)]  \nonumber \\
+ E_2\sin(\omega_2 t+\delta)
[a\cos(\omega_1 t) + 2b\cos(2\omega_1 t)]~~~~~~~~~~~~~~~
\end{eqnarray}
was also examined in Ref. \cite{gdr06prl}.
Consider $\omega_1/\omega_2$ to be a rational number, $p/q$. The period of the function $E(t)$  is  $T = qT_1=pT_2$,
where $T_1=2\pi/\omega_1$ and $T_2=2\pi/\omega_2$. Time-shift transformations, $t \rightarrow t+T/2$,
change phase variables, $t_1 \rightarrow t_1 + q\pi$ and $t_2 \rightarrow t_2 + p\pi$. It is easy to check that the function $E(t)$
possesses shift-symmetry (\ref{Sh-SymmFunc}) whenever $q$ is even and $p$ is odd.
Therefore, in the limit of vanishing dissipation, when  $q$ is
even and $p$ is odd,  symmetry $\hat{S}_x$, Eq.~(\ref{2.9a}), holds and the ratchet current should disappear.
Assume now that $q$ is odd. Symmetry  $\hat{S}_x$ is already broken but the time-reversal symmetry, $\hat{S}_t$, Eq.~(\ref{2.9a}), remains.
Its validity  depends on the phase $\delta$: It holds when $q\delta = (n+1/2)\pi$, with $n$ integer. The current is expected to show the familiar
sine-like dependence on $q\delta -\pi/2$, perhaps with an additional lag  accounting for the decoherence-induced shift effect.
In the quasiperiodic limit, as before, we can think of two independent variables, $t_1=\omega_1 t$ and $t_2=\omega_2 t$.
The function (\ref{tricky}) is shift-symmetric with respect to $t_2$, i. e. it changes sign under the transformation
$t_2 \rightarrow t_2+\pi$. Therefore, the ratchet transport should disappear in the limit $pq \rightarrow \infty$. The
results of experiments \cite{gdr06prl},  Fig.~\ref{renzoni4},  confirmed that indeed the rectification is lost  in the quasiperiodic limit
due to the restoration of the shift-symmetry of $E(t)$.

\section{Directed transport in two and three spatial dimensions}
\label{sec4}
In this section we  generalize the symmetry analysis to the case of two- and three-dimensional potentials.
Two new features can be expected right from the start.
First, the current is  a vector now and it has a certain length and is pointing at certain direction.
Second, the new angular dependence may lead to a rotational motion, i.e. to a vorticity. Both linear and
vortex currents may become nonzero. Moreover, they may be intricately coupled to each other in the
manner similar to the  spin-orbit coupling effect \cite{griff}.
Quantum optics also gives practical reasons to extend the symmetry analysis in this direction.  Nowadays,  by using several laser beams,
experimentalists can  routinely fabricate two- and three-dimensional optical potentials of very different
symmetries and shapes \cite{Morsch2006, Windpassinger2013} that can be driven in addition \cite{Grynberg2001}.
These advances served a  testing ground to study weakly-dissipative ratchets with  two- and three-dimensional
periodically driven potentials.

In this section we consider the dynamics of a classical particle in a
$d=\{2,3$\}-dimensional space-periodic potential, which is driven by a zero-mean
ac-field. The particle may contribute
to a directed current along a certain direction, and, at the same
time, it could perform a rotational vortex-like motion. The main issue we want to address here is how to
simultaneously control  two different transport modes. By using the symmetry analysis, we will first identify the symmetries which
ensure that either translational or vortex components along certain
directions are strictly zero. Then, by using a $2d$ model, we will show how,
by breaking these symmetries one by one, we can
control the particle motion generating either directed or vortex currents or both simultaneously.
Finally, we will review an experimental realization of this idea with cold atoms and two-dimensional optical potentials \cite{lr09}.

\subsection{Symmetry analysis}
\label{Symmetry analysis_2d}

We start with a  model of a classical particle  exposed
to a potential force and fluctuations produced by the coupling to a heat bath,
\begin{equation}
m \ddot{{\bf r}} + \gamma \dot{\bf r} = {\bf g}({\bf r},t)+
\boldsymbol{\xi}(t),
~~{\bf g}({\bf r},t)=-\boldsymbol{\nabla} V({\bf r},t)\\
\label{1sys}
\end{equation}
Here ${\bf r}=\{x,y,z\}$ is the coordinate vector of the particle.
The force ${\bf g}({\bf r},t)=\{g_\alpha({\bf
r},t)\}$, $\alpha=x,y,z$, is time- and space-periodic:
\begin{equation}
{\bf g}({\bf r},t)={\bf g}({\bf r},t+T)={\bf g}({\bf
r}+\mathbf{L}_\alpha,t),~\alpha=x,y,z \label{sp_per}.
\end{equation}
$\mathbf{L}_\alpha$ are the
lattice vectors.
The absence of dc-components of the force vector implies
\begin{equation}
\langle {\bf g} ({\bf r} ,t) \rangle_{\mathbf{L},T}\equiv
\int_{0}^T\int_{\mathbf{L}} {\bf g} ({\bf r} ,t) \, dt \, dx dy dz=0
\label{aver}
\end{equation}
where the spatial integration is performed over the unit cell of the lattice.

The current is a vector now, ${\bf J} = \{ J_\alpha\}$, with the components given by the corresponding components
of the average velocity vector $\mathbf{v}=\dot{\mathbf{r}}$
of the particle,
\begin{equation}
 J_\alpha=  \lim_{t \rightarrow \infty} \frac{v_{\alpha}}{t}.
\label{current_2_3d}
\end{equation}
The conditions for the current to be absent, i. e. ${\bf J} \equiv {\bf 0}$, and the corresponding symmetries can be obtained
by  direct generalizations of the symmetry analysis  for the $1$d case, discussed in the previous section.
Namely, there are two fundamental symmetry transformations,
\begin{eqnarray}
  &&{\hat {\bf S}}_{\bf r}[{\boldsymbol \Xi}, \tau]:~ {\bf r} \rightarrow -{\bf r} + {\boldsymbol \Xi},  \quad t \rightarrow t+\tau
\, ,\label{Space2d} \\
  &&{\hat {\bf S}}_t[{\boldsymbol \Xi},\tau]:~ {\bf r} \rightarrow {\bf r} + {\boldsymbol \Xi}  \, , \quad t \rightarrow -
t+\tau \, , \label{Time2d}
\end{eqnarray}
parametrized by a shift vector  ${\boldsymbol \Xi} = \{\chi_{\alpha}\}$ and scalar parameter $\tau$, with both depending on the
shape of the force function ${\bf g} ({\bf r} ,t)$. Any of the transformations that change the sign of ${\bf J}$, Eq.~(\ref{current_2_3d}),
such that when the system (\ref{1sys}) is invariant under either of the transformations, yield a zero of the average corresponding ratchet current.

Consider as an example the additive case,
\begin{eqnarray}
\label{rocking2d}
{\bf g}({\bf r},t)=-\boldsymbol{\nabla}V({\bf r})
+\mathbf{E}(t)\equiv \mathbf{f}({\bf r}) +\mathbf{E}(t).
\end{eqnarray}
Similar to the one-dimensional case,
symmetry ${\hat {\bf S}}_{\bf r}$ holds when the potential force is
\textit{anti-symmetric},
$\mathbf{f}(-\mathbf{r}+\mathbf{r}')=-\mathbf{f}(\mathbf{r})$,
 and the driving function is \textit{shift-symmetric},
$\mathbf{E}(t+T/2)=-\mathbf{E}(t)$. The symmetry ${\hat {\bf S}}_t$
holds at the Hamiltonian limit, $\gamma=0$, if the driving force is
\textit{symmetric}, $\mathbf{E}(-t+t')=\mathbf{E}(t)$. Finally, the symmetry ${\hat {\bf S}}_t$ holds at the overdamped
limit, $m=0$ , if the potential force is shift-symmetric,
$\mathbf{f}(\mathbf{r}+{\boldsymbol \Xi})=-\mathbf{f}(\mathbf{r})$
and the driving force is anti-symmetric,
$\mathbf{E}(t+t')=-\mathbf{E}(-t)$. All of the above can be re-formulated in terms of the potential, $V({\bf r}) = V(x,y)$, taking into
account the relations between the symmetries of the function and its derivative/integral, see Sec.
\ref{Symmetries of periodic functions}.

The system can be coupled to a heat bath in the standard way, by adding to the rhs of Eq.~(\ref{1sys}) a random Gaussian vector,
$\boldsymbol{\xi}(t)=\{\xi_{x},\xi_{y},\xi_{z}\}$,  $\langle \xi_{\alpha}(t) \xi_{\eta}(t') \rangle = 2\gamma k_B\mathcal{T} \delta
(t-t') \delta_{\alpha \eta}$,  $\alpha,\eta=\{x,y,z\}$. The statistical description of the system evolution is provided by  the
Fokker-Planck equation (FPE) \cite{risken1989}:
\begin{equation}
\{\frac{\partial}{\partial t} + \nabla_{\mathbf{r}}\cdot \frac{\mathbf{p}}{m}-
{\mathbf{\nabla}}_{\mathbf{p}} \cdot [\gamma
\mathbf{v}- \mathbf{g}(\mathbf{r},t)]- \gamma m D
\triangle_{\mathbf{p}}\}P({\bf{r}},\mathbf{p},t)=0,
\label{fpe_under}
\end{equation}
where $\mathbf{p}=m\dot {\bf r}$ and $D = k_B \mathcal{T}$. The respective FPE for $m=0$
reads
\begin{equation}
\{\gamma \frac{\partial}{\partial t}+\nabla_{\mathbf{r}}\cdot
{\bf{g}}({\mathbf{r}},t)-  D \triangle_{\bf{r}}\}P({\bf{r}},t)=0.
\label{fpe_over}
\end{equation}
Similar to their one-dimensional counterparts, Section \ref{From Langevin dynamics to the Fokker-Planck equation},
these linear differential equations  have
unique, space- and time-periodic attractor solutions, $P_A({\bf{r}},\mathbf{p},t)$  and $P_A({\bf{r}},t)$,
correspondingly \cite{risken1989}. The asymptotic currents can then be written as
\begin{eqnarray}
&&{\mathbf{J}}=\langle \mathbf{\mathbf{v}}\cdot
P_A({\bf{r}},\mathbf{p},t)\rangle_{T,\mathbf{L}},~~m=1~,
\label{cur_under}\\
&&{\mathbf{J}}={\gamma}^{-1}\langle {\bf g}({\bf r},t) \cdot
P_A({\bf{r}},t)\rangle_{T,
\mathbf{L}},~~m=0~.\label{cur_over}
\end{eqnarray}
 It is easy to see that the FPE for the underdamped case, Eq.~(\ref{fpe_under}),
inherits all the symmetries of the corresponding equations of motions, Eq.~(\ref{1sys}), including the Hamiltonian limit,
$\gamma \rightarrow 0$.  The time-reversal symmetry for the overdamped case, Eq.~(\ref{fpe_over}), can be explained in the same manner
used for the $1$d case, Eqs.~(\ref{ls} - \ref{series}).

The is a new possibility to control the current.
Namely, we can align  the ratchet current ${\bf{J}}$ along a desirable direction. This can be achieved
by setting the adjacent components of the current vector to zero.
For that one has to define the symmetry operations
which change the signs of adjacent current component only.
If such a symmetry is present then the current along the corresponding directions is absent.
As an illustration, consider  a generalization of the symmetry ${\widehat S}_{x}$, which acts in the following manner:
\begin{eqnarray}
  &&{\hat {\bf S} }_{x}[{\boldsymbol \Xi}, \tau]:~  x \rightarrow -x + {\chi_x}, ~~y \rightarrow y + {\chi_y} , \quad t \rightarrow t+\tau
\label{SSx}.
\end{eqnarray}
Whenever equation (\ref{1sys}) is invariant under this transformation, the $x$-component of the current is absent,  $J_x = 0$. If
there are no further symmetries then a ratchet current will appear along the $y$-direction.
The extension of this idea to the $3$d case, with the symmetries of the type ${\widehat {\bf S}}_{x,y}$, is straightforward.

As an illustration, we consider a $2$d rocking ratchet,
\begin{eqnarray}
H(r,p,t) = \frac{{\bf p}^2}{2} + V(x,y) + {\bf E}(t)\cdot {\bf r},
\label{Add2d}
\end{eqnarray}
with the potential
\begin{eqnarray}
V(x,y) = \cos(x)[1 + \cos(2y)].
\label{Va_xy}
\end{eqnarray}
The rocking force, ${\bf E}(t)$, has components
 \begin{eqnarray}
E_{x,y}(t) = E^{(1)}_{x,y}\sin(t) + E^{(2)}_{x,y}\sin(2t + \theta_{x,y}).
\label{Ea_xy}
\end{eqnarray}
When $E^{(2)}_{x} = E^{(1)}_{y} = 0$ and $\theta_y = 0$, the symmetry ${\hat {\bf S} }_{x}[{\boldsymbol 0}, \pi]$ holds.
This implies that  $J_x = 0$ and the directed transport is confined to the $y$-axis.

At variance to the one-dimensional case, particles
in two and three dimensions can perform rotational motion.
Even in the
case when the directed current is absent, $\mathbf{J}=0$,
the particle  can still perform unbiased diffusion in the
coordinate space. In order to distinguish between directed transport
and spatial diffusion on one side, and \textit{rotational currents} on the
other side, we use the angular velocity
\begin{equation}
{\bf \Omega}(t)= [\dot {\bf r}(t) \times \ddot {\bf r}(t)]/\dot {\bf
r}^2(t),
 ~~{\bf
J}_{\Omega}=\langle {\bf \Omega}(t) \rangle_t \label{angular2}~,
\end{equation}
as a quantifier for the particle vorticity. ${\bf \Omega}(t)$ is
invariant under translations in space and time, and describes the
speed of rotation with which the velocity vector ${\bf {\dot r}}(t)$
encompasses the origin.

We  now search for symmetry
operations which leave the equation of motion, Eq.~(\ref{1sys}), invariant, while changing the sign of
the angular velocity ${\bf \Omega}(t)$.  The sign  can be
inverted by (i) the time-reversal transformation, $t\rightarrow -t$,
together with an optional space-inversion, $\mathbf{r} \rightarrow
\pm\mathbf{r}$. The corresponding symmetry transformation is
\begin{eqnarray}
{\hat {\bf R}}^{\pm}_{t} [{\boldsymbol \Xi},\tau]:~ {\bf r} \rightarrow \pm{\bf r} + {\boldsymbol \Xi}  \, , \quad t \rightarrow -
t+\tau \,\label{Rt}.
\end{eqnarray}
Note that the symmetry ${\hat {\bf R^{+}} }_{t}$ formally is identical to the symmetry ${\hat {\bf S} }_{t}$, Eq.~(\ref{Time2d})
There are two extra symmetries for $d = 2$. Namely, the following operations change  sign of ${\bf \Omega}(t)$:
(ii) the permutation of the variables, $ \{x,y\} \rightarrow \{y,x\}$,
and (iii) the mirror reflections, $\hat{\sigma}_{x}: \{x,y\} \rightarrow \{-x,y\}$ or $\hat{\sigma}_{y}: \{x,y\} \rightarrow \{x,-y\}$.
The corresponding symmetries are:
\begin{eqnarray}
{\hat {\bf R} }_{P} [{\boldsymbol \Xi},\tau]:~ {\bf r} \rightarrow {\widehat {\cal P}} {\bf r} + {\boldsymbol \Xi}  \, , \quad t \rightarrow -
t+\tau \,\label{Rp}.
\end{eqnarray}
\begin{eqnarray}
{\hat {\bf R} }_{x,y} [{\boldsymbol \Xi},\tau]:~ {\bf r} \rightarrow \hat{\sigma}_{x,y} {\bf r} + {\boldsymbol \Xi}  \, , \quad t \rightarrow -
t+\tau \,\label{RSig}.
\end{eqnarray}

The rotational symmetries can also be expressed in terms of the potential, $V(x,y)$,
and driving vector, ${\bf E}(t)$.  As an illustration, we  consider the Hamiltonian version of the two-dimensional ratchet with
additive driving, Eq.~(\ref{Add2d}).
If the driving function is symmetric,
${\bf E}(-t+\tau) = {\bf E}(t)$, then, in the Hamiltonian limit, symmetry ${\hat {\bf R}}^{+}_{t}$ holds, independent
of the symmetry properties of $V(x,y)$. However, symmetry ${\hat {\bf R}}^{-}_{t}$ holds only when the drive function is anti-symmetric,
${\bf E}(-t+\tau) = -{\bf E}(t)$, and the potential is symmetric, $V(x,-y+\chi_y) = V(-x+\chi_x,y) = V(x,y)$. Note that this  symmetry
does not forbid the ratchet  current  since the corresponding transformation does not change the sign of ${\mathbf{J}}$.
Symmetry ${\hat {\bf R} }_{P}$ holds when both functions, potential and drive vector, posses the permutation symmetry,
$V(x,y)=V(y,x)$ and $E_x(t) = E_y(t)$. The analysis for the dissipative and overdamped limit can be performed in the similar manner.
The final outcome of the symmetry analysis is presented with Table \ref{tabl3}, where we list the conditions for the
functions $V(x,y)$ and ${\bf E}(t)$ have to satisfy for the corresponding rotational symmetries be present.
We leave it as an exercise to the reader to figure out the rotational symmetries for the $2$d version of the multiplicative
set-up, Eq.~(\ref{2.22}).
\begin{table}
\begin{center}
\includegraphics[angle=0,width=0.99\textwidth]{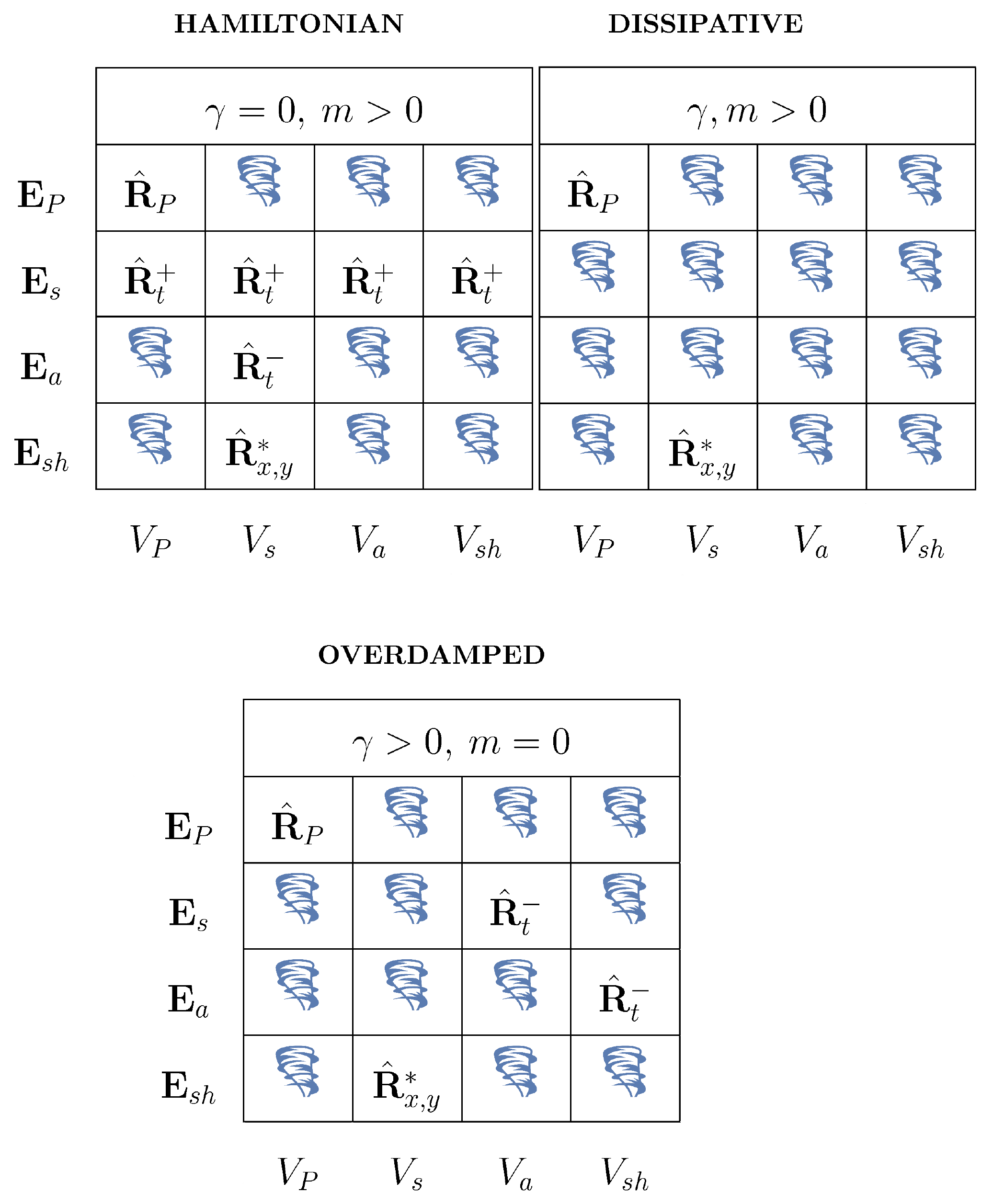}
\caption{(color online) Conditions for  the rotational
symmetries, Eqs.~(\ref{Rt}-\ref{RSig}), to hold for the two-dimensional ratchet with additive driving, Eq.~(\ref{Add2d}).
Asterix indicates that one of the symmetries
${\hat {\bf R} }_{x,y}$, Eq.~(\ref{RSig}), holds when the potential $V(x,y)$ is symmetric
with respect to the relevant variable, $x$ or $y$. Tornado icon indicates the cases when all symmetries are broken
and the nonzero rotational current, Eq.~(\ref{angular2}), can be expected.
Note that any other combination, not included in the tables, corresponds to the case when all the symmetries are broken.
}
\label{tabl3}
\end{center}
\end{table}

\subsection{Applications: numerical studies}
\label{Applications_2d}

As an illustration,  we consider a two-dimensional version of the rocking ratchet,
Eqs.~(\ref{Add2d} - \ref{Ea_xy}).  The potential (\ref{Va_xy}) is shift-symmetric, with
${\boldsymbol \Xi} = \{\pm\pi,0\}$. The symmetry ${\hat {\bf S} }_{\bf r}$ is already
broken since the driving vector ${\bf E}$ is not shift-symmetric. Therefore  we
expect $\textbf{J} \neq 0$ for the case of nonzero dissipation, $\gamma > 0$.

In Fig.~\ref{fig:conv1} we show the results of numerical integration of
equations (\ref{fpe_under}, \ref{fpe_over}). By applying operations ${\hat {\bf S} }_{r}$ and
$\theta \to \theta+\pi$, one can see that
$\textbf{J}(\theta+\pi)=-\textbf{J}(\theta)$, which allows for the
inversion of the current direction by shifting $\theta$, see Fig.~\ref{fig:conv1}(a).
In the overdamped limit $m=0$, symmetry
${\hat {\bf S} }_{t}$ is restored for $\theta=0,\pm\pi$, and therefore
$\textbf{J}(-\theta)=-\textbf{J}(\theta)$ (thick lines in
Fig.~\ref{fig:conv1}(a)). Upon approaching the Hamiltonian limit,
$\gamma \rightarrow 0$, the points where ${\bf J}=0$ shift from
$\theta=0,\pi$ to $\theta=\pm \pi/2$, where the symmetry ${\hat {\bf S} }_{r}$ is restored again
(thin lines in Fig.~\ref{fig:conv1}(a)). In the
underdamped regime, the asymptotic current can be approximated with the target function
\begin{eqnarray}
J_\alpha
\propto  J_\alpha^{0} \sin[\theta-\theta_0^\alpha(\gamma)], ~\alpha = \{x,y\}.
\label{lag_2d}
\end{eqnarray}
The phase lag is equal to $\theta_0^{x,y} = \pi/2$
and  $\theta_0^{x,y} = 0$ in the Hamiltonian and overdamped
limits, respectively,  Fig.~\ref{fig:conv1}(c).

\begin{figure}[t]
\begin{center}
\includegraphics[width=0.75\linewidth,angle=0]{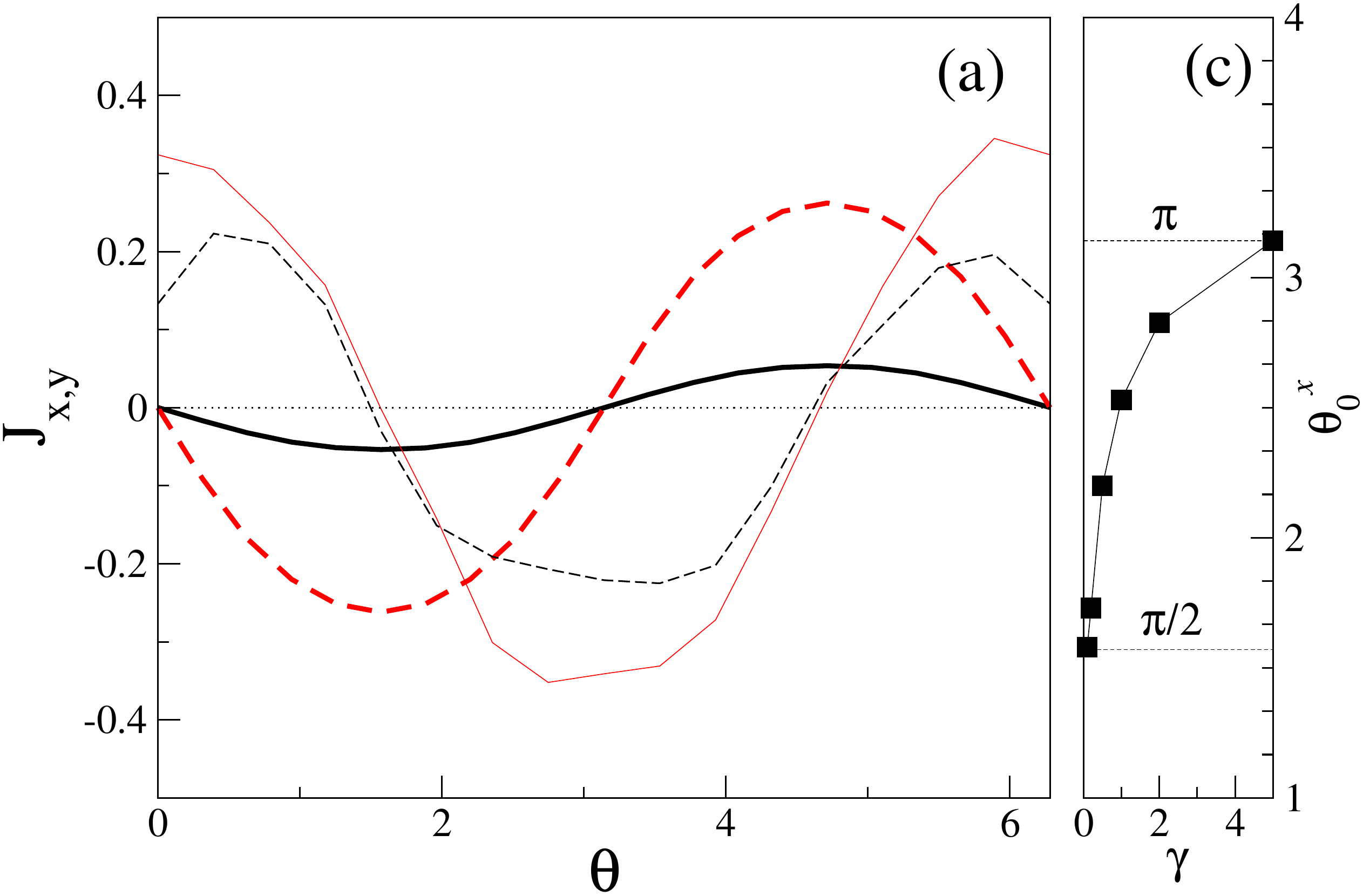}
\caption{(color online)  (a) Dependencies of the current components, $
J_{x}$ (solid line) and $J_{y}$ (dashed line), on $\theta$ for the system
(\ref{Add2d} - \ref{Ea_xy}). Curves correspond to the overdamped ($ m = 0, \gamma = 1 $, thick lines) and
underdamped ($ m = 1, \gamma = 0.1 $, thin lines) cases,
respectively; (c) The phase lag $\theta^{(0)}_x$ as a
function of the dissipation strength $\gamma$. The parameter are $D = 1$,
$E_{x}^{(1)}=-E_{x}^{(2)}=2$, $E_{y}^{(1)}=-E_{y}^{(2)}=2.5$.
Adapted from Ref. \cite{dzfy08prl}.} \label{fig:conv1}
\end{center}
\end{figure}

The realization of the current direction control is presented
with Figure \ref{fig:conv2}.
For the system given by Eqs.~(\ref{Add2d} - \ref{Ea_xy}) with $E_x^{(2)}=E_y^{(1)}=0$,
$\theta=0$, the symmetry  ${\hat {\bf S} }_{x}[{\boldsymbol 0}, \pi]$ implies that the current along the
$x$-direction is absent, $J_x=0$, and the directed transport is
happening along the $y$-axis only, see Fig.~\ref{fig:conv2}, curve (i).


To illustrate the control of the rotational current, we use the tilting setup with the following
potential and drive functions:
\begin{eqnarray}
&&V(x,y)=[-3\left (\cos x+\cos y\right )+  \cos x \cos y]/2,~~ \label{potent3}\\
&&E_x(t)=E_x^{(1)}\cos t~,~~ E_y(t)=E_y^{(1)}\cos(t+\theta)~.
\label{force3}
\end{eqnarray}
We have numerically integrated the corresponding stochastic equations of motions, Eq.~(\ref{1sys}), and
averaged the results over $N=10^5$ different
realizations. Fig.~\ref{fig:conv3} shows the
dependence of the rotational current (\ref{angular2}) on the
relative phase $\theta$. The system is invariant under the
transformation ${\hat {\bf S} }_{\bf r}$,  therefore the directed
current is absent, ${\bf J} = 0$. However, for the underdamped case, $\gamma \neq
0$, all the relevant symmetries, Eqs.~(\ref{Rt}-\ref{RSig}), are violated,
and the resulting rotational current (\ref{angular2}) is nonzero.
Note that the symmetry
${\widehat {\bf R}}_t$ is restored when $\theta=0,\pm \pi$, thus the
current disappears in the Hamiltonian and overdamped limits for
these values of $\theta$. The right inset on
Fig.~\ref{fig:conv3} shows a single stochastic trajectory, which shows how the particle is acquiring an average
nonzero angular momentum.

%
\begin{figure}[t]
\begin{center}
\includegraphics[width=0.75\linewidth,angle=0]{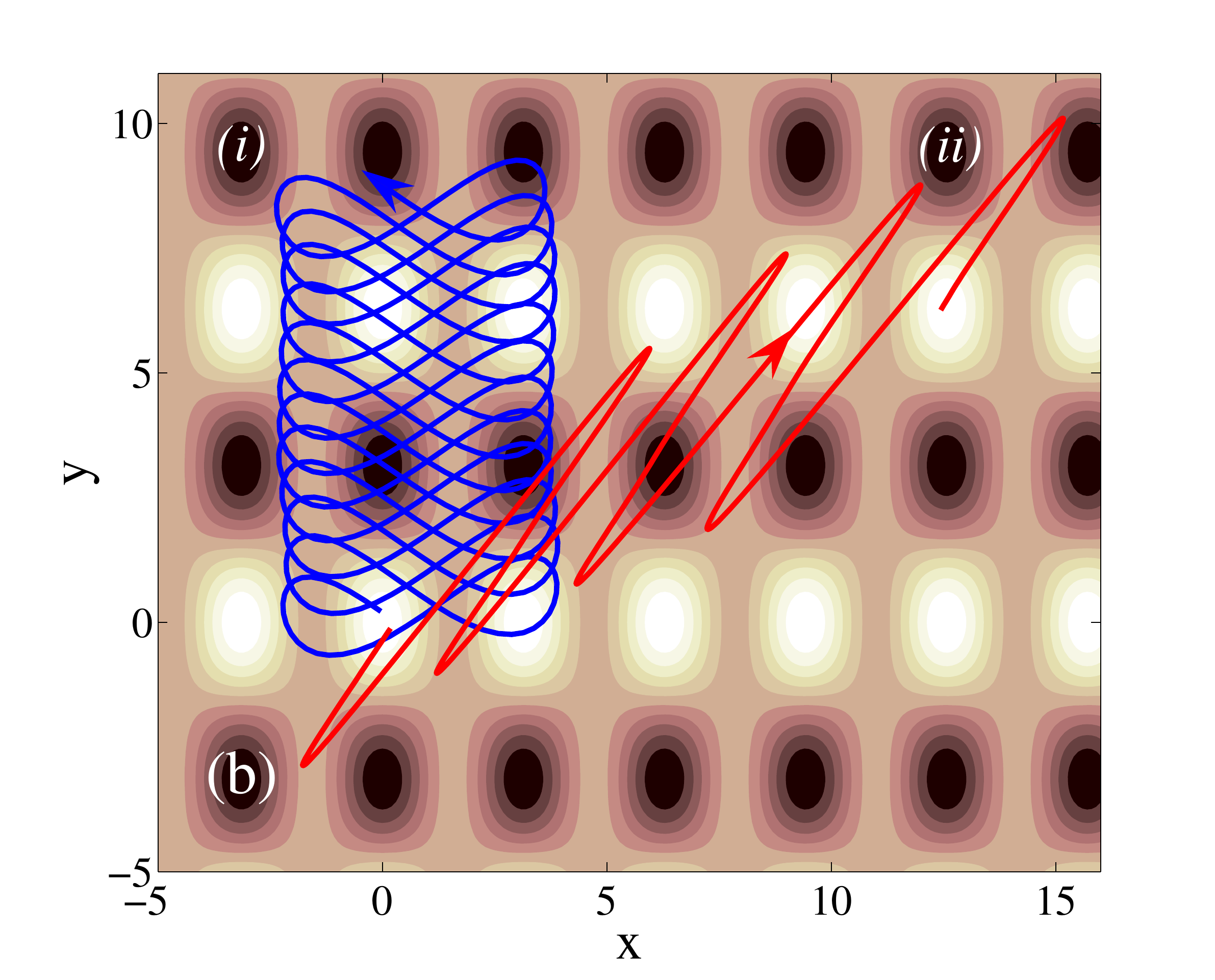}
\caption{(color online)  Time evolution of the mean particle positions
$\bar{\mathbf{r}}(t) = \int {\mathbf r} {\mathcal F}_A({\mathbf r}, {\mathbf v},
t) d{\mathbf r} d{\mathbf v}$, for the system given by Eqs.~(\ref{Add2d} - \ref{Ea_xy}). The trajectories are
superimposed on the contour plot of the potential (\ref{Va_xy}).
Curve (\textit{i}) corresponds to the parameters $E_x^{(1)}=3$,
$E_x^{(2)}=E_y^{(1)}=0$, $E_y^{(2)}=3.5$, $\theta=0$, and curve (\textit{ii})
to the set of parameters of Fig. \ref{fig:conv1}(a). Both trajectories were obtained by averaging over $10^6$ independent realizations.
The other parameters are $D = m
= 1, \gamma = 0.1 $. Adapted from Ref. \cite{dzfy08prl}.} \label{fig:conv2}
\end{center}
\end{figure}


The  overdamped limit, $m=0$, is singular for the function
(\ref{angular2}) since the velocity of a Brownian particle, ${\bf \dot
r\textrm{(t)}}$, is a nowhere differentiable function. The
overdamped limit can be approached by increasing $\gamma$ at a fixed
$ m = 1$.  Numerical simulations show that when $ \gamma/m \geq 5$  the
rotational current completely reflects the symmetries corresponding
to the overdamped case, see the dependence $\theta_0(\gamma)$ depicted on
Fig.~\ref{fig:conv3} (b).

We did not address the exact Hamiltonian limit, $\gamma = 0$, here.
The phase-space dimension of the system (\ref{1sys}) equals
five for $d=2$ and seven for $d=3$. At the same time invariant tori have the dimension three for
$d=2$ and four for $d=3$ \cite{suz92}. It is known that there are no topological constraints on the Hamiltonian evolution
in such  cases \cite{Arnold1964, Liberman1982}.
Already in the $2d$ case, the Hamiltonian dynamics of an ac-driven particle, even initiated within the region of low kinetic energy,
is not restricted to a finite-volume chaotic manifold
\cite{Liberman1982, Arnold1964}.
Therefore, an unbounded, possibly very  slow,  diffusion in the momentum
subspace may take place  \cite{Arnold1964, Bolotin1999, Mather2004}.
Direct numerical integrations of the equations of motion are  not very conclusive
since the time scale of the  momentum diffusion could be huge \cite{Mather2004}.
The sum rule \cite{sokd01prl} also does not apply here  because of the absence of compact invariant  manifold  over
which the system can self-average itself.  The relevant phase space structures and
the evolution of ac-driven $2$- and $3$-d Hamiltonian systems  are less explored up to date
\cite{Mather2004, Delshams2011}, and effects caused by symmetry breaking in such systems certainly deserves
further studies\footnote{Here recent computational advances could be of help \cite{Seibert2011}.}.

\begin{figure}[t]
\begin{center}
\includegraphics[width=0.75\linewidth,angle=0]{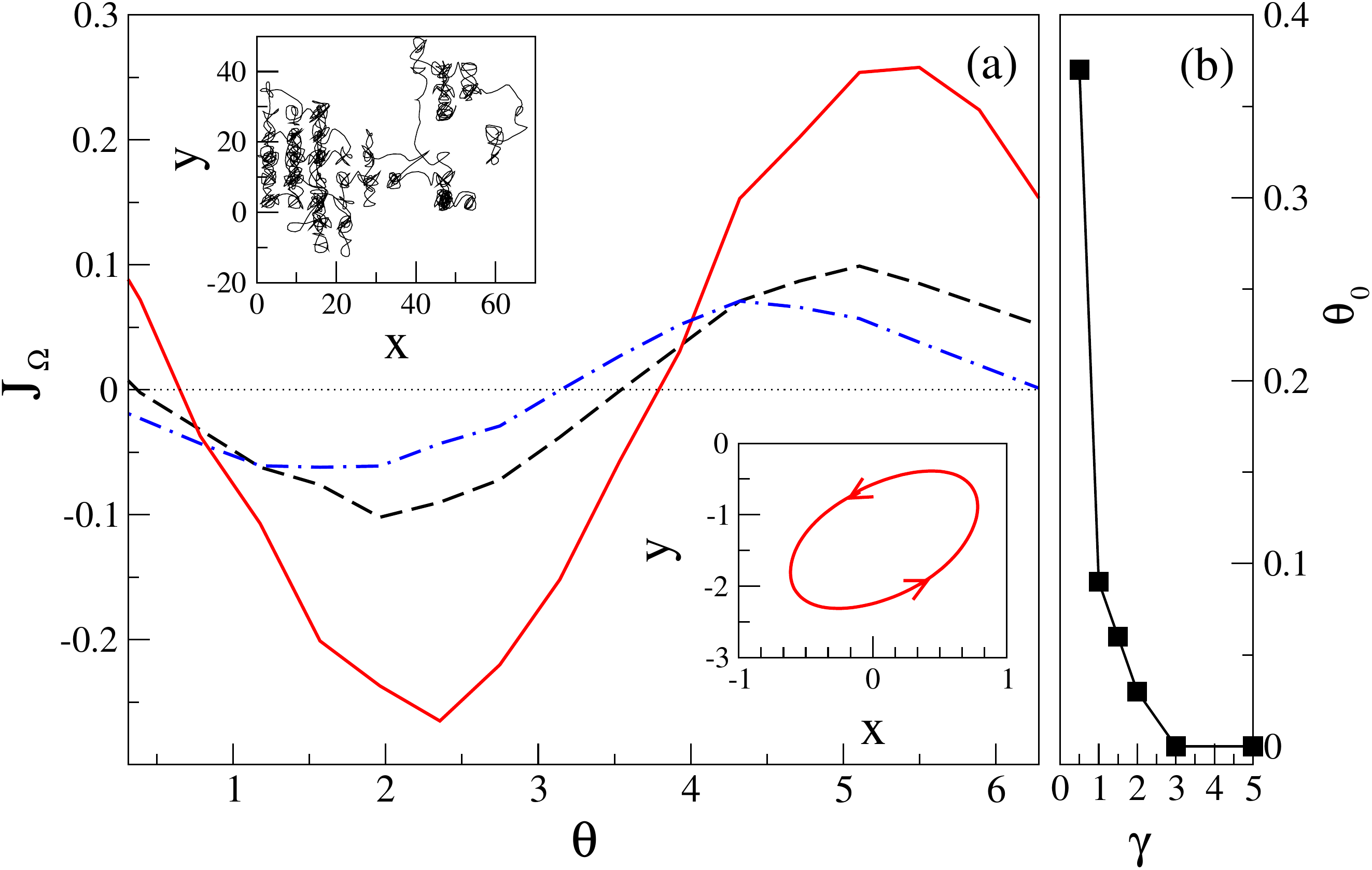}
\caption{(color online) (a) Dependence $J_{\Omega}(\theta)$,
Eq.~(\ref{angular2}), for (\ref{1sys}),
(\ref{potent3})-(\ref{force3}), with $m=1$, $D=0.5$,
$E_x^{(1)}=0.4$, $E_y^{(1)}=0.8$ and $\gamma=0.2$ (solid line),
$\gamma=0.05$ (dashed line), and $\gamma=2$ (dashed-dotted line).
Insets: the trajectory (left insert) and the corresponding attractor
solution, $\bar{\mathbf{r}}(t)$, (right inset) for the case
$\gamma=0.2$ and $\theta=\pi/2$. We performed an averaging over  $N=10^{5}$ independent
stochastic realizations in order to obtain the attractor; (b) The
phase lag $\theta_0$ as a function of the dissipation strength
$\gamma$. Adapted from Ref. \cite{dzfy08prl}.} \label{fig:conv3}
\end{center}
\end{figure}

To conclude this section, we briefly overview other existing models.
The case of two-dimensional stochastic rocking ratchets under the influence of
a colored noise has been studied in Ref.~\cite{gk03pre}. Since equivalent,
in a statistical sense, colored noises,
$\zeta_x(t)$ and $\zeta_y(t)$, have been used as driving forces, the
symmetry ${\widehat {\bf R}}_P$ (\ref{Rp}) can be violated only by an
asymmetric potential. Yet all potentials considered in Ref.
\cite{gk03pre} were invariant under the permutation transformation
$\widehat {\cal P}$. As a consequence, vortex structures for the
 velocity field presented in Ref. \cite{gk03pre} are
completely symmetric (clockwise vortices are mapped into
counterclockwise ones by ${\widehat  P}$) and, therefore, the
average rotation for any trajectory would be equal to zero.

There are some studies
of directed motion of particles, both quantum
and classical, in two-dimensional arrays of asymmetric scatterers \cite{da98pre,orjn01prl}.
The symmetries  were broken  along one direction only. Naturally, a directed
current occurred in the direction with broken symmetries. In the series of  papers
\cite{cs05prb,cs05pre,em06prb,c06epjb,cems07}, particle transport in periodic arrays of
scattering semi-disks and obstacles, under the additional influence of an
ac-drive of zero mean, has been considered. It has been
shown that by tuning the polarization direction of the
ac drive it was possible to change the direction of the current.
In the papers \cite{emrj03prl,er04pre,be07pre} the dynamics of colloidal suspensions of ferromagnetic
particles, placed in an external time-periodic magnetic
field has been studied. It was shown, both theoretically \cite{er04pre,be07pre} and
experimentally \cite{emrj03prl}, that due to the symmetry breaking induced by
the shape of the time-dependent magnetic field, particles perform directed
rotations. These rotations cause a non-zero macroscopic torque of the carrier
liquid, which effect was measured in experiment \cite{emrj03prl}.

Finally, in Ref. \cite{cs06} evolution of electrons in spatially elongated quantum dots,
modeled either by a elliptically-symmetric  single-well potential  or by the
Bunimovich stadium, under the influence of a linearly polarized microwave
radiation was studied. The radiation field plays the role of   ac drive,
${\bf E}(t)= \{E_x(t), E_y(t)\} =\{A \cos \varphi, A \sin \varphi \} \cos \omega t$, with the angle
$\varphi$ describing the polarization with respect
to the symmetry axis of the dot. Note, that such a system lacks spatial periodicity, and,
therefore does not fall into the class of directed transport  models, described by
Eq.~(\ref{1sys}). However, it supports unidirectional
rotation of electrons due to the symmetry breaking of the rotational transport component, induced by the
field when its direction does not coincide with the symmetry axes of the
stadium, e.g. $\varphi \neq 0, \pi/2$. As a result, a non-zero magnetization, caused by the unidirectional
rotations, appears.

\subsection{Experiments with cold atoms}

The creation of two- and three-dimensional  optical potentials  is a well-developed procedure in experimental physics of cold atoms,
and since the late 1990s  it is routinely performed in many laboratories \cite{Grynberg2001,verkek}. The corresponding technique is based
on an extension of the idea used to create one-dimensional optical potentials, see Sec. \ref{Experiments with cold atoms}.
An atom is immersed into an electromagnetic field, created by $N$ lasers of the same wavelength, $\lambda = 2\pi/\omega_L$,
\begin{eqnarray}
{\bf E}({\bf r}, t) = \sum_{j=1}^N E^0_j{\bf e}_j \mathsf{Re}\{\exp[-i(\omega_L t - {\bf k}_j\cdot {\bf r} -\phi_j)]\},
\label{EM_wave}
\end{eqnarray}
where $E^0_j$, ${\bf e}_j$, ${\bf k}_j$ are the intensity, polarization vector and wave-vector of the $j$-th beam. The electric
dipole interaction between the atom and the field lead to the creation of a spatially-dependent potential,
\begin{eqnarray}
V({\bf r}) \propto \overline{{\bf E}^2}({\bf r}, \phi_j),
\label{EM_potential}
\end{eqnarray}
where $\overline{...}$ stands for the averaging over the fast phase $\omega_L t$.
\begin{figure}
\begin{tabular}{cc}
\includegraphics[width=0.5\linewidth,angle=0]{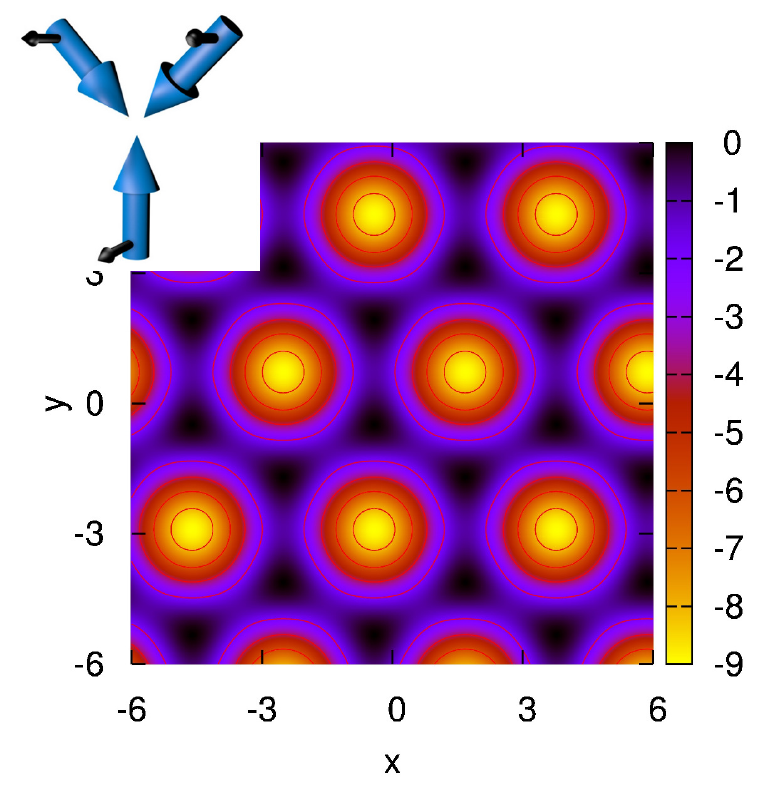}
\includegraphics[width=0.5\linewidth,angle=0]{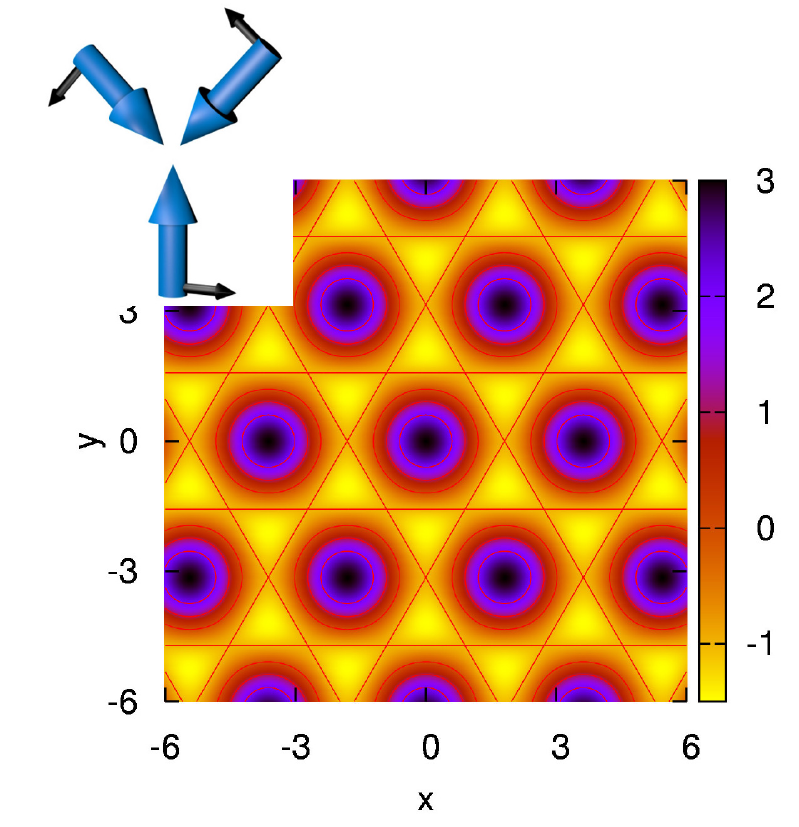}
\end{tabular}
\label{Figure13} \caption{(color online) Three-beam configurations and the corresponding two-dimensional optical potentials.
Big blue arrows indicate the directions of the beam wave-vectors
while the small black arrows denote the polarization of the beams. In both cases the beams are confined to $x-y$ plane.
When the polarization vectors point in the $z$-direction, the corresponding optical potential appears as a triangular lattice.
When the polarization vectors  span  $x-y$ plane in (counter)clockwise manner,
the resulting potential takes form of a hexagonal honeycomb lattice.} \label{2dlaser}
\end{figure}
Three laser beams of equal intensities are already enough to create  two-dimensional periodic potentials of tunable topology and
strength \cite{Grynberg2001}. Two most popular set-ups are shown on Fig.~\ref{2dlaser}.  In both cases  beams  intersect in
the $x - y$ plane at an angle of $120^\circ$ to each other. When the polarization vectors of the beams (black arrows)
point along the $z$-direction, the resulting potential forms a triangular optical lattice\footnote{The topography of a potential is defined
by the structure of the potential minima.}, left panel of Fig.~\ref{2dlaser}. When the beams are polarized in the $x-y$ plane, in a
clockwise or counter-clockwise manner, a hexagonal potential appears \cite{verkek},  right panel of
Fig.~\ref{2dlaser}. The relative beam phases $\phi_j$ are key parameters. For a three-beam setup, the number of relative phases
is two, i. e.  $\phi_1 - \phi_2$ and $\phi_2 - \phi_3$. The number of independent space translations for a $2$-d periodic
potential is also two. Hence, variations of the phases can only induce a global shift of the potential but cannot change the potential
topography \cite{Grynberg2001}. Therefore one can realize a two-dimensional rocking ratchet, Eq.~(\ref{Add2d}).
Namely, periodic modulations of the relative phases allows to introduce rocking forces, both in  $x$ and $y$ directions,
in a controllable manner. The idea is similar to that  used for the experimental realization $1$-d cold atom ratchets,
Sec. \ref{Experiments with cold atoms}: by shaking
the potential with a  time-periodic phase, $a_\alpha(t+T) = a_\alpha(t)$, $\alpha = \{x ,y\}$, one introduces a rocking force
along the $\alpha$-direction, $f_\alpha(t) \propto \ddot{a}_\alpha(t)$, in the co-moving frame, $\tilde{\alpha} = \alpha - a_\alpha(t)$.
The asymptotic current is the same in both frames, laboratory and co-moving, $J_{\tilde{\alpha}} = J_{\alpha}$.
In an experiment, the relative phases can be controlled by using three acousto-optical modulators \cite{Grynberg2001}.


\begin{figure}[t]
\begin{center}
\includegraphics[angle=0,width=0.55\textwidth]{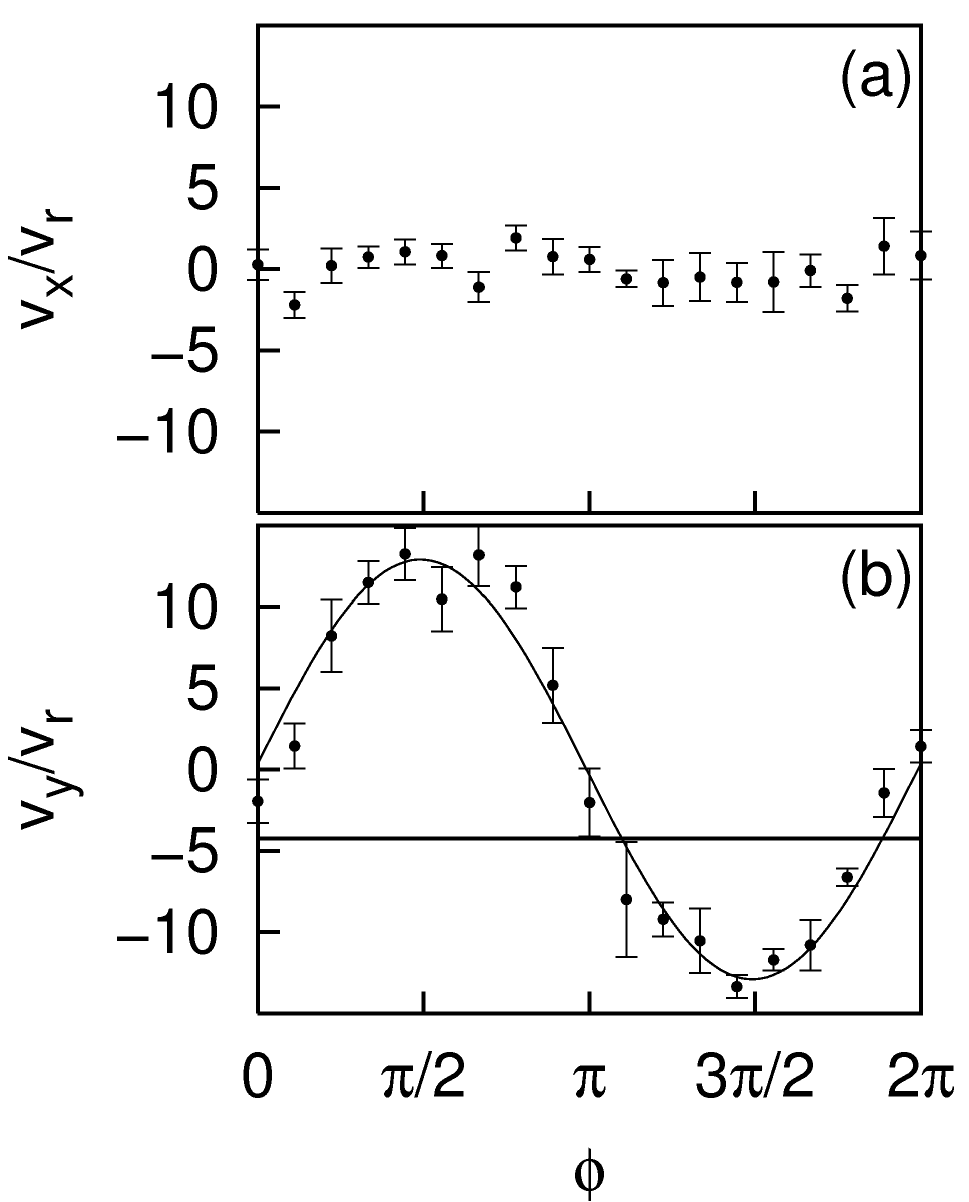}
\caption{Components of the average velocity of the atomic cloud center of mass
as a function of  $\theta_y$ (denoted by $\phi$  here) measured in experiments
 with a two-dimensional optical rocking ratchet-like potential, Eqs.~(\ref{Add2d}, \ref{Ea_xy}), and cold cesium atoms.
The driving vector has components $E^{(2)}_{x} = E^{(1)}_{y} = 0$. The solid line is a
 fit of the data with the function  $v_y = v_{\rm max}\sin(\theta + \theta_0)$.
Adapted from Ref. \cite{lr09}.} \label{fig2REn1}
\end{center}
\end{figure}

This idea has been realized by the group of F. Renzoni \cite{lr09}.
The driving vector was of
the standard bi-harmonic form, Eq.~(\ref{Ea_xy}), with amplitudes $E^{(1,2)}_{x,y}$  and phases $\theta_{x,y}$ as tunable parameters.
An optical potential $V(x,y)$ of the hexagonal topology, right panel of
Fig.~\ref{2dlaser}, was used in the experiments. The potential is symmetric in both direction, $x$ and $y$.
For the driving vector with $E^{(2)}_{x} = E^{(1)}_{y} = 0$, symmetry
${\widehat {\bf S}}_{x}[{\bf 0},\pi]$, Eq.~(\ref{SSx}), forbids directed current in the $x$-direction.
In the Hamiltonian limit, transport of atoms along the $y$-direction is controlled by the symmetry ${\widehat {\bf S}}_{t}[{\bf 0},\pi]$, such that $J_y = 0$
whenever $\theta_y = 0, \pm \pi$.  The experimental results shown on Fig.~\ref{fig2REn1}
perfectly followed the prediction of the symmetry analysis.

A more subtle control of the current direction was realized by using the following idea.
Consider a
bi-harmonic drive vector (\ref{Ea_xy}) with phases $\theta_{x,y} = \pi/2$
and fixed ratios between the harmonic amplitudes, $E^{(1)}_{x}/E^{(2)}_{x} = E^{(1)}_{y}/E^{(2)}_{y}=3/4$. This setup breaks all
relevant symmetries and therefore a directed current should appear.  The only control parameter left is the ratio between the harmonic amplitudes
in mutually orthogonal directions, $q = E^{(1)}_{x}/E^{(2)}_{x}$.  By tuning $q$ to zero (infinity) it was possible to direct the current along the
$x$($y$)-direction. Intermediate values of $q$ correspond to the atomic  transport along diagonal directions, see Fig.~\ref{fig2REn2}.
Note that both transport modes, along $x$ and $y$ directions,  are not independent but dynamically coupled. Therefore the overall two-dimensional transport
cannot be considered as a `product' of two mutually independent, one-dimensional rectification processes.


\begin{figure}[t]
\begin{center}
\includegraphics[angle=0,width=0.75\textwidth]{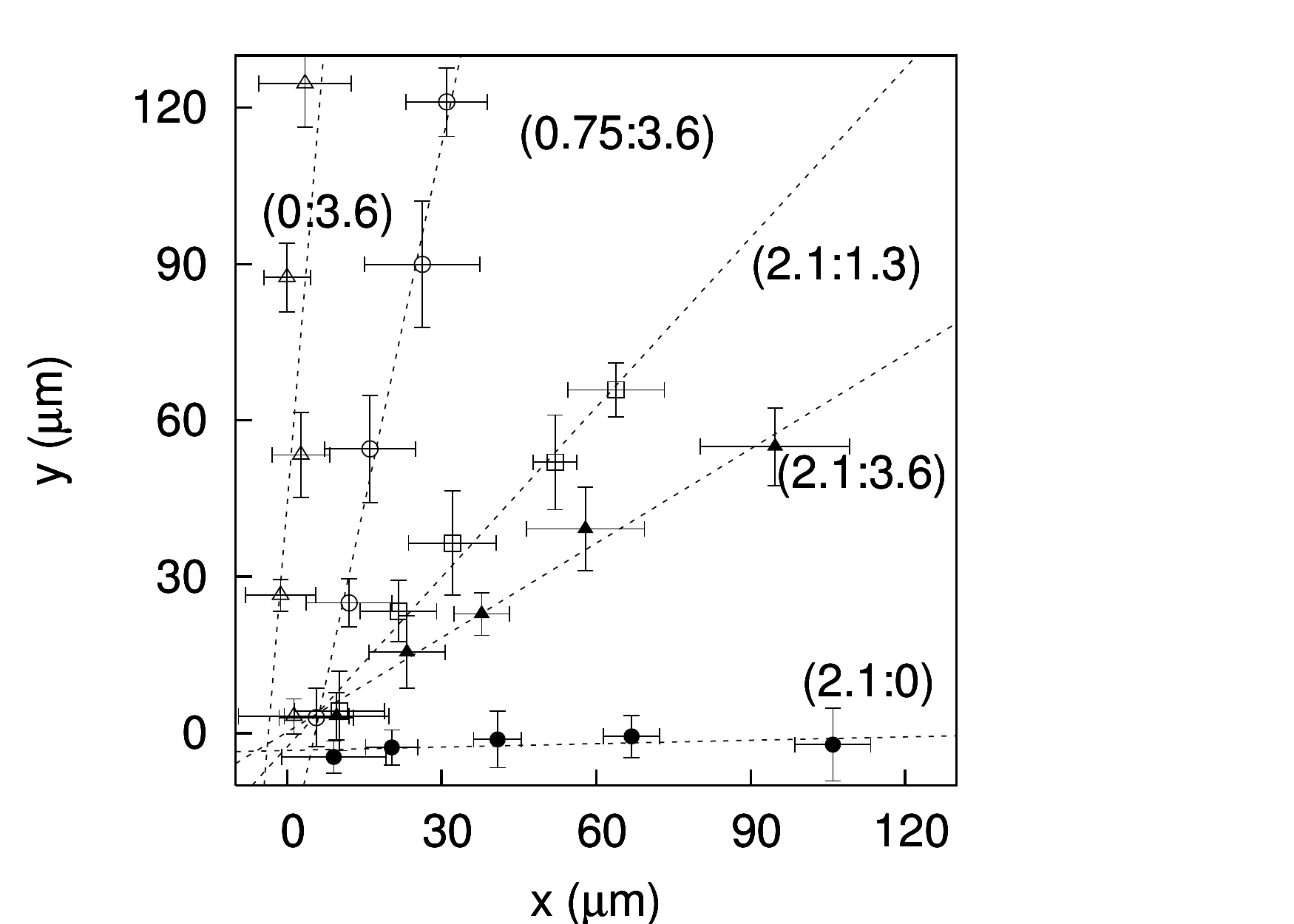}
\caption{Performance of a two-dimensional cold atom rocking ratchet, Eqs.~(\ref{Add2d}, \ref{Ea_xy}).
The plot shows positions of the atomic cloud center-of-mass
at different instants of time. The different data sets correspond to
different ratios between the driving amplitudes, $q = E^{(1)}_{x} : E^{(2)}_{x}$.
Points were obtained by taking  fluorescence images of the atomic cloud, see Fig.~\ref{flour},
at intervals of $0.5$ ms. The lines are the best fits of
the data with  linear functions $y=(\upsilon_y/\upsilon_x)x$.
Adapted from Ref. \cite{lr09}.} \label{fig2REn2}
\end{center}
\end{figure}

Realizations of a peculiar ratchet-like mechanism with cold atoms moving in two- and three-dimensional optical lattices were
reported in Ref. \cite{spdjnsk06}. There were no ac-driving forces, and the rectification effect appeared
due to decoherence-induced transitions of atoms between two different optical potentials.
If the optical potentials were shifted in
such a way that their relative phases in $x,y$ and $z$-directions were different from $k\pi$,
$k=0,\pm1$, the presence of an optical pumping mechanism, which induces transitions of atoms between  potentials,
led to the appearance of a net motion of the atomic cloud.
By tuning the relative phases, it was possible to shuttle atoms into desirable directions.

\section{Quantum ratchets}
\label{sec5}

Experimental realizations of ac-driven ratchets with cold atoms
discussed in the previous two sections \cite{ss-prg03prl,jgr05prl,gbr05prl,r05contp, lr09}
are in a good agreement with the numerical results obtained for the classical models.
There are two key factors responsible for this accordance. First of all, it is the initial state of the atomic cloud used in experiments, namely its temperature.
The de Broglie wavelength of a cesium atom at the temperature of the order of tens $\mu$K is of the order of $10$ $n$m while
the spatial period of the optical potentials $L$ is of the order of $\mu$m.
The atom therefore can be well approximated by a point-like particle.  Even at several $\mu$K  the
corresponding de Broglie wavelength is of the order of $100$ $n$m, which is still smaller than $L$.
Only by cooling atoms down to $n$K temperatures, i. e. by entering into the Bose-Einstein condensation regime, one could
get atomic wavelengths of the order of ten $\mu$m, which are now larger than the  potential
period. Below this temperature atoms  start behave themselves as genuine quantum objects.
At the same time the kinetic energy of the atom becomes of the order of the recoil energy $E_R = \hbar k_L^2/2m$,
which means that the atom evolves within the potential range and feels the potential shape.
Last but not least factor responsible for downgrading the atomic evolution to the classical limit was  the decoherence  \cite{gbr05prl}.
Strong interaction between the laser field and atoms induced a  chain
of stochastic adsorption/emission events with a characteristic rate comparable to the period of the driving
which process destroyed the coherence of the quantum evolution.

Therefore, there are two conditions which have to be respected in order to grasp  quantum
ratchets in experiment.  First,  the
atomic cloud has to be cooled down to the BEC-transition
temperature so that the distribution of the atom momenta is narrow with respect to the energy scale set by the potential height,
and  the atomic de Broglie wavelength becomes of the order (or even larger than)  the optical potential period $L$.
After that the tunneling between potential wells becomes essential and  starts contribute to the transport
process.
Second, a further detuning of the optical potential is needed \cite{Morsch2006} in order to make decoherence effects
negligible on the time scale of experiment.

Both demands are  within the reach of modern  ultra-cold atom optics  \cite{Bloch2008}, and the
creation of an atomic Bose-Einstein condensates (BECs) followed by the consecutive loading of the BEC into an optical confinement,
is  an almost everyday  routine in many laboratories across the world.
Quite naturally,  a realization of ac-driven coherent quantum ratchets  was reported in 2009 \cite{weitzScience}.

In this section we  present an extension of the  symmetry-analysis to the case of coherent quantum ratchets \cite{dmfh07},
outline the peculiarities  of the  quantum rectification process, and review the first experimental realizations of a multiplicative quantum ratchet
with a Bose-Einstein  condensate of rubidium atoms \cite{weitzScience}.

\subsection{Symmetry analysis}

We start with a quantum particle moving in a time and space periodic potential.
The Hamiltonian is
\begin{equation}
H(\hat{x},\hat{p},t)=\frac{\hat{p}^{2}}{2}+V(\hat{x},t),
\label{eq:Qham}
\end{equation}
where
$V(\hat{x},t) = V(\hat{x}+\hat{L},t) = V(\hat{x},t+T)$, and  $\hat{L}$ is the translation operator
over the distance $L$.
The corresponding  Schr\"{o}dinger equation reads
\begin{equation}
i \hbar \frac{\partial}{\partial t} |\psi(t)\rangle =
H(t)|\psi(t)\rangle. \label{eq:Schrodin}
\end{equation}
The Hamiltonian  (\ref{eq:Qham}) is formally identical to its classical version,
with the only  difference that the position and momentum  are operators now  not numbers.
Despite this similarity, there are two technical issues that make  the symmetry analysis in the quantum case different from the previous classical case.

The first  issue concerns the definition of the  quantum current.
It is not a variable of the equation of motion like in the classical limit
but an expectation value of a certain operator.
Namely, the current for a given initial wave function
is quantified through the instantaneous momentum expectation value,
\begin{equation}
J(t; t_0)= \frac{1}{m}\langle
\psi(t,t_0)|\hat{p}|\psi(t,t_0)\rangle.\label{eq:current1}
\end{equation}

Second,  the evolution of the quantum system is governed by the equation which includes  the
potential $V(\hat{x},t+T)$ while the evolution of the classical system is governed by the equation (\ref{2.1})
that has a force as an input, i. e. the derivative of the potential $g(x,t) = -\partial_x V(x,t)$.
In the case of a multiplicative setup, Eq.~(\ref{2.22}),
both functions, potential and force, are time- and space-periodic. So are the corresponding equations of motion, and there is no visible
difference with respect to the symmetry analysis.
However, in the case of additive setup, Eq.~(\ref{2.21}), where the classical force function is time-space periodic,
the original potential,
\begin{equation}
V(\hat{x},t) = V(\hat{x}) - \hat{x}E(t), \label{eq:qqadd}
\end{equation}
is not. This problem can be resolved by resorting to
a gauge transformation  \cite{Henneberger1968}, $|\psi \rangle \rightarrow \exp(-\frac{i}{\hbar}x\int_{t_0}^{t}E(t')dt') |\psi \rangle$, which  yields  a new Hamiltonian,
\begin{equation}
H(\hat{x},\hat{\tilde{p}},t)=\frac{[\hat{\tilde{p}}- A(t;t_0)]^{2}}{2} + V(\hat{x}).
\label{eq:ham2}
\end{equation}
The new momentum operator is time-dependent now, $\hat{\tilde{p}} = \hat{p} + A(t;t_0)$, with
a time-periodic vector potential, $A(t;t_0) = A(t+T;t_0) = -\int_{t_0}^{t} E(t')dt' + A(t_0)$ where the constant
$A(t_0)$ is chosen such that the time average of $A(t)$ over one temporal period of the drive
vanishes. Since the vector potential  is of zero mean, the time averaged expectation
values of the momentum operators, $\hat{p}$ and $\hat{\tilde{p}}$, are identical.
Note  the explicit parametric dependence of the vector potential on the parameter $t_0$ (`starting time' henceforth) which determines the
initial phase of the driving field, $E(t_0)$. As we will show, this parameter plays an important role in the dynamics of coherent
quantum ratchets.

The solution of the  Schr\"{o}dinger equation (\ref{eq:Schrodin}) for a given initial state $|\psi(t_0)\rangle$
can be formally written as $|\psi(t+t_0)\rangle =
U(t,t_{0})|\psi(t_0)\rangle$, where $U(t,t_{0})$ is the propagation or evolution operator.
Hamiltonian (\ref{eq:Qham}) is periodic in time with period $T$, and the system evolution can be evaluated
by using eigenfunctions of the Floquet operator, $U(T,t_0)$, which
propagates the system over one period of the driving, $ |\psi_{\alpha}(t)\rangle= e^{-i\frac{E_{\alpha}}{\hbar}t}
|\phi_{\alpha}(t)\rangle$, $|\phi_{\alpha}(t+T)\rangle=|\phi_{\alpha}(t)\rangle$  \cite{Shirley1965,Sambe1973,Grifoni1998}.
The eigenfunctions can be obtained by solving the eigenvalue
problem for the Floquet operator,
\begin{equation}
U(T,t_{0})|\phi_{\alpha}(t_{0})\rangle = e^{-i
E_{\alpha}T/\hbar}|\phi_{\alpha}(t_{0})\rangle, \label{eq:Floquet}
\end{equation}
with the eigenvalues
$e^{-i E_{\alpha}T/\hbar}$. Their phase are known as quasienergies,
$E_{\alpha} \in [-\hbar\omega/2,  \hbar\omega/2]$.
When $\hbar \ll 1$, Floquet states of the quantum  system can be associated with different invariant manifolds
of its classical version, see Fig.~\ref{Fig:husi}.
\begin{figure}[t]
\begin{center}
\begin{tabular}{cc}
\includegraphics[angle=0,width=0.44\textwidth]{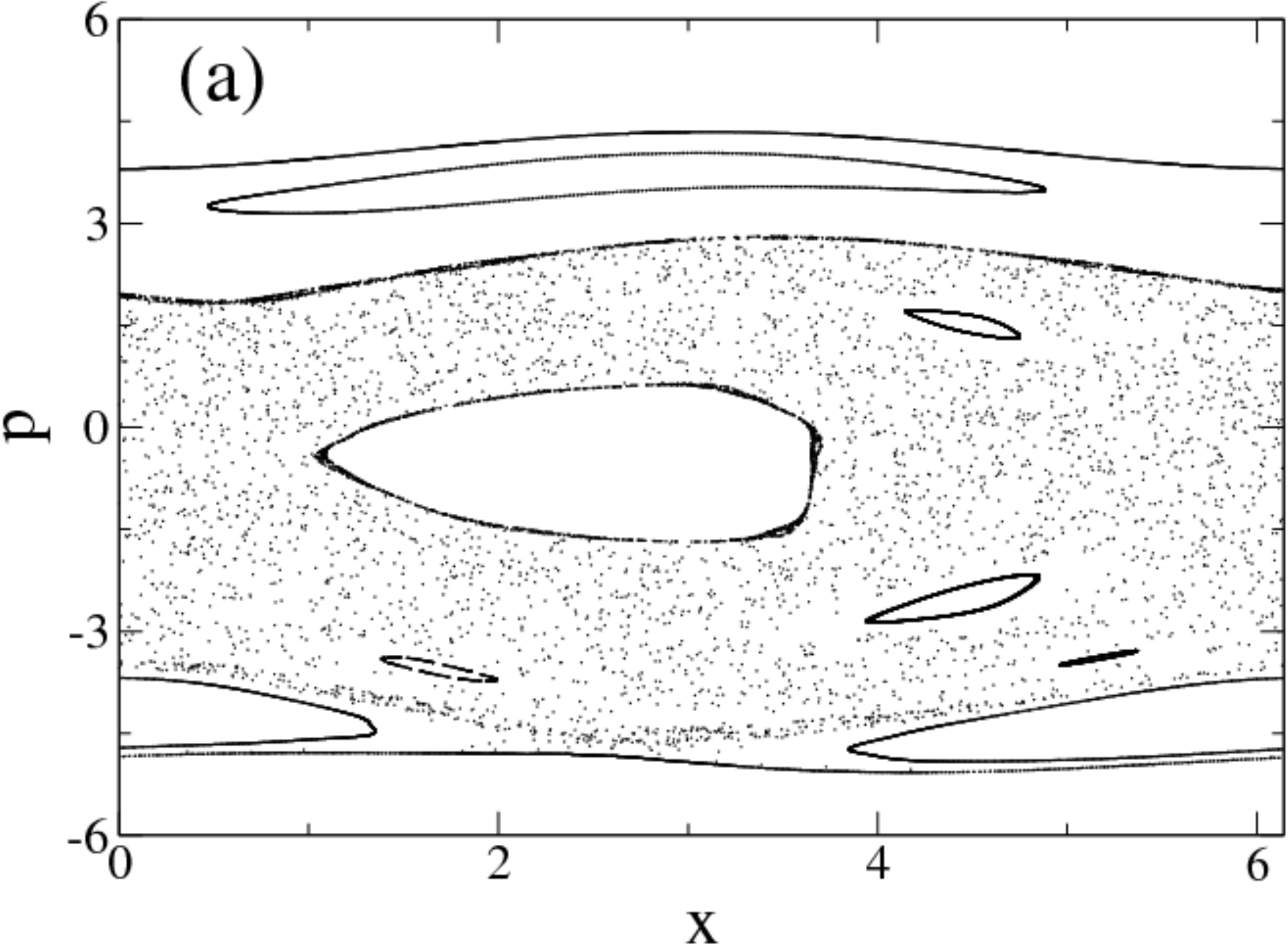}
\includegraphics[angle=0,width=0.44\textwidth]{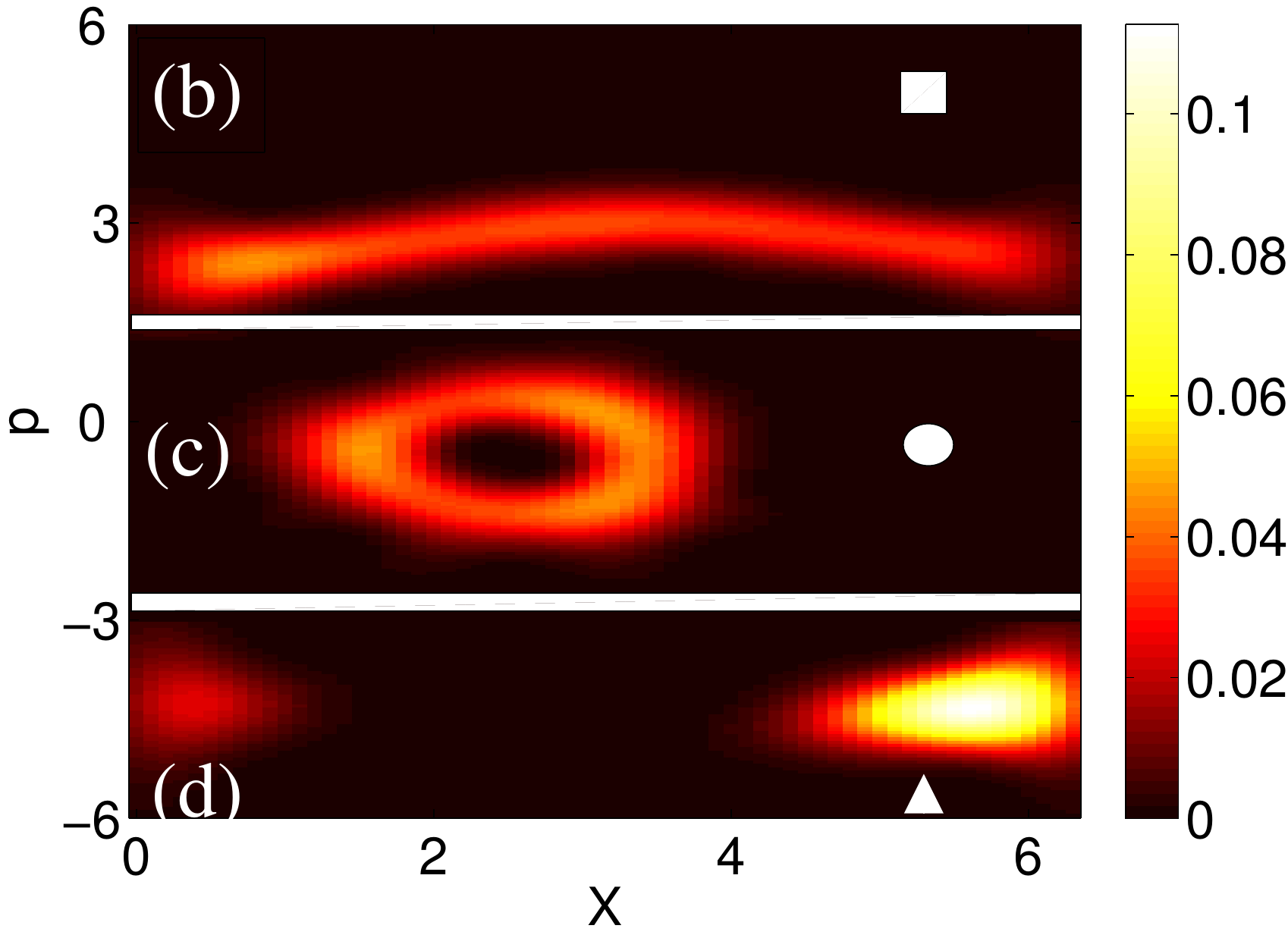}
\end{tabular}
\label{Figure1d} \caption{(color online) (a) Poincar\'{e} section of the Hamiltonian system (\ref{eq:QQpotential}, \ref{eq:QQdriving})
in  the classical limit and (b-d) Husimi distributions \cite{Husimi1} of different Floquet
states of the quantum version,  $\hbar=0.2$.
The parameters are $E_{1}=E_{2}=2$, $\omega=2$,
$\theta=-\pi/2$ and $t_{0}=0$. Adapted from Ref. \cite{dmfh07}} \label{Fig:husi}
\end{center}
\end{figure}
The set of Floquet states, $\{|\phi_{\alpha}(t)\rangle\}$, forms a complete orthonormal basis
and the state of the quantum system at any instant of time is
\begin{equation}
|\psi(t; t_0)\rangle = \sum_{\alpha} C_{\alpha}(t_{0}) e^{-it
E_{\alpha}t/\hbar}|\phi_{\alpha}(t_{0} + t)\rangle, \label{eq:expans}
\end{equation}
where the coefficients $\{C_{\alpha}(t_0)\}$ are given by the expansion of the
initial state over the Floquet basis,
\begin{equation}
|\psi(t_0)\rangle = \sum_{\alpha} C_{\alpha}(t_{0})|\phi_{\alpha}(t_{0})\rangle.
\end{equation}
The expansion  coefficients explicitly depend on  $t_0$ so that
even  when the shape of the initial wave function, $|\psi(t_{0})\rangle$, is fixed, the expansion over the
Floquet basis still depends on the starting time.
This is because Floquet states, although
retaining their initial shapes after one round of the driving, are time-dependent vectors.
Due to the discrete translational invariance of the Hamiltonian in $x$-space, all Floquet states can be arranged into  subsets,
characterized by different values of  quasimomentum  $\kappa$:
$|\phi_{\alpha}(x+2\pi)\rangle = {\rm e}^{i \hbar \kappa}
|\phi_{\alpha}(x)\rangle$, $\kappa \in [-\pi/L, \pi/L]$.
The quasimomentum subspaces are invariant under the evolution of the Schr\"{o}dinger equation.

We start the analysis with the case $\kappa=0$ (the general case $\kappa \neq 0$ will be considered later on).
The case of zero quasimomentum corresponds either to (i) an initial state which
was uniformly smeared over the whole, strictly  speaking, infinite
potential $U(x,t=t_0)$ or (ii) to the case  when the evolution of the system is confined a potential in a form of a ring of the length $L$.
The first situation can be well approximated with an initial cloud of atoms
distributed over a distance much larger than the spatial period of the potential.

Consider now transport properties of Floquet states. Similar to the invariant manifolds of classical deterministic systems,
every Floquet state can be labeled by its averaged velocity, $\bar{\upsilon}_{\alpha} = (1/mT) \int_0^T p_{\alpha}(t) dt$,
where $p_\alpha (t) = \langle \phi_{\alpha}(t)|\hat{p}| \phi_{\alpha}(t)\rangle$.
Combining this definition with Eqs.~(\ref{eq:current1}) and (\ref{eq:expans}), we arrive at the following expression
for the current,
\begin{equation}
J(t; t_0)=  \frac{1}{m}\sum_{\alpha,\alpha'} C_{\alpha}(t_0) C^{*}_{\alpha'}(t_0) e^{-i
      (E_{\alpha}-E_{\alpha'})t/\hbar} \langle
\phi_{\alpha'}(t+t_0) |\hat{p} |\phi_{\alpha}(t+t_0) \rangle.\label{eq:currentQ1}
\end{equation}
It can be represented as the sum of diagonal and off-diagonal contributions,
\begin{eqnarray} \nonumber
J(t; t_0)=   \frac{1}{m}\sum_{\alpha}| C_{\alpha}(t_0)|^2 p_{\alpha}(t+t_0) + ~~~~~~~~~~~~~~~~\\
\frac{1}{m}\sum_{\alpha \neq\alpha'} C_{\alpha}(t_0) C^{*}_{\alpha'}(t_0) e^{-i
      (E_{\alpha}-E_{\alpha'})t/\hbar} \langle
\phi_{\alpha'}(t+t_0) |\hat{p} |\phi_{\alpha}(t+t_0) \rangle.\label{eq:currentQ2}
\end{eqnarray}

Now we turn to the asymptotic averaged current,
\begin{equation}
\tilde{J}(t_0) =
\lim_{t \rightarrow \infty} \frac{1}{t} \int_{t_0}^t J(\tau; t_0) d\tau. \label{eq:currentQQQQQ1}
\end{equation}
We assume  that the quasienergy spectrum of the system
has no degeneracies, i. e. there are no pairs $\{\alpha, \alpha' \neq \alpha\}$
such that $E_{\alpha} = E_{\alpha'}$. This is a well-justified assumption  supported by the extension of the von
Neumann-Wigner theorem \cite{Neumann1929} to the case of periodically modulated Hamiltonians
\footnote{Nodegeneracy can be  absent in some specific models, e. g. when the dynamics of a quantum ratchet is restricted
to several \textit{fixed} bands, with an enforced crossing between a pair of them \cite{heimsoth2010,creffield2011}}.

Note that heretofore we consider  the case of vanishing quasimomentum $\kappa=0$ only. The symmetry action is more
involved for nonzero quasimomentum, and this case will be addressed later on. 
The immediate consequence of this assumption is that as time grows, all interference terms
collected into the second sum on the rhs of Eq.~(\ref{eq:currentQ2}) are averaging themselves out. Thus, in the asymptotic  limit
$t \rightarrow \infty$, the diagonal part is only left,
\begin{equation}\label{eq:quantum_as_current}
\tilde{J}(t_0) =   \sum_{\alpha}| C_{\alpha}(t_0)|^2 \upsilon_{\alpha}.
\end{equation}

This result  demarcates coherent quantum Hamiltonian ratchets from their classical counterparts in a clear-cut manner.
While it was possible to assign a unique  current value to the chaotic layer of the classical Hamiltonian ratchet, $J = \upsilon_{ch}$,
see also Eq.~(\ref{sum_rule}), it is impossible to do so in the quantum limit.
The stochasticity, which takes care of erasing memory on initial conditions
and induces averaging over the chaotic region, is absent in the  quantum world.
The evolution of a coherent quantum system is governed by the linear equation, Eq.~(\ref{eq:Schrodin}),
so that the system keeps memory on its initial state imprinted into the set of
coefficients $C_{\alpha}(t_0)$. Even in the asymptotic limit, the quantum ratchet current still
depends not only on the shape of the initial wave function but also on the starting time $t_0$, because of the explicit time-dependence
of the Floquet basis. A quantum coherent ratchet can be
thought as a set of moving conveyer belts -- the Floquet states --
and a cloud -- the initial wave packet -- is distributed among them so that every belt gets a certain load, see Fig.~\ref{tercera}(a).
The total current, Eq.~(\ref{eq:quantum_as_current}),  is a sum of the belt velocities weighted with the corresponding loads.
A classical Hamiltonian ratchet can be thought of as  a turbulent stream, which drags all particles
with the same average velocity, $\upsilon_{ch}$, independently of their initial positions,
see Fig.~\ref{tercera}b. The same picture holds true when the classical system is coupled to a heat bath.
The asymptotic solution of the corresponding Fokker-Planck equation,
Eq.~(\ref{fpe1_under}) or Eq.~({\ref{fpe1_over}, consists of a single Floquet state only, $P_A(...,t)=P_A(...,t+T)$.
All the other Floquet states have
multipliers with nonzero imaginary parts and therefore die out in the course of time \cite{Jung}.
The asymptotic current does not depend on  initial conditions in this case as well.

\begin{figure}[t]
\begin{center}
\begin{tabular}{cc}
\includegraphics[angle=0,width=0.45\textwidth]{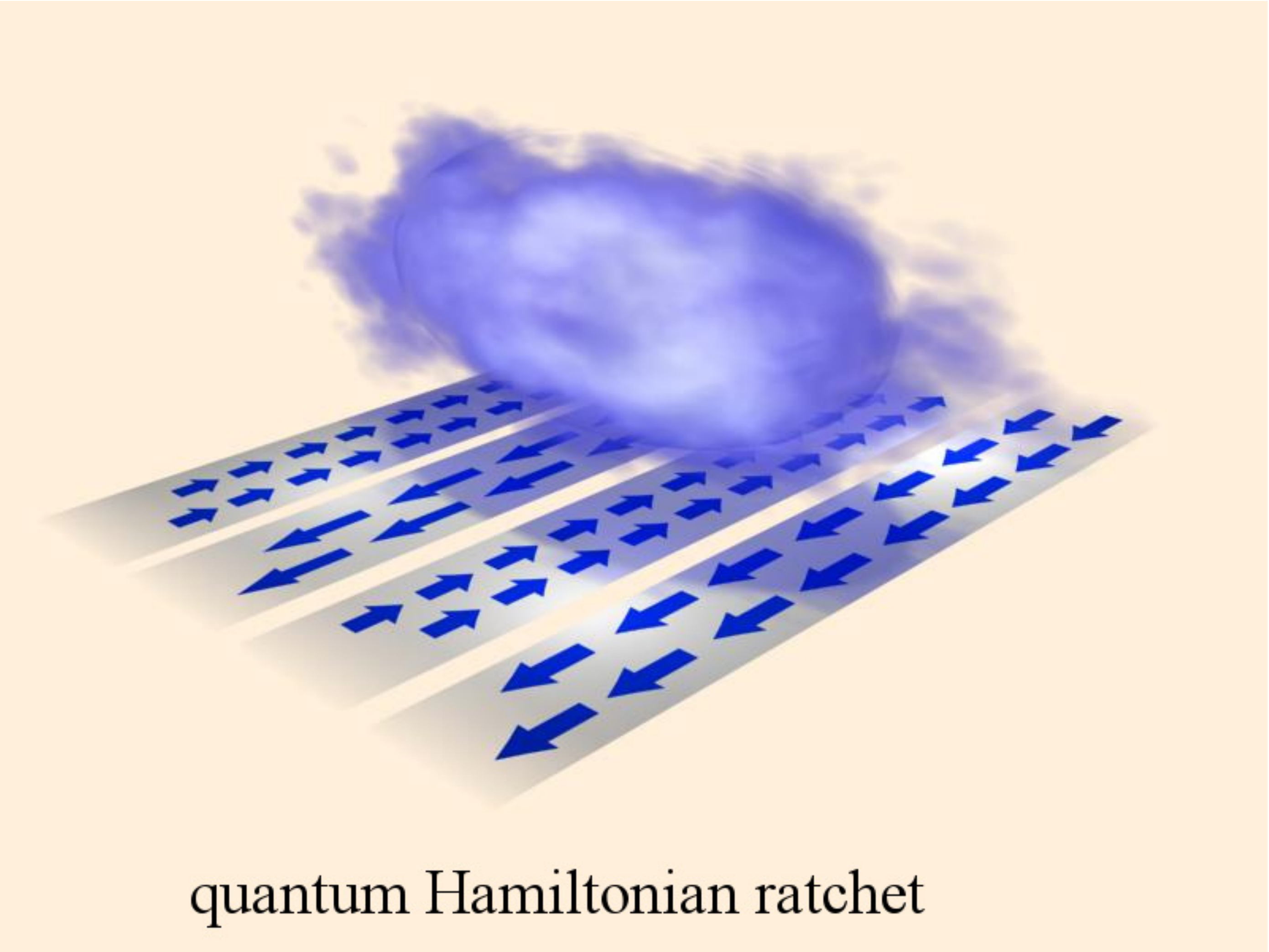}
\includegraphics[angle=0,width=0.45\textwidth]{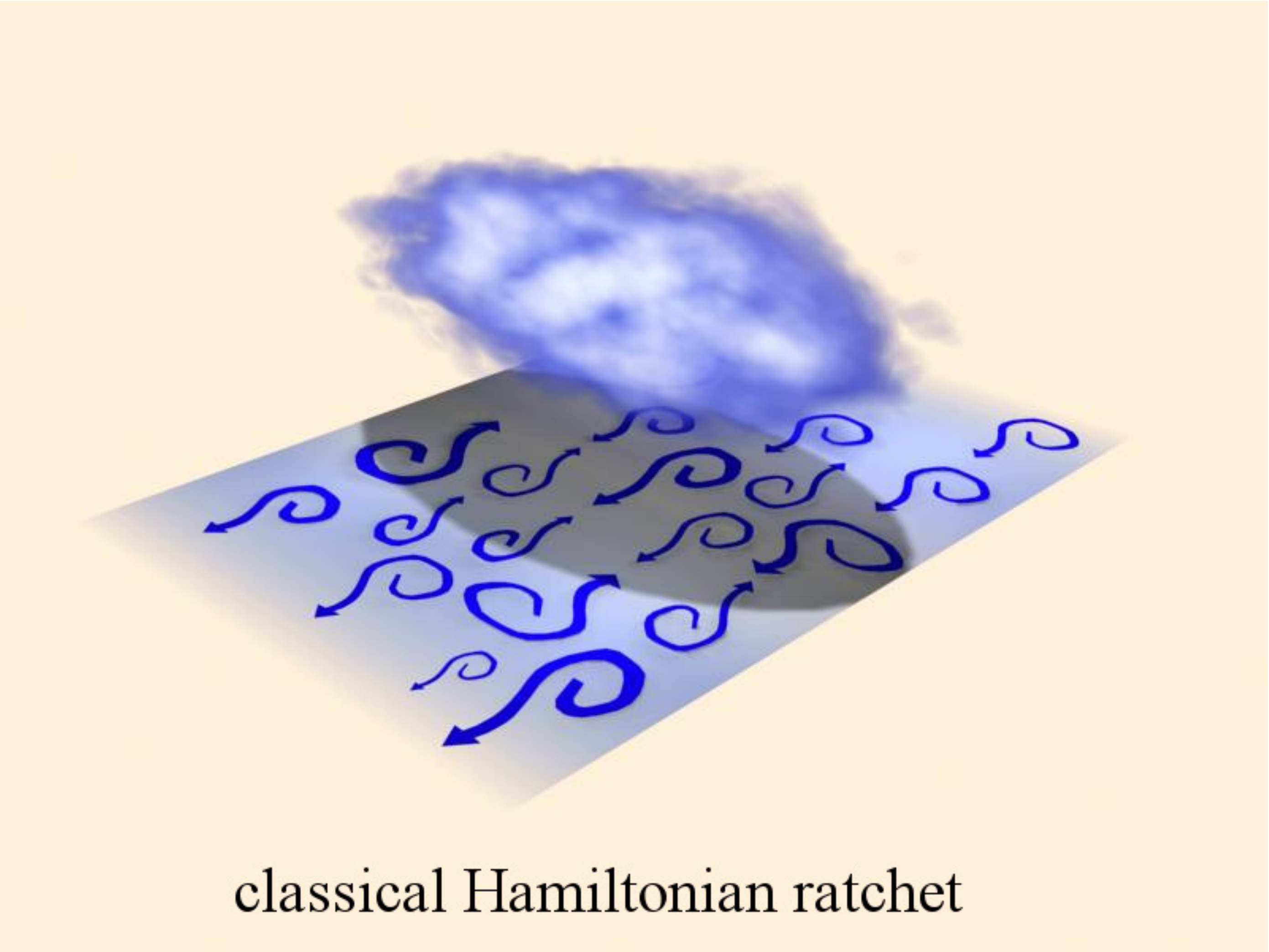}
\end{tabular}
\end{center}
\caption{(color online) Performance of a quantum (left) and classical (right) Hamiltonian ratchets.
In the quantum limit an ac-driven ratchet can be represented as a set of conveyer belts (Floquet states), each moving
with a constant velocity. The asymptotic current,  Eq.~(\ref{eq:quantum_as_current}),  depends on the way an initial
wave packet is distributed among the belts. The asymptotic current in the chaotic layer of a classical Hamiltonian  ratchet is
of the same value for all initial conditions chosen within the layer, see Eq.~(\ref{sum_rule}).}\label{tercera}
\end{figure}

Having now the definition of the quantum current,  we can turn to the symmetry analysis.
Remarkably, the relevant symmetries are similar to the classical case  \cite{dmfh07, Denisov2007_2}.  Namely,
there are two basic symmetries,
Eq. (\ref{eq:Schrodin}),
\begin{equation}
{\hat S}_x[\hat{\chi},\tau]: \{\hat{x}, \hat{p}, t\} \rightarrow \{-\hat{x}+\hat{\chi}, -\hat{p}, t+\tau \}, \label{eq:Sxq}
\end{equation}
and
\begin{equation}
{\hat S}_t[\hat{\chi},\tau]: \{\hat{x}, \hat{p}, t\} \rightarrow \{\hat{x}+\hat{\chi}, -\hat{p}, -t+\tau\},  \label{eq:Stq}
\end{equation}
where $\hat{\chi}$ is the translational operator over the distance $\chi$.
However, the ways the symmetries affect the system dynamics in the quantum limit are very different from
how they realized  themselves in classical ratchets.
A symmetry, when present,  is imprinted in the  corresponding  Floquet operator, $U(T,t_0)$, and thus influences
transport properties of its eigenstates \cite{dmfh07}.

The presence of symmetry ${\hat S}_x$, also called `generalized parity` \cite{Grifoni1998}, implies that the
corresponding Floquet states  obey:
\begin{equation}
|\phi_{\alpha}(-x, t)\rangle = \sigma_{\alpha}|\phi_{\alpha}(x, t + \tau)\rangle, \label{eq:parity_X}
\end{equation}
where $\sigma_{\alpha} = \pm 1$. The space reversal changes the sign of the momentum operator,  $\hat{p} = -i\hbar\partial/\partial x$.
The expectation values of the instantaneous velocity of a Floquet state at instants of times separated by $\tau$  are
symmetry related,
 $\upsilon_{\alpha}(t) = -\upsilon_{\alpha}(t+\tau)$. Therefore, the average velocity, $\bar{\upsilon}_\alpha$, of any Floquet state
is equal to zero.

The symmetry ${\hat S}_t$  involves complex conjugation and the
Floquet states of the corresponding system obey:
\begin{equation}
|\phi_{\alpha}(x, T-t)\rangle = \sigma_{\alpha}|\phi^*_{\alpha}(x, t)\rangle, \label{eq:parity_t}
\end{equation}
Because of the complex conjugation operation,  the momentum operator, $\hat{p} = -i\hbar\partial/\partial x$,  changes its sign under the
symmetry transformation. From Eq.~(\ref{eq:parity_t}) it follows that $\upsilon_{\alpha}(t) = -\upsilon_{\alpha}(T-t)$, so that
the average velocity of any Floquet state is again zero, $\bar{\upsilon}_\alpha = 0$.

When all Floquet states are non-transporting, the asymptotic average current (\ref{eq:quantum_as_current})
vanishes \textit{for any initial wave function $\psi(t_0)$ and any choice of starting time} $t_0$
\footnote{Note that heretofore we consider  the case of vanishing quasimomentum $\kappa=0$ only. The symmetry action is more
involved for nonzero quasimomentum, and this case will be addressed later on.}}.
However, the transient, finite-time
current $\tilde{J}(t, t_0) =  1/t \int_{t_0}^t J(\tau; t_0) d\tau$ can still be detected -- even when one of the symmetries is present.
It is the result of the interference between different Floquet states, see the second sum on the rhs of Eq.~(\ref{eq:currentQ2}).
The transient time depends on the values of the terms $\langle
\phi_{\alpha'}(t) |\hat{p} |\phi_{\alpha}(t) \rangle$ and the statistics of the splittings, $\vartriangle E_{\alpha, \alpha'}
= E_{\alpha} - E_{\alpha'}$. The effect can be very long-lasting in the absence of level repulsion as for
the Poisson level-spacing distribution of integrable and near-integrable quantum systems \cite{Stockmann1999}.

Finally, we address the case of $\kappa \neq 0$\footnote{It is noteworthy that there was no need for that in the classical case.
As long as only the averaged current is concerned, it is enough  to solve the Fokker - Planck
equation for the periodic boundary  conditions, i. e. with $\kappa =0$ \cite{r02pr}.} \cite{Zhan2011}.
In this case we deal with the set of Floquet bands $\epsilon_{\alpha}[\kappa]$, which  extend over the Brillouin zone $\kappa \in [-\pi/L, \pi/L]$.
By virtue of the Hellmann-Feynman theorem \cite{Sambe1973},
the average velocity of the state is equal to the local slope of the corresponding Floquet band \cite{sokd01prl},
\begin{equation}\label{Eq:sambe}
\bar{\upsilon}_{\alpha,\kappa} = \hbar^{-1}\frac{\partial E_{\alpha} (\kappa)}{\partial\kappa}.
\end{equation}
The generalization of the basic symmetries is straightforward, and so we get
\begin{equation}
{\hat S}^{\kappa}_x[\hat{\chi},\tau]: \{\hat{x}, \hat{p}, t, \kappa\} \rightarrow \{-\hat{x}+\hat{\chi}, -\hat{p}, t+\tau, -\kappa\}, \label{eq:SKxq}
\end{equation}
and
\begin{equation}
{\hat S}^{\kappa}_t[\hat{\chi},\tau]: \{\hat{x}, \hat{p}, t, \kappa\} \rightarrow \{\hat{x}+\hat{\chi}, -\hat{p}, -t+\tau, -\kappa\}. \label{eq:SKtq}
\end{equation}
In words, both symmetries reverse  quasimomentum values and map every Floquet band into itself, a
negative branch onto a positive one and vice versa, $E_{\alpha}[\kappa] = E_{\alpha}[-\kappa]$. Therefore all Floquet bands
are symmetric at $\kappa = 0$ when at least one of the symmetries holds. From this it  immediately follows that all bands are flat
at  $\kappa = 0$ and the corresponding Floquet states are non-transporting. The symmetries transformations map these states
onto themselves since $\kappa = -\kappa = 0$. However, a Floquet state with $\kappa \neq 0$
may possess a nonzero velocity $\upsilon_{\alpha}[\kappa]$ even in the case of symmetry
(yet there is always a symmetry-related  state with the opposite velocity on the  same band
at $\kappa' = -\kappa$).  Both fundamental symmetries involve time transformations, ${\hat S}^{\kappa}_t$ for sure and
${\hat S}^{\kappa}_x$ in same cases. This means that even though the initial wave packet can be perfectly symmetric $\kappa=0$,
$\psi(t=0;\kappa) = \psi(t=0;-\kappa)$, this fact alone does not guarantee the absence of the asymptotic current in the symmetric cases. The contributions from
the symmetry-related Floquet states to a symmetric wave packet  are not equal in general, $C_{\alpha,-\kappa}(t_0) \neq C_{\alpha,\kappa}(t_0)$.
Only the averaging over the initial phase $t_0$ or the presence of specific symmetry ${\hat S}^{\kappa}_x[\hat{\chi},\tau=0]$
can guarantee the absence of the asymptotic current.

\subsection{Applications: numerical studies}

We start with a quantum version of the rocking ratchet, Section \ref{Additive driving}.
The potential
\begin{equation}
U(x) = 1+\cos(x), \label{eq:QQpotential}
\end{equation}
is driven by the bi-harmonic driving force,
\begin{equation}
E(t;t_0) = E_1 \cos(\omega [t-t_0]) + E_2 \cos(2\omega [t-t_0] + \theta).\label{eq:QQdriving}
\end{equation}
All information on transport properties of the quantum ratchet
is encoded in the eigenpectrum and eigenstates (i. e., Floquet states) of the corresponding Floquet operator, $U(T,t_0)$.
The typical dependence of the quasienergies  on the asymmetry parameter $\theta$ is shown on
Fig.~\ref{Fig:spectrum} \cite{dmfh07}.  The spectrum exhibits two symmetries,
$E_{\alpha}(\theta)=E_{\alpha}(-\theta)$ and
$E_{\alpha}(\theta)=E_{\alpha}(\theta +\pi)$, which are consequences
of the bi-harmonic setup (\ref{eq:QQdriving}). While some bands show strong dependence on
$\theta$, others have rather flat profiles. The former host the Floquet states lying within the potential range, so the state energies
$\langle\langle \phi_{\alpha}|H(t)|\phi_{\alpha}\rangle\rangle_T$ are comparable to $V_0$.
These states are strongly affected by the variations of $\theta$. The latter bands bear Floquet
states which either (i)  have small energies  and are mostly localized  at the bottoms of the potential wells or (ii) have large energies and lie
much above the potential. States of both kinds remain almost unaffected by variations of $\theta$.

Fig.~\ref{Fig:husi} (b-d) shows Husimi distributions
\cite{Husimi1}
of  several Floquet states, whose  quasienergies are marked
by corresponding symbols  in Fig.~\ref{Fig:spectrum}.
\begin{figure}
\begin{center}
\includegraphics[angle=0,width=0.65\textwidth]{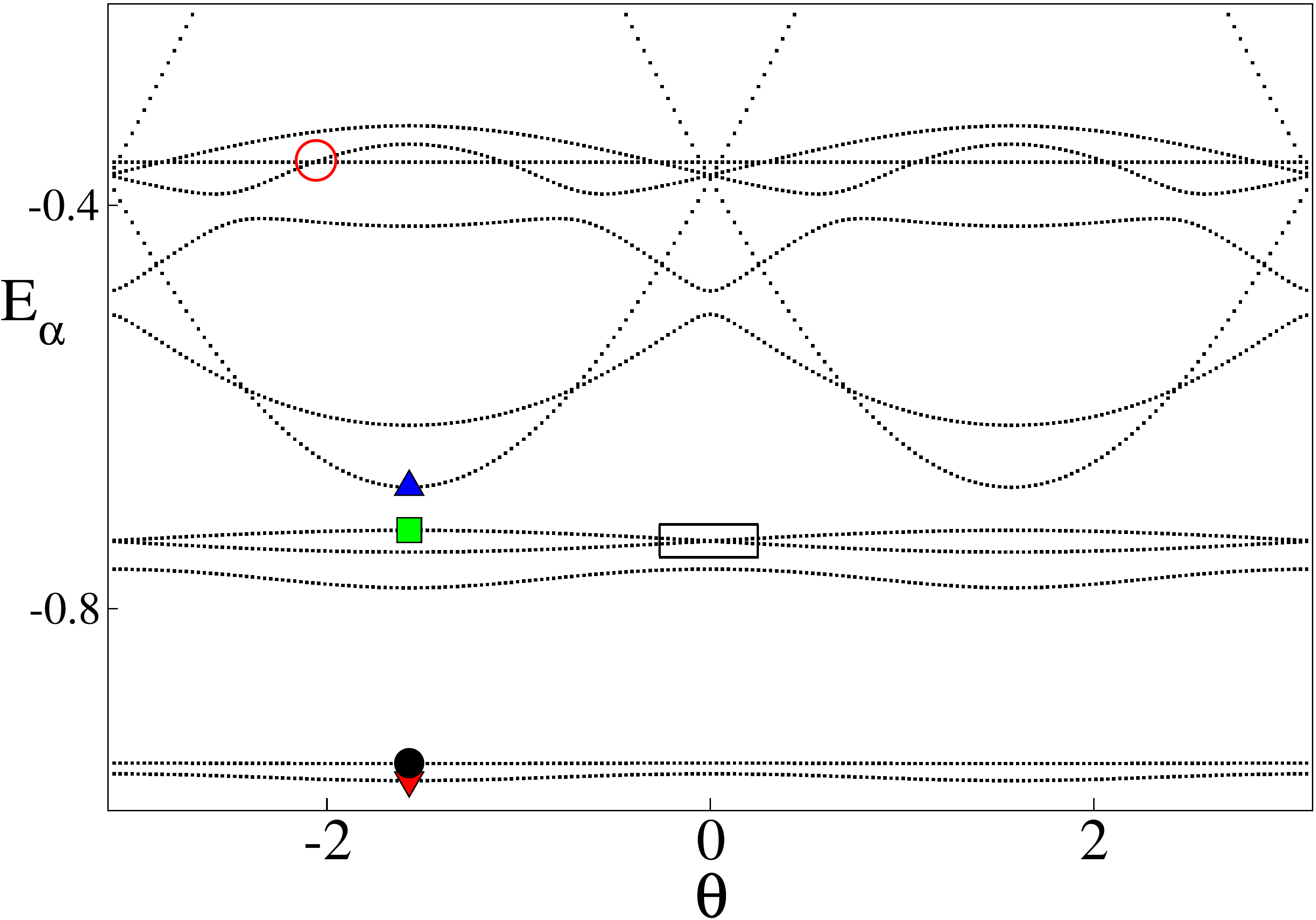}
\caption{(color online) A part of the quasienergy  spectrum  vs $\theta$.
The symbols indicate the corresponding Floquet
states shown in Fig.~\ref{Fig:husi} (b-d). The empty red circle
indicates  an avoided crossing between two
eigenstates. Adapted from Ref. \cite{dmfh07}.}
\label{Fig:spectrum}
\end{center}
\end{figure}
For the symmetric case $\theta=0,\pm\pi$, the set of
Floquet states splits into two non-interacting symmetric and antisymmetric subsets,
depending on the sign of $\sigma_{\alpha}$ in Eq.~(\ref{eq:Sxq}). Symmetric and antisymmetric Floquet bands
alternate in the quasienergy spectrum and are separated by finite gaps.

For $\theta\neq 0,\pm \pi$, all the Floquet states become asymmetric.  Upon deviation from the symmetry points, Floquet states acquire  nonzero
average velocities thus becoming transporting. Now one can detect a nonzero quantum current even by starting with
an initial state of zero velocity. We restrict the consideration
to the initial wave function  in the form of a zero-momentum plane wave,  $|\psi(t_{0})\rangle = |0\rangle
=\frac{1}{\sqrt{2 \pi}}$\footnote{One can use as the initial wave function the Bloch ground state of the undriven potential. Such a choice
is relevant in the context of experiments \cite{Salger2013}.}. This initial state mainly overlaps with
Floquet states of small kinetic energies, which in the quasiclassical limit, $\hbar \ll 1$, have their
Husimi distributions localized in the chaotic layer region of the corresponding classical system. We will
first discuss the results obtained for the current $J(t_0)$, Eq.~(\ref{eq:quantum_as_current}), averaged over
the starting time, $J=1/T \int_{0}^{T}J(t_{0})dt_{0}$, and then address the  dependence of ratchet characteristics on $t_0$.
Fig.~\ref{Fig:current}(a) shows the dependence of the average current on the phase
$\theta$. The average current $J$ shows the expected symmetry properties,
$J(\theta) = -J(\theta+\pi) = -J(-\theta)$, and looks like a smooth curve with several sharp peaks on top of it.
These peaks are produced by avoided crossing (resonances) between two Floquet eigenstates, see the insets on Fig.~\ref{Fig:current}(a).
One state, which we denote $|1 \rangle$, has relatively small kinetic energy and overlaps substantially with the initial wave function $|\psi(t_{0})\rangle$.
The other state, $|2 \rangle$, has large kinetic energy and  overlaps weakly with the initial state off the resonance point.
At the resonance,  these eigenstates `mix' \cite{Timberlake1999}, see Fig.~\ref{Fig:current}(b).
This results in the appearance of a new hybrid shape for the state $|1 \rangle$, which acquires a  tangible velocity now.
Its overlap with the initial state remains mostly unaffected, so  the avoided crossing finally results in a strong current enhancement.
This resonance-like effect can be taken as a quantum analog of the overlap between the chaotic layer
and a ballistic resonance in the phase space of  the classical Hamiltonian ratchet, Fig.~\ref{figHam2}(b).
The width and position of the quantum resonance are tunable.
By changing the amplitude of the second harmonics, $E_2$, it is possible to split the resonance
peak into two  and then move two peaks apart \cite{dmfh07}.

Without additional averaging over the starting time, the asymptotic current depends on $t_0$.
However, the variation of the starting time will affect only the overlap of the Floquet states with the initial wave function.
The  structure of resonance peaks, being determined by avoided crossings  between Floquet bands,
is independent of the parameter $t_0$. On  Fig.~\ref{Fig:temp} we show the asymptotic (however still initial-time dependent) current
versus both $\theta$ and
$t_0$. As expected, the  resonant current enhancement is present for all values of $t_0$ and the variations of the starting time
only  modify the current amplitude.

\begin{figure}
\begin{center}
\includegraphics[angle=0,width=0.75\textwidth]{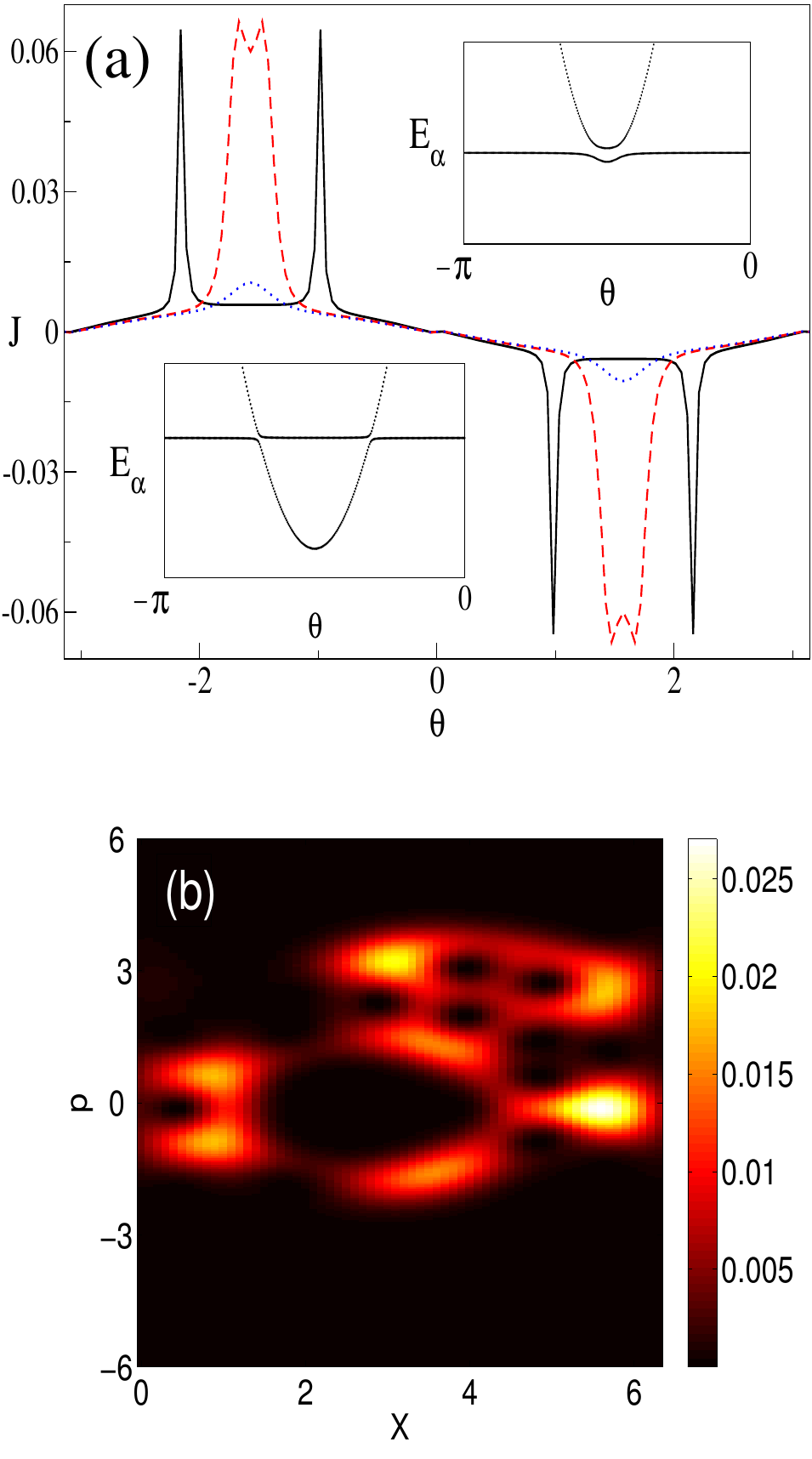}
\caption{ (color online) (a) The average current $J$ (in units of the recoil momentum)
vs $\theta$ for different  values of the second harmonic amplitude,
$E_{2}$: $0.95$ (pointed line), $1$ (dashed line) and $1.2$ (solid
line). Insets: the part of the quasienergy spectrum where the resonance takes place,
for $E_{2}=1$ (top right) and $E_{2}=1.2$ (bottom left).
The parameters are $E_{1}=3.26$ and
$\omega=3$.
(b) Husimi function for the upper eigenstate that appears in (a) (top right inset) with $\theta=-\pi/2$.
Adapted from Ref. \cite{dmfh07}.} \label{Fig:current}
\end{center}
\end{figure}

\begin{figure}
\begin{center}
\includegraphics[angle=0,width=0.65\textwidth]{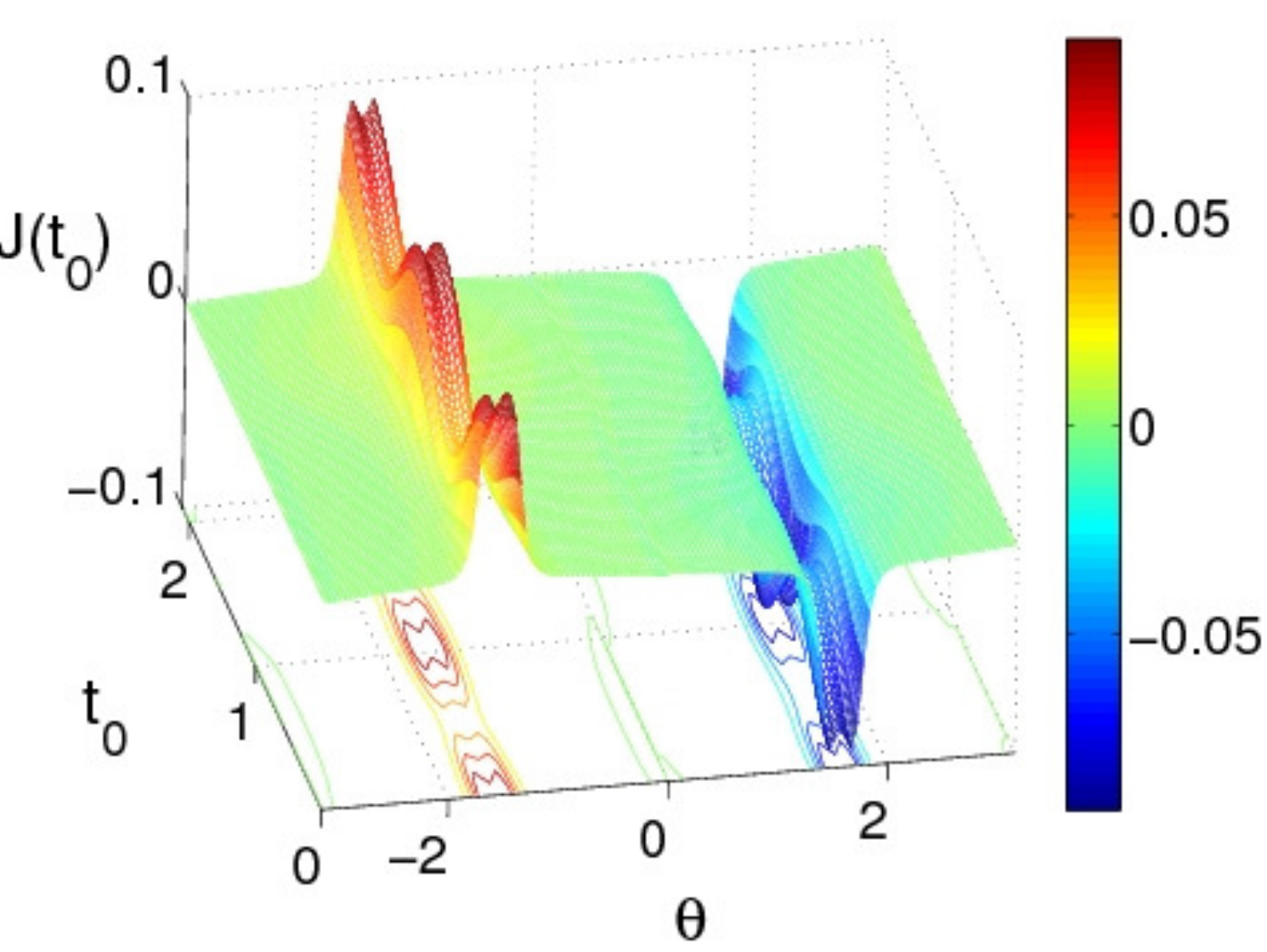}
\caption{(color online) Current dependence on initial time $t_0$ and phase $\theta$.
The  parameters are the same as in Fig.~\ref{Fig:current}. Adapted from Ref. \cite{dmfh07}.}
\label{Fig:temp}
\end{center}
\end{figure}

\begin{figure}
\begin{center}
\includegraphics[angle=0,width=0.65\textwidth]{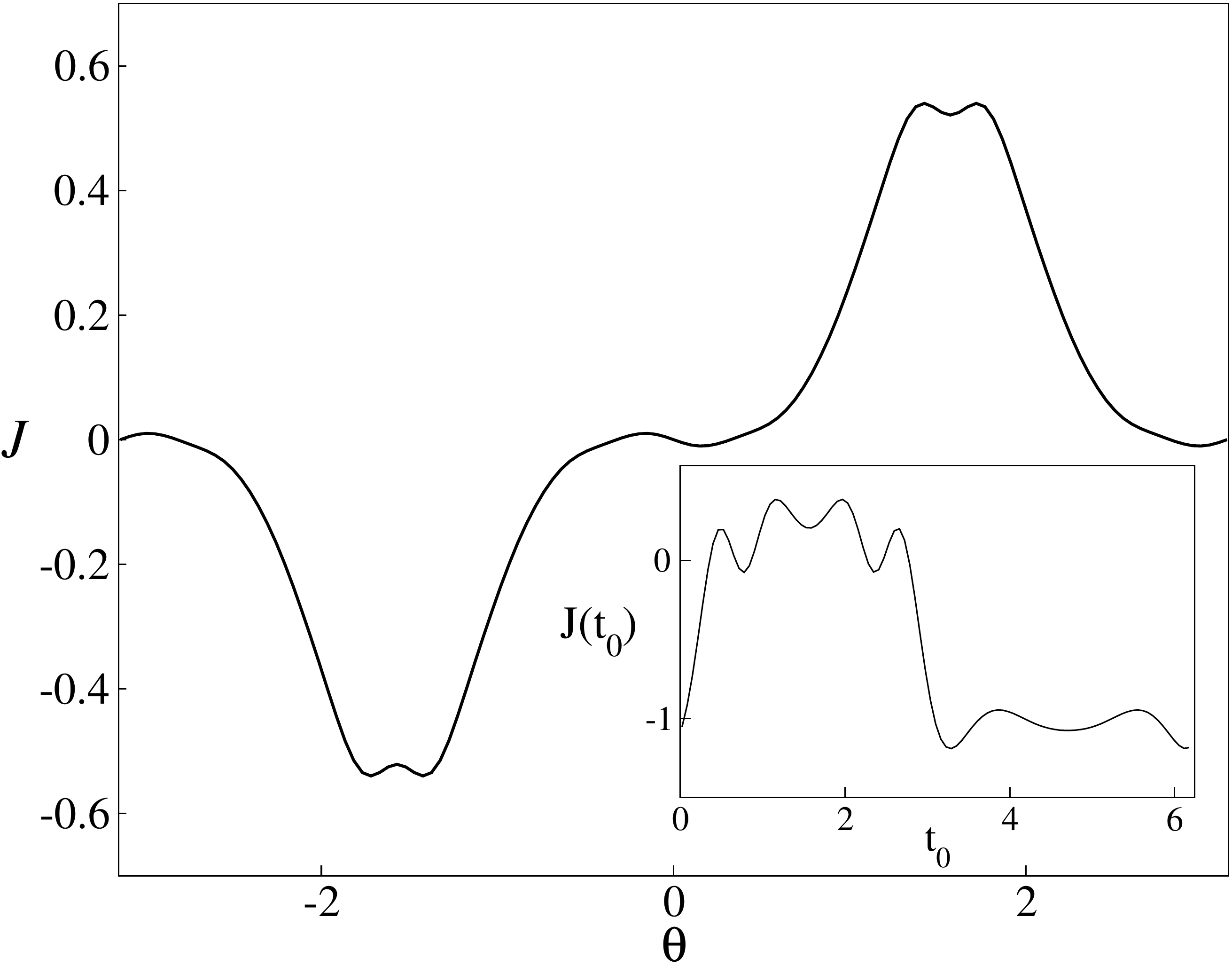}
\caption{The average current $J$ vs $\theta$ for the quantum flashing ratchet,
Eqs.~(\ref{eq:hamil-flashing}, \ref{eq:potential}). Inset:
Asymptotic current as a function of the starting time $t_{0}$ at
$\theta=-\pi/2$. The parameters are $E_{1}=2$, $E_{2}=1.5$,
$\hbar=1$, $\omega=1$, $K=1.5$, $s=0.25$, $\phi=\pi/2$. Adapted from Ref. \cite{dmfh07}.
}
\label{Fig:multiQ}
\end{center}
\end{figure}

Finally, we consider a  quantum ratchet with multiplicative driving \cite{dmfh07},
\begin{equation}\label{eq:hamil-flashing}
H(x,p,t)=\frac{p^{2}}{2}+U(x)E(t-t_{0}),
\end{equation}
where $U(x+L)=U(x)$ and $E(t+T)=E(t)$.
Here, we use the potential
\begin{equation}\label{eq:potential}
U(x)=K\left[\cos(x)+s \cos(2x+\phi)\right],
\end{equation}
where $\theta$ is the parameter which destroys the potential symmetry
away from $\theta = 0, \pm \pi$. The relevant symmetries for
the classical limit of the Hamiltonian (\ref{eq:hamil-flashing}) are
listed in Table 2, Section \ref{Multiplicative driving}.
Neither the bi-harmonic potential nor the bi-harmonic driving alone is sufficient to break
both fundamental symmetries. With a single-harmonic drive  symmetry $\hat{S}_t$ survives while
in the case of single-harmonic potential symmetry $\hat{S}_t$ is still present.
For the above potential we again use the
function (\ref{eq:QQdriving}) for a drive, which choice ensures that for $\phi\neq 0,
\pm\pi$ and $\theta\neq 0, \pm\pi$ all the relevant symmetries
 are violated.

Fig.~\ref{Fig:multiQ}   shows the average current $J$ versus $\theta$ for
the asymmetric potential, $\phi=-\pi/2$, at the deep quantum
limit $\hbar=1$. The current dependence possesses the symmetry
$J(\theta) = -J(\theta+\pi) = -J(-\theta)$,
and similar to the previously considered case
of rocking quantum ratchets, we observe broad quantum resonances and a
dependence of the asymptotic current on the initial time $t_{0}$,
see inset in Fig.~\ref{Fig:multiQ} (a).

\subsection{Experiments with Bose-Einstein condensates}
\label{Experiments with Bose - Einstein condensate}

\begin{figure}
\begin{center}
\includegraphics[angle=0,width=0.85\textwidth]{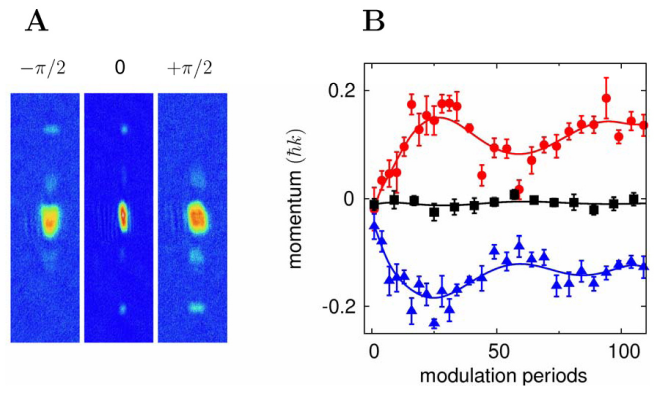}
\caption{(color online) (a) Time-of-flight images showing the atomic velocity distribution after 26 modulation periods
of the ratchet potential for a relative phase between temporal harmonics  $\theta = -\pi/2$ (left),
$\theta = 0$ (middle), and $\theta = \pi/2$ (right);  (b) Mean atomic momentum as a function of
time for $\theta = -\pi/2$ (blue triangles), $\theta = 0$ (black squares) and $\theta = \pi/2$ (red dots).
Adapted from Ref.~\cite{weitzScience}.
}
\label{Fig:weitz1}
\end{center}
\end{figure}

The first realization of a  quantum Hamiltonian ratchet with multiplicative driving was
reported in Ref.~\cite{weitzScience}.
The setup used in the experiments corresponded to the model described by Eqs.~(\ref{eq:hamil-flashing},
\ref{eq:potential}). The $\lambda/2$ spatial component of the optical potential was created by using the
two counter propagating laser beams of wavelength $\lambda$, see Section \ref{Experiments with cold atoms}.
The second optical lattice with period $\lambda/4$ was realized by using dispersive properties of
four-photon Raman transitions between ground state sub-levels of rubidium atoms \cite{rgscw06,Geckeler2007}.
Finally, the temporal modulations were induced by periodically changing the intensity of the beams.
A Bose-Einstein condensate of rubidium atoms was first loaded
into the optical lattice potential, and after finite-time exposition
to the driving (from 26 to 100 periods of the drive function) the atoms were released and freely expanded during  15 $ms$.
After this time,  an absorption image was recorded, see Fig.~\ref{Fig:weitz1} (a).
The sharp diffraction peaks serve as evidence that the evolution of the atomic ensemble was  coherent during the experiment.
The mean velocity of the atomic cloud was calculated as $\bar{v}=\bar{p}/m_{Rb}$, with $\bar{p}=2\hbar k\sum_s s |c_s|^2$,
where $|c_s|^2$ denotes the fraction of atoms in the $s$-th order momentum state, $|2s\hbar k\rangle$.
Time evolution of the mean velocity of the cloud, shown on Fig.~\ref{Fig:weitz1}(b), reveals two important features. First, it  validates
the predictions of the symmetry analysis. Namely, the current is absent when $\theta = 0$, so that symmetry ${\hat S}_t$, Eq.~(\ref{eq:Stq}), holds,
and (ii) inversion of  $\theta$ reverses the current. The detected oscillations of the cloud mean  velocity in time are attributed to the contribution from
the interference between different Floquet states of the driven quantum system, see second term on the rhs of Eq.~(\ref{eq:currentQ2}), thus serving as  another
evidence of the coherence of the system evolution.

The dependencies of the atomic cloud velocity on the  phases $\theta$ and $\phi$ are shown in Fig.~\ref{Fig:weitz2}.
As predicted by the theory, when  $\theta = 0$  or $\phi = 0$, the velocity is nearly zero. The velocity  reaches  maximum values when the values of both phases
are close  to $\pm \pi/2$, at which values  the desymmetrization of Floquet states is expected to be  maximal. An important feature of quantum ratchets, which has been observed before
in numerical studies, is the dependence of the current on the starting time $t_0$. This dependence is another trademark of the coherent quantum
evolution. The absence of decoherence allows the system to maintain memory about the initial shape of the potential so that the latter is imprinted in the asymptotic characteristics.
Fig.~\ref{Fig:weitz3} presents the measured dependence of the cloud velocity on $t_0$, which clearly corroborated this theoretical prediction.
By tuning $t_0$ in the experiment it was possible to more than double the directed transport velocity.  Finally, a resonant-like dependence
of the atomic current on the frequency of the drive has been observed. Namely, a sharp maximum was detected for the value of driving frequency  close to the recoil frequency $\omega_R$
and a smaller resonance peak was observed at $\omega \approx 2\omega_R$. Outside the resonances current value was almost negligible. Altogether the observed effects
reproduced the full spectrum of features predicted for ac-driven quantum ratchets.

\begin{figure}
\begin{center}
\includegraphics[angle=0,width=0.85\textwidth]{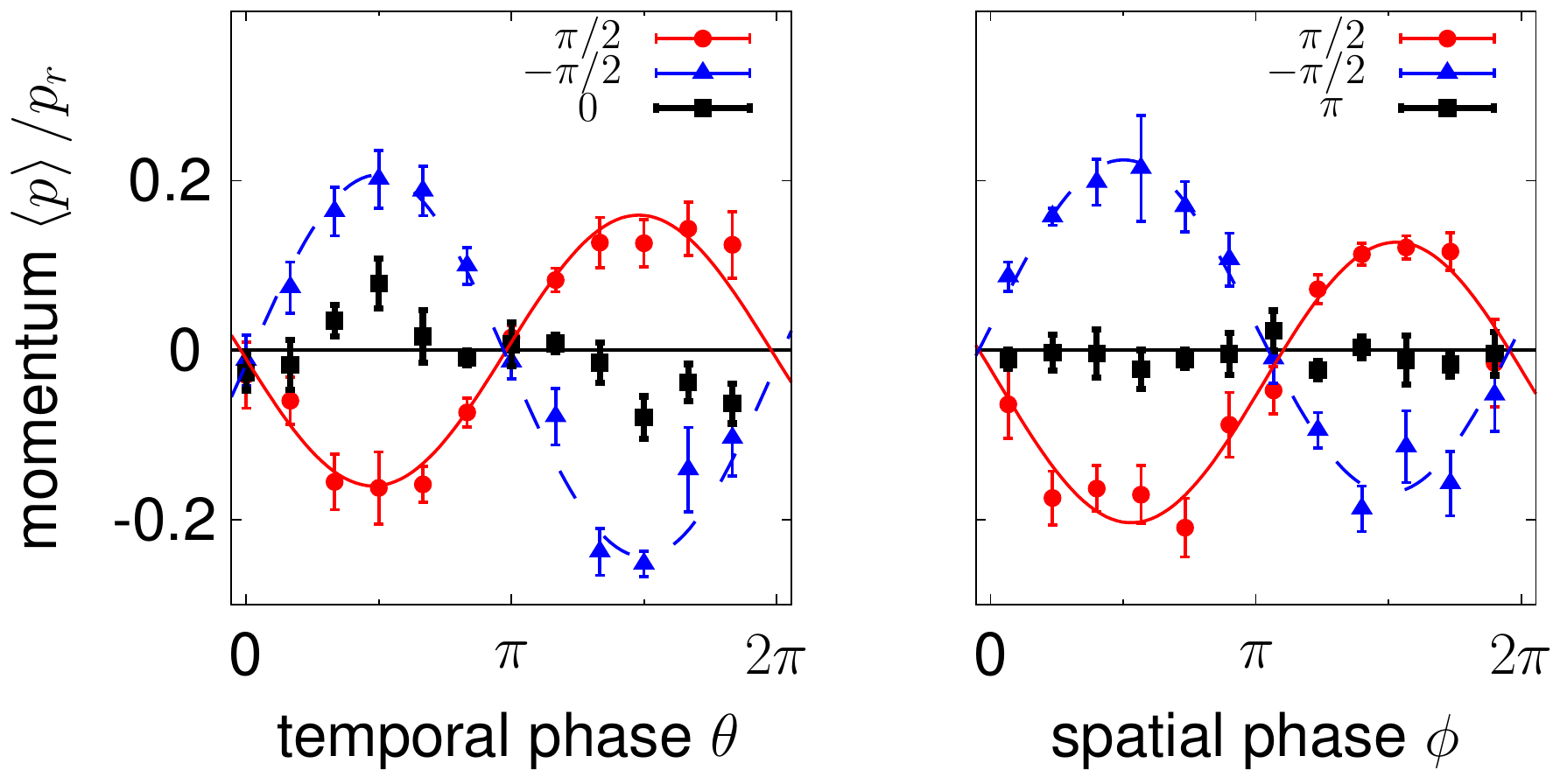}
\caption{(color online) Mean atomic momentum versus  temporal phase $\theta$ (left panel) and spatial phase $\phi$ (right panel).
The values of the complementary phase are indicated in the right corner of each figure.
Courtesy of Martin Weitz.}
\label{Fig:weitz2}
\end{center}
\end{figure}

\begin{figure}
\begin{center}
\includegraphics[angle=0,width=0.65\textwidth]{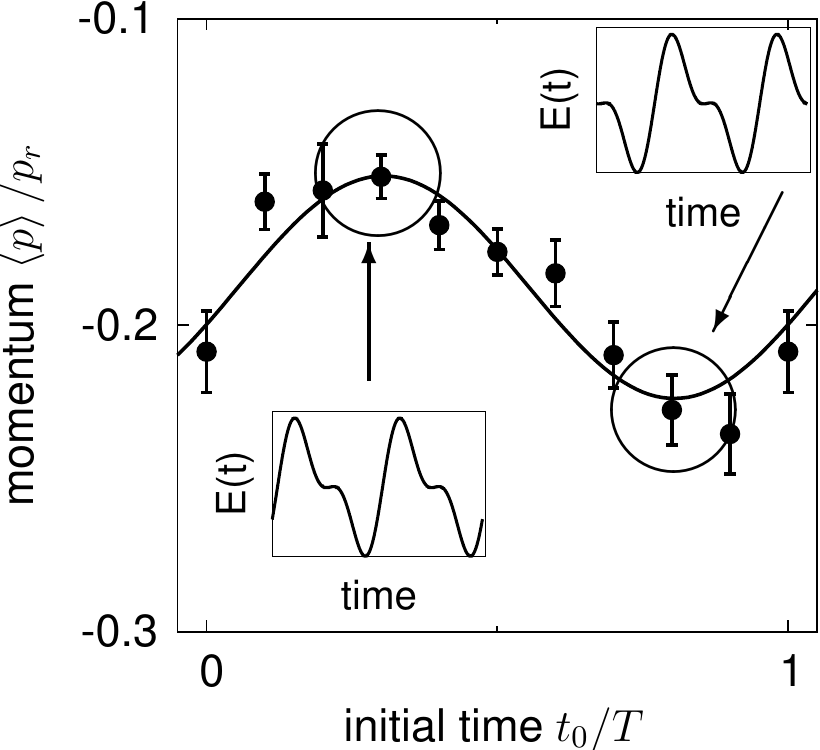}
\caption{Mean  momentum of the BEC cloud as a function of  starting time $t_0$.
Insets show the initial profiles of the driving field.
Adapted from Ref. \cite{weitzScience}.}
\label{Fig:weitz3}
\end{center}
\end{figure}

Finally, a realization of a kicked-rotor version of quantum multiplicative ratchet
with a BEC of Rubidium atoms, was reported in Ref. \cite{Dana2008}.

\section{Extensions and outlook}
\label{sec6}

In this review we focused on classical and quantum Hamiltonian single-particle ratchets and their
realizations with cold and ultracold atomic gases. Many-particle effects and their role in the performance
of classical Hamiltonian ratchets have been studied in Ref.~\cite{Liebchen012}.
To present date not much results on interacting quantum ratchets are collected \cite{creff1}.
Both on the mean-field level of the Gross-Pitaevsky equation \cite{Pitaevskii2003,Giorgini2008}
and on the genuine quantum many-body level \cite{Bloch2008},
we anticipate intriguing novel results \cite{DPoletti2009,creffield2009,motor1}. Interactions between particles may not only  modify the performance of the ratchets qualitatively
but, as well, can also introduce a new  symmetry operation --
\textit{permutation} -- which may  discriminate between fermionic and bosonic ratchets. An additional
symmetry operation may further enhance the overall symmetry and prevent a particle current in situations
where it exists in the single-particle regime. Another interesting property of many body systems is the appearance of
two separately conserved quantities, namely particle number and energy. Both can be assigned with a current.
Ratchets can therefore be designed to promote any of the two current types or even both simultaneously.
In particular, the Mott insulator state \cite{Greiner2002,Greiner2002a}  might prevent a flow of atoms, but allow for a directed energy transport.

As we already discussed in Sec.~\ref{Symmetry analysis_2d}, in higher space dimensions vortex currents can be generated. This effect
mimics a magnetic field acting on a charged particle. The symmetry-based approach could be relevant for the creation of artificial
magnetic fields for neutral and spinless particles, e.g. for atoms \cite{Dalibard2011} or even photons \cite{Fang2012}.
The symmetry analysis has already been used to synthesize non-Abelian gauge fields with rocking two-dimensional optical potentials \cite{Hauke2012}.
The same idea can be implemented within other concepts like the spin-orbit coupling in condensed atoms \cite{Lin2011},
where the coupling strength can be tuned by ac-modulations of the laser intensity \cite{Zhang2013}.
Floquet topological insulators  and the fractional quantum Hall effect \cite{Hasan2010,Kitagawa2010,Lindner2011}
provide further experimental candidates for an implementation of the symmetry analysis.

Notably, many of these essentially quantum effects can also be realized with pure optical setups, by using
engineered networks of  optical waveguides \cite{Onoda2004, Longhi2009,Yin2013}. Different optical realizations of one-dimensional ac-driven quantum ratchets have
been proposed ~\cite{gdf06,longhi09}. In the setup suggested in Ref.~ \cite{longhi09}, the role of time is played by the direction of light propagation
and a time-dependent potential is created by spatial modulations of the refractive index along the paraxial direction. The time evolution of a
`wave-packet' can be visualized by looking into the spatial intensity of the beam inside the slab. An experimental realization of this idea has already been reported  in Ref. \cite{Dreisow2013}. Most interestingly, the analog of time-reversal breaking in an optical waveguide array with periodic modulations of refractive index has been
implemented rather recently for the creation of  photonic Floquet topological insulators \cite{Rechtsman2013}. Finally, optical versions of two-dimensional quantum
ratchets can be built with two-dimensional photonic crystals \cite{fo03,photonic,Boguslawski2012}, cf. Fig.~\ref{fig:Denz}.

An alternative experimental test bed for the ratchet concept with ultracold atoms is provided by the `atomic wire' or the `atomic chip' technology \cite{Fortagh2007}.
A current-carrying wire produces a
magnetic field curling around it, and a potential  minimum of the field can be created above the wire by applying an additional dc magnetic component.
As a result, a pipe-like potential appears on top of the wire which confines atoms in the transverse direction.
The potential  follows the  wire. Upon   slightly meandering this wire it produces a spatially modulated potential.
Next, ac-modulations of the flowing current can create the needed ac-modulations of the potential in the transverse direction. The classical variant of a
ratchet-based magnetic micro-pump was proposed in Ref.~\cite{lade10} where the corresponding symmetry analysis has been presented.

\begin{figure}[t]
\begin{center}
\includegraphics[angle=0,width=0.95\textwidth]{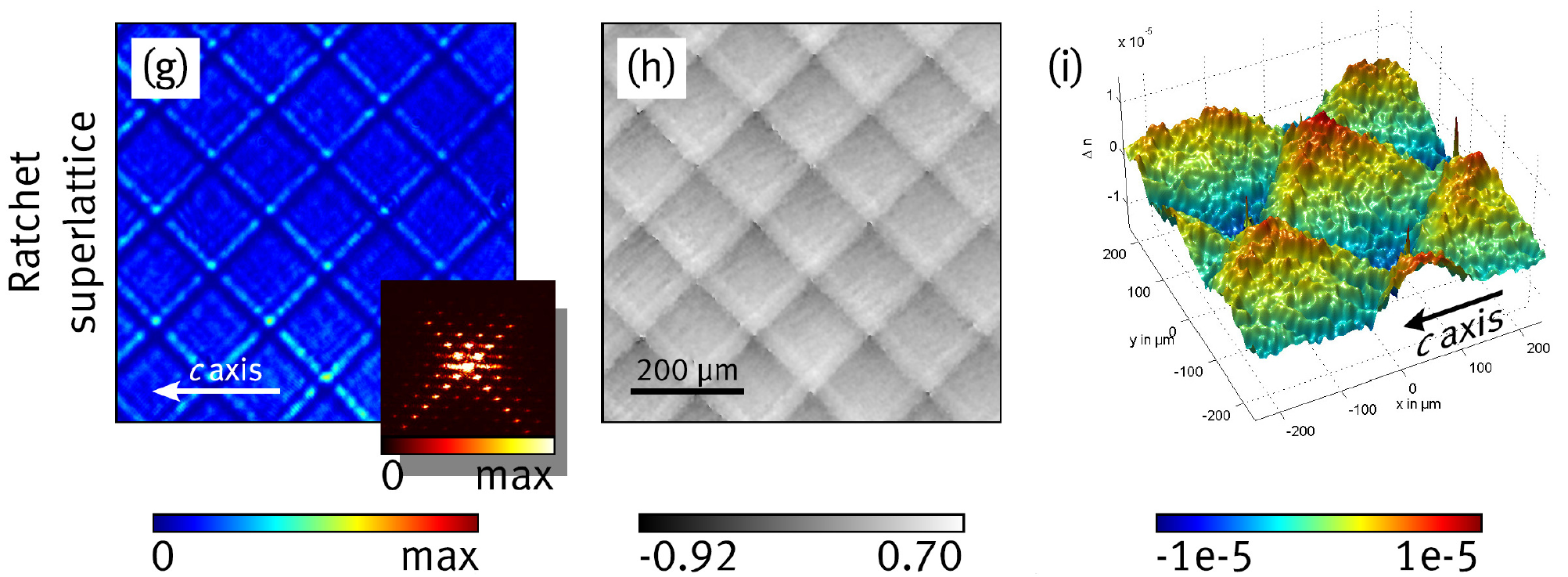}
\caption{(color online) Two-dimensional photonic superlattice with an imprinted ratchet potential.
Right panel shows the corresponding refractive index landscape. Adapted from Ref.~\cite{Boguslawski2012}} \label{fig:Denz}.
\end{center}
\end{figure}

When traversing from classical to quantum ratchets, as detailed  in Sec.~5, the  current had to be re-defined via  the
Schroedinger equation. Other wave equations can be considered as well, and the ratchet concept can be adapted likewise to any other sort of physical current.
For instance, the symmetry analysis was applied to the energy current generated by a nonlinear field equation which
describes  a long Josephson junction \cite{Zolotaryuk2002, Salerno2002, Fistul2003}.
The predicted symmetry breaking was linked to the presence of solitons, and the corresponding  soliton ratchet was subsequently
successfully realized experimentally with a fluxon trapped in an annular Josephson junction which was driven by a microwave field \cite{Ustinov2004}.
A promising  extension of this idea is the realization of soliton ratchets with matter-wave solitons in a dense Bose-Einstein condensate,  in the regime of either attractive
(ratchets with bright solitons) \cite{Poletti2008,Abdullaev2010}
or repulsive (ratchets with dark solitons) \cite{Burger1999,Becker2008} interatomic interactions. A symmetry-based control of
the directed  motion of  domain walls in ferromagnets has been studied in Ref.~\cite{Zolotaryuk2011} by using the Landau-Lifshitz field equation as a model.

\begin{figure}
\begin{center}
\begin{tabular}{cc}
\includegraphics[width=0.4\linewidth,angle=0]{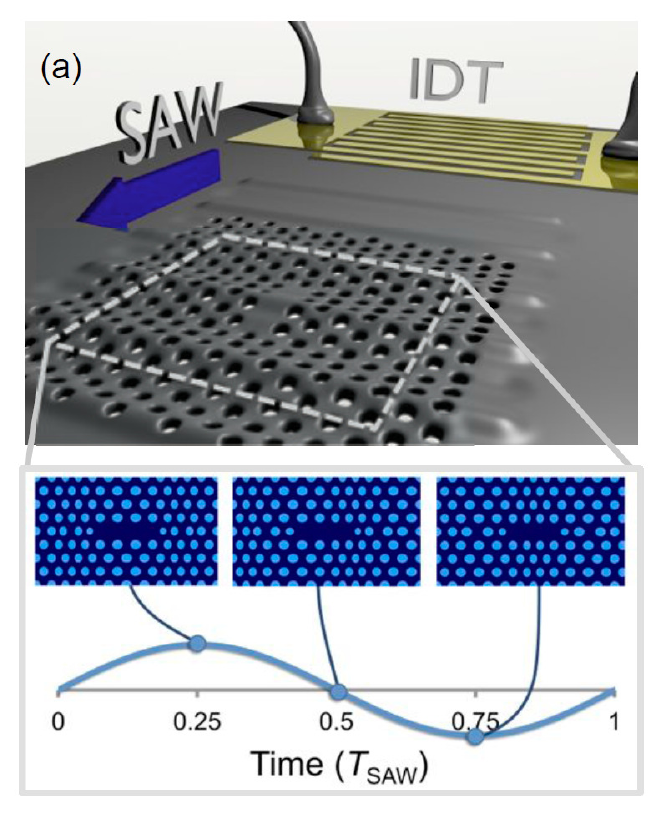}
\includegraphics[width=0.4\linewidth,angle=0]{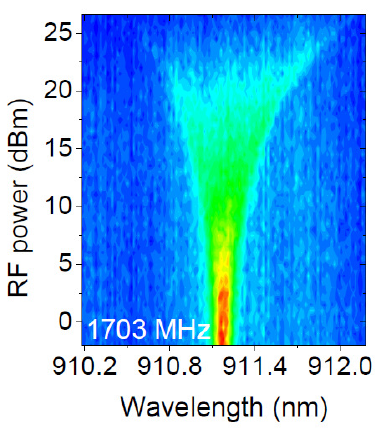}
\end{tabular}
\label{Figure11} \caption{(color online) Left panel: Photonic crystal
nanocavity modulated by a surface acoustic wave. The wave is generated by a radio-frequency
pulse applied to an interdigital transducer. Lower sketch shows deformations  of the nanocavity at
different times during one modulation period. Right panel: Cavity emission spectra as
a function of the pulse power for the driving frequency $1.7$ GHz. Adapted from Ref.~\cite{Fuhrmann2011}.} \label{fig:wir}
\end{center}
\end{figure}

The symmetry approach is not restricted to the case of directed transport only. It  can be also applied to any expectation
value where zeroes in asymptotic regimes emerge that are due to some symmetries, see the general recipe given in Sec.~2.
A proper engineered ac-driving  can then lift such symmetries so that the system yields a nonzero expectation value of a designated
observable.
A good example is the induced transverse magnetization of diluted $s=1/2$ impurity spins
\cite{Arimondo1968}. Here, the applied magnetic field assumed both static and time-periodic components, which however were all perpendicular
to the generated magnetization direction.
The theoretical `prediction' and explanation of this experimental observation  was obtained independently $33$ years later \cite{Flach2001, Flach2002}
\footnote{One of us (SF) is indebted to Ennio Arimondo for a clarifying dinner discussion on the theoretical
work with the final statement ``\textit{Your theory has been experimentally verified 33 years ago}'', and
for the supplying the corresponding reference, Ref.~\cite{Arimondo1968}.}.
Symmetry analysis can also serve some purposes that are not directly related
to a directed transport or polarization of an observable. For example, an interesting application
of Hamiltonian ratchets for patterned deposition of particles was proposed in Refs~\cite{petri2010,liebchen2011}.

A perspective venue for the symmetry analysis is the developing field of tunable metamaterials \cite{Board2010}. The ability to change their properties
in response to the exposition to voltage \cite{Chen2008} and light \cite{Padila2006,Shadirov2012} makes these materials appealing candidates
for applications in many areas. The responses can be tangibly nonlinear \cite{Zheludev2012} and reveal memory effects \cite{Driscoll2009}.
Thus ac-modulations of the controlling field can produce a whole new spectrum of tunable non-equilibrium responses that are absent
in the stationary limit. Recent experiments with photonic crystal nanocavities  driven by surface acoustic waves show how ac modulations indeed  modify the spectrum of the light emitted from the cavity \cite{Fuhrmann2011}, see Fig.~\ref{fig:wir}. The experimental setup resembles the traveling potential ratchet discussed in Sec.~~\ref{Traveling potentials}.
The observed effect of an increasing asymmetry of the spectral broadening with the increase of the driving strength poses a  question about the
origin of this dynamical symmetry-breaking phenomenon. Another  intriguing issue concerns the possibility to observe
the consequences of  bi-harmonic driving
with tunable asymmetry (see in  Eq.~(\ref{e2})) in the emission spectrum of the cavity.

In order to properly address the experimental reality, especially of photonics (see the right panel on Fig.~\ref{fig:Denz}) and atomic chip
technologies \footnote{Diminutive fluctuations in wire's
meandering and small roughness of the wire surface lead to a strong disorder in the created magnetic potential \cite{Fortagh2007}.},  the issue of disorder
has to be taken into account. The role of a static potential disorder in the ac-driven ratchet transport is presently an almost unexplored territory \footnote{Though some steps in this
directions have been made in the classical corner of the research, see Refs.~\cite{Marchesoni1997,Popescu2000,afs-b01pre}.}.
We believe that further progress in this direction, especially in the quantum area, will create an intriguing research field at the interface
between the Anderson localization \cite{Anderson1958,Roati2008,Sanchez-Palencia2010,Segev2013}, Floquet formalism \cite{PRKohler05,Shirley1965,Sambe1973,Grifoni1998}
and quantum chaos \cite{Stockmann1999}.

Finally, the authors share the hope that this  overview will inspire
and invigorate readers to embark on this timely theme and enrich it with pursuing their own research.

\section{Acknowledgements}

It is a pleasure to thank our colleagues and collaborators Yaroslav Zolotaryuk, Oleg Yevtushenko, Ferruccio Renzoni, Martin Weitz,
Sigmund Kohler, Tobias Salger and Christopher Grossert for  helpful discussions, suggestions and comments. We acknowledge the financial support
of the German Research Foundation (DFG) through the grants, HA1517/31-2, HA1517/35, DE1889/1.



\end{document}